\documentclass[a4paper,11pt]{article}
\usepackage{jheppub} 
\usepackage{lineno}
\nolinenumbers

\usepackage{orcidlink}
\usepackage{adjustbox}
\usepackage{multirow}
\usepackage{lmodern}

\allowdisplaybreaks[4]


\arxivnumber{2604.19612} 

\title{\boldmath QCD-factorization amplitudes from flavour symmetries: beyond the $SU(3)$ symmetric case}

\author[a]{Wen-Sheng Fang\,\orcidlink{0009-0001-8411-3875},}
\author[b]{Tobias Huber\,\orcidlink{0000-0002-3851-0116},}
\author[b,1]{Xin-Qiang Li\,\note{Corresponding author.}\orcidlink{0000-0002-3962-3577},}
\author[b,2]{Eleftheria Malami\,\note{Address after September 30, 2025: DAMTP, University of Cambridge, Centre for Mathematical Sciences, Wilberforce Road, CB3 0WA, Cambridge, United Kingdom}\orcidlink{0000-0001-9479-0151},}
\author[b]{and Gilberto~Tetlalmatzi-Xolocotzi\,\orcidlink{0000-0001-8083-0335}}

\affiliation[a]{Institute of Particle Physics and Key Laboratory of Quark and Lepton Physics~(MOE), Central China Normal University, Wuhan, Hubei 430079, China}
\affiliation[b]{Theoretische Physik 1, Center for Particle Physics Siegen (CPPS),  Universit\"at Siegen, Walter-Flex-Stra{\ss}e 3, D-57068 Siegen, Germany}

\emailAdd{1284913895@mails.ccnu.edu.cn}
\emailAdd{huber@physik.uni-siegen.de}
\emailAdd{xqli@mail.ccnu.edu.cn}
\emailAdd{em2088@cam.ac.uk}
\emailAdd{gtx@physik.uni-siegen.de}

\abstract{Using experimental information on branching ratios as well as direct and mixing-induced CP asymmetries, we perform a data-driven analysis of charmless non-leptonic $B \to PP$ decays, where $P$ is any of the light pseudoscalar mesons. Implementing flavour-$SU(3)$ breaking at the level of transition form factors, decay constants and phase space factors, we find a good fit to the current experimental data. Our best-fit point materializes in QCD-factorization amplitudes whose central values resemble many features of the dynamical predictions obtained within the QCD factorization framework. Moreover, we do not find any strong indications that the size of annihilation amplitudes is numerically enhanced beyond the naive $\Lambda_{\textrm{QCD}}/m_b$ scaling. Subsequently, we address a number of phenomenological applications, among which are various flavour puzzles that have been persisting in non-leptonic $B$ decays for quite some time.}

\begin{document}
\maketitle
\flushbottom

\section{Introduction}
\label{sec:intro}

The phenomenon of charge-parity (CP) violation is crucial for the evolution of the Universe into the state in which it is observed today. Moreover, it represents one of the rare instances observed in Nature of a broken discrete symmetry. Up to the present day, the quark flavour sector of the Standard Model (SM) contains the only established source of CP violation in terms of the single phase of the Cabibbo-Kobayashi-Maskawa (CKM) matrix. Here in turn, the charmless non-leptonic decays of $B_{(s)}$ mesons play a prominent role in quantifying the amount of CP violation and --~more in general~-- in investigating the CKM mechanism of quark flavour mixing, due to the interference between various decay topologies. As an important feature, these decays offer numerous observables such as branching ratios, direct and mixing-induced CP asymmetries, polarization fractions, and suitably chosen combinations thereof, which entail a rich and interesting phenomenology, in particular since precision is constantly increasing on both the experimental and the theoretical side~\cite{ParticleDataGroup:2024cfk,HeavyFlavorAveragingGroupHFLAV:2024ctg,FlavourLatticeAveragingGroupFLAG:2024oxs}. 

On the experimental side, the two-body charmless non-leptonic $B_{(s)}$ decays have been central to the programs of the two $B$ factories~\cite{BaBar:2014omp}, Belle~II~\cite{Belle-II:2018jsg} and experiments at hadron colliders~\cite{LHCb:2018roe}. Still, further efforts for new measurements at the upgraded LHC and at Belle~II will be required since some decay channels and observables -- especially CP asymmetries and channels with $\eta^{(\prime)}$ mesons -- have not yet been measured or are only poorly constrained~\cite{ParticleDataGroup:2024cfk,HeavyFlavorAveragingGroupHFLAV:2024ctg}.

On the theoretical side, the charmless non-leptonic $B_{(s)}$ decays are described within the framework of the effective weak Hamiltonian, which gives rise to weak phases from the CKM matrix elements and strong phases from the matrix elements between purely hadronic initial and final states. Whilst the latter are still challenging to compute on the lattice, modern techniques mainly build on the topological diagram decomposition~\cite{Chau:1990ay,Gronau:1994rj,Gronau:1995hn,Chiang:2004nm}, factorization methods~\cite{Beneke:1999br,Beneke:2000ry,Beneke:2001ev,Keum:2000ph,Keum:2000wi,Lu:2000em,Bauer:2000yr,Bauer:2001yt,Beneke:2002ph,Beneke:2002ni}, and the approximate flavour symmetry of the light quarks under strong interaction~\cite{Zeppenfeld:1980ex,Savage:1989ub,Grinstein:1996us}.\footnote{For other alternatives, we refer the readers to refs.~\cite{Khodjamirian:2000mi,Khodjamirian:2002pk,Khodjamirian:2003eq,Khodjamirian:2005wn} for light-cone sum rule (LCSR), and \cite{Cheng:2004ru,Cheng:2005bg,Atwood:1997iw} for long-distance final-state interactions. In the exact $SU(3)$ limit, the topological diagram approach (TDA) has also been shown to be equivalent to the irreducible representation approach~\cite{He:2018php,He:2018joe,Wang:2020gmn}.} Each of these approaches has its benefits and shortcomings. For instance, the QCD factorization (QCDF) approach~\cite{Beneke:1999br,Beneke:2000ry,Beneke:2001ev,Beneke:2003zv} or the soft-collinear effective theory (SCET)~\cite{Bauer:2000yr,Bauer:2001yt,Beneke:2002ph,Beneke:2002ni}, which is the field theoretical formulation of the former, consistently disentangles short- from long-distance physics in the heavy quark limit and as such allows to include higher-order perturbative corrections to the hard scattering kernels~\cite{Beneke:2005vv,Kivel:2006xc,Pilipp:2007mg,Beneke:2006mk,Jain:2007dy,Li:2005wx,Li:2006jb,Bell:2007tv,Bell:2009nk,Beneke:2009ek,Kim:2011jm,Bell:2014zya,Bell:2015koa,Huber:2015bva,Huber:2016xod,Bell:2020qus,Beneke:2020vnb,Beneke:2021jhp}. On the other hand, no consistent factorization framework that allows to compute corrections at sub-leading power in $\Lambda_{\rm QCD}/m_b$ from first field-theoretical principles could be achieved so far (for recent progress in non-leptonic decays see ref.~\cite{talkBoeerCKM2025}, and for conceptual work on refactorization in SCET$_{\textrm{II}}$ see refs.~\cite{Beneke:2018wjp,Liu:2018czl,Beneke:2019slt,Liu:2019oav,Liu:2020tzd,Liu:2020wbn,Bell:2022ott,Feldmann:2022ixt,Liu:2022ajh,Cornella:2022ubo,Hurth:2023paz,Bell:2024bxg,Delto:2025epy,Cornella:2026zkd}). One of the advantages of flavour-$SU(3)$ or its subgroups is the fact that it makes basically no assumptions about the various scales of QCD effects that play a role in non-leptonic decays. However, flavour $SU(3)$ and its $SU(2)$ subgroups of U- and V-spin (corresponding to the exchange of $d\leftrightarrow s$ and $u \leftrightarrow s$, respectively) are rather badly broken by the splitting between the up/down and strange quark masses. While certain progress has been made in quantifying the amount of flavour breaking~\cite{Gronau:1995hm,Jung:2009pb,Cheng:2012xb,Grossman:2012ry,Muller:2015lua}, the situation cannot be regarded as resolved and a rigorous implementation is still lacking.

In the field of two-body non-leptonic $B_{(s)}$ decays, people have over the years come up with a number of observables that are rather clean on the theoretical and experimental sides, some of which are in colour-allowed tree-level decays to open-charm final states~\cite{Beneke:2000ry,Huber:2016xod,Bjorken:1988kk,Neubert:1997uc}. In charmless final states, the most prominent ones are in the $K\pi$ channels, notably the difference $\Delta A_{CP}$~\cite{Gronau:2005kz} of the direct CP asymmetries which still persists as the so-called $K\pi$ puzzle (see \textit{e.g.} refs.~\cite{Buras:2003yc,Chiang:2004nm,Mishima:2004um,Cheng:2004ru,Fleischer:2007mq,Bell:2015koa,Fleischer:2018bld}), and the null-test sum rule~\cite{Gronau:1998ep,Gronau:2005kz,Bell:2015koa} which combines several $B \to K\pi$ branching ratios and direct CP asymmetries. While the latter agrees with its SM prediction~\cite{Belle-II:2023ksq,Bell:2015koa}, there are certain tensions between theoretical predictions and experimental measurements, for instance in the branching ratios of $\bar{B}_{(s)}^0 \to D^{(\ast)+}_{(s)} \{K^-,\pi^-\}$ decays~\cite{Huber:2016xod,Bordone:2020gao,Iguro:2020ndk,Cai:2021mlt,Bordone:2021cca,Fleischer:2021cct,Fleischer:2021cwb,Piscopo:2023opf,Meiser:2024zea,Atkinson:2024hqp} as well as in the observable $\Delta A_{CP}$ that has been persisting for more than a decade. Moreover, the decay-width ratio $\Gamma(\bar{B}^0 \to K^+K^-)/\Gamma(\bar{B}^0_{s} \to \pi^+\pi^-)$ between pure annihilation channels seems to be anomalous~\cite{Wang:2013fya,Chang:2014yma,Bobeth:2014rra}, but is harder to assess theoretically, at least in the QCDF approach.

A meaningful method to shed light on the various flavour puzzles are (global) fits to data, which have been performed rather extensively in the past (see \textit{e.g.} refs.~\cite{Huber:2021cgk,Berthiaume:2023kmp,Bhattacharya:2025wcq,Bhattacharya:2025rrv,BurgosMarcos:2025xja}), under various assumptions and with rather different outcomes. The present work aims at determining the QCDF amplitudes from a fit to all the available $B\to PP$ data ($P$ is a charmless pseudoscalar meson), with the goal of accessing the size of annihilation amplitudes and of obtaining a comprehensive picture about whether experimental data can be explained in the QCDF parametrization. It goes beyond our previous work~\cite{Huber:2021cgk} which was done in the $SU(3)$ limit by implementing the $SU(3)$ breaking at the level of transition form factors, decay constants and phase space. It also goes beyond ref.~\cite{BurgosMarcos:2025xja} in that we include channels with $\eta^{(\prime)}$ mesons in the flavour-broken scenario. Moreover, we fit to a different linear combination of amplitudes that we find more suitable for disentangling individual QCDF amplitudes in view of information from dynamics at the next-to-leading order (NLO) in $\alpha_s$~\cite{Beneke:2001ev,Beneke:2003zv}. Let us anticipate already here that in our setup we find a very good fit to data, which is in line with ref.~\cite{BurgosMarcos:2025xja} but differs from recent findings by other groups (see \textit{e.g.} refs.~\cite{Berthiaume:2023kmp,Bhattacharya:2025wcq,Bhattacharya:2025rrv}). In part of the latter references the number of independent amplitudes got reduced via the use of electroweak-penguin-tree relations. We argue that these relations do not withstand quantum or power corrections (see the derivation in~\cite{Shi:2025eyp}), and therefore advocate to use the complete set of independent amplitudes for phenomenology as identified at several places in the literature~\cite{He:2018php,beneketalk,Shi:2025eyp}. As part of our phenomenological study, we will put under scrutiny the electroweak-penguin-tree relations based on our fit results. We must also emphasize that, to easily see the equivalence between the TDA and QCDF parametrizations, we should have the same starting point, \textit{i.e.}, starting from the effective weak Hamiltonian.

This paper is organized as follows. We start with our theoretical framework in section~\ref{sec:thframework}, and present the physical amplitudes and observables in section~\ref{sec:physampsandobservables}. The fit to experimental data is then described in section~\ref{sec:fit}. We present our fitting results in section~\ref{sec:results} and some relevant phenomenological discussions in section~\ref{sec:pheno}. Finally, we conclude in section~\ref{sec:conclusion}. Explicit expressions for the decay amplitudes in terms of the amplitude coefficients are relegated to appendix~\ref{app:ampcoeffs}.

\section{Theoretical framework}
\label{sec:thframework}

\subsection{Effective weak Hamiltonian}

The effective weak Hamiltonian for charmless non-leptonic $B_{(s)}$ decays consists of a sum of local operators $Q_i$ multiplied by the short-distance Wilson coefficients $C_i$ and products of the CKM matrix elements $\lambda^{(D)}_{p}=V_{pb} V_{pD}^\ast$, with $D=d, s$ and $p=u, c, t$. Depending on the decay channels considered, we can decompose the effective weak Hamiltonian into two parts, $\Delta S=1$ and $\Delta S=0$, with each part being composed of the dimension-six four-quark operators $Q_{1}, \ldots, Q_{10}$.\footnote{Here we do not write down the electromagnetic ($Q_{7\gamma}$) and chromomagnetic ($Q_{8g}$) dipole operators explicitly, as their contributions can be absorbed into the penguin sector~\cite{Beneke:2003zv}.} For the $\Delta S=1$ part, we have~\cite{Buchalla:1995vs}
\begin{equation} \label{eq:Heff}
{\cal H}_{\rm eff}^{\Delta S=1}=\frac{G_{F}}{\sqrt{2}}\left[\lambda^{(s)}_{u}
\left(C_{1}Q^{(u)}_{1}+C_{2}Q^{(u)}_{2}\right)+
\lambda^{(s)}_{c}\left(C_{1}Q^{(c)}_{1}+C_{2}Q^{(c)}_{2}\right)-
\lambda^{(s)}_{t}\sum_{i=3}^{10}C_{i}Q_{i}\right] + \text{h.c.},
\end{equation} 
where $Q_{1}$ and $Q_{2}$ are the ``current-current'' or ``tree'' operators, and they are given, respectively, by
\begin{align}
Q_{1}^{(u)} & = (\overline{s}_{i}u_{j})_{V-A}
(\overline{u}_{j}b_{i})_{V-A}, &
Q_{1}^{(c)} & = (\overline{s}_{i}c_{j})_{V-A}
(\overline{c}_{j}b_{i})_{V-A}, \nonumber \\
Q_{2}^{(u)} &= (\overline{s}_{i}u_{i})_{V-A}
(\overline{u}_{j}b_{j})_{V-A}, &
Q_{2}^{(c)} &= (\overline{s}_{i}c_{i})_{V-A}
(\overline{c}_{j}b_{j})_{V-A},
\end{align}
while $Q_{3},\ldots, Q_{6}$ are the ``QCD penguin'' operators, with
\begin{align}
Q_{3} & = (\overline{s}_{i}b_{i})_{V-A}
\sum_{q=u,d,s,c,b}(\overline{q}_{j}q_{j})_{V-A}, &
Q_{4} & = (\overline{s}_{i}b_{j})_{V-A} 
\sum_{q=u,d,s,c,b}(\overline{q}_{j}q_{i})_{V-A}, \nonumber \\
Q_{5} & = (\overline{s}_{i}b_{i})_{V-A}
\sum_{q=u,d,s,c,b}(\overline{q}_{j}q_{j})_{V+A}, &
Q_{6} & = (\overline{s}_{i}b_{j})_{V-A} 
\sum_{q=u,d,s,c,b}(\overline{q}_{j}q_{i})_{V+A}.
\end{align}
The four ``electroweak penguin (EWP)'' operators $Q_{7},\ldots, Q_{10}$ are given, respectively, by
\begin{align}
Q_{7} & = \frac{3}{2} (\overline{s}_{i}b_{i})_{V-A}
\sum_{q=u,d,s,c,b}e_{q}(\overline{q}_{j}q_{j})_{V+A}, &
Q_{8} & = \frac{3}{2} (\overline{s}_{i}b_{j})_{V-A} 
\sum_{q=u,d,s,c,b}e_{q}(\overline{q}_{j}q_{i})_{V+A}, \nonumber \\
Q_{9} & = \frac{3}{2} (\overline{s}_{i}b_{i})_{V-A}
\sum_{q=u,d,s,c,b}e_{q}(\overline{q}_{j}q_{j})_{V-A}, &
Q_{10} & = \frac{3}{2} (\overline{s}_{i}b_{j})_{V-A} 
\sum_{q=u,d,s,c,b}e_{q}(\overline{q}_{j}q_{i})_{V-A}.
\end{align}
Here $(\overline{\alpha}_i\beta_{i(j)})_{V \pm A}=\overline{\alpha}_i\gamma_\mu(1 \pm \gamma_5)\beta_{i(j)}$, and $i,j$ are the $SU(3)$ colour indices; $e_q$ is the electric charge of the quark $q$, given in terms of that of the positron. 

The effective weak Hamiltonian for $\Delta S=0$ part can be obtained from eq.~\eqref{eq:Heff} with the replacement $s\to d$. Usually, we make use of the unitarity relations of the CKM matrix elements, $\lambda_{u}^{(D)}+\lambda_{c}^{(D)}+\lambda_{t}^{(D)}=0$, to simplify eq.~\eqref{eq:Heff}, which results in the so-called ``$\lambda_u-\lambda_t$'' and ``$\lambda_u-\lambda_c$'' conventions. 

\subsection{Amplitudes in the topological diagram approach}

The amplitudes for $\bar{B} \to M_1 M_2$ decays can be compactly expressed in terms of traces over flavour matrices. To this end, let us collect the three $B$-meson states into a row vector $B=\left(B^-, \bar{B}^0, \bar{B}_s\right)$, and represent the final-state pseudoscalar mesons by the matrix
\begin{equation}
M=\left(\begin{array}{ccc}
\frac{\pi^0}{\sqrt{2}}+\frac{\eta_q}{\sqrt{2}}+\frac{\eta_q^{\prime}}{\sqrt{2}} & \pi^{-} & K^{-} \\
\pi^{+} & -\frac{\pi^0}{\sqrt{2}}+\frac{\eta_q}{\sqrt{2}}+\frac{\eta_q^{\prime}}{\sqrt{2}} & \bar{K}^0 \\
K^{+} & K^0 & \eta_s+\eta_s^{\prime}
\end{array}\right)\,,
\end{equation}
where $\eta_q=(u \bar{u}+d \bar{d})/\sqrt{2}$ and $\eta_s=s \bar{s}$ in the quark-flavour basis, and they mix into the physical mesons $\eta$ and $\eta^\prime$.

Following the same treatment as in refs.~\cite{He:2018php,Shi:2025eyp}, we decompose the EWP operators $Q_{7,\ldots, 10}$ according to the following rule:
\begin{align}
    (\bar{D}b)_{V-A} \sum_{q=u,d,s,c,b}e_{q}(\overline{q}_{j}q_{j})_{V\pm A} = (\bar{D}b)_{V-A} (\bar{u}u)_{V\pm A} - \frac{1}{3} (\bar{D}b)_{V-A} \sum_{q=u,d,s,c,b}(\overline{q}_{j}q_{j})_{V\pm A}\,,
\end{align}
where the first term leads to the tree-like penguin topologies denoted by $P_T$, $P_C$, $P_{TA}$, $P_{TE}$, $P_{AS}$ and $P_{ES}$ to be introduced later, while the second one can be incorporated into the QCD penguin sector. In full generality, we can then parameterize the amplitudes of two-body charmless $B$-meson decays as 
\begin{equation} \label{eq:AmpTDA}
  \mathcal{A}^{\mathrm{TDA}} = i\,\frac{G_F}{\sqrt{2}} \left[\mathcal{T}^{\mathrm{TDA}} + \mathcal{P}^{\mathrm{TDA}}\right],
\end{equation}
with the tree sector given by
\begin{align}
\mathcal{T}^{\mathrm{TDA}} =\;&
T\, B_i (M_1)^i_j \, \bar{H}^{jl}_k (M_2)^k_l
+ C\, B_i (M_1)^i_j \, \bar{H}^{lj}_k (M_2)^k_l
+ A\, B_i \, \bar{H}^{il}_j (M_1)^j_k (M_2)^k_l
\nonumber\\
&+ E\, B_i \, \bar{H}^{li}_j (M_1)^j_k (M_2)^k_l
+ T_{ES}\, B_i \, \bar{H}^{ij}_l (M_1)^l_j (M_2)^k_k
+ T_{AS}\, B_i \, \bar{H}^{ji}_l (M_1)^l_j (M_2)^k_k
\nonumber\\
&+ T_S\, B_i (M_1)^i_j \, \bar{H}^{lj}_l (M_2)^k_k
+ T_{PA}\, B_i \, \bar{H}^{li}_l (M_1)^j_k (M_2)^k_j
+ T_{P}\, B_i (M_1)^i_j (M_2)^j_k \, \bar{H}^{lk}_l
\nonumber\\
&+ T_{SS}\, B_i \, \bar{H}^{li}_l (M_1)^j_j (M_2)^k_k + (M_1 \leftrightarrow M_2)\,,
\end{align}
and the penguin sector by
\begin{align}
\mathcal{P}^{\mathrm{TDA}} =\;&
P\, B_i (M_1)^i_j (M_2)^j_k \, \tilde{H}^k
+ P_T\, B_i (M_1)^i_j \, \tilde{H}^{jl}_k (M_2)^k_l
+ S\, B_i (M_1)^i_j \, \tilde{H}^j (M_2)^k_k
\nonumber\\
&+ P_C\, B_i (M_1)^i_j \, \tilde{H}^{lj}_k (M_2)^k_l
+ P_{TA}\, B_i \, \tilde{H}^{il}_j (M_1)^j_k (M_2)^k_l
+ P_A\, B_i \, \tilde{H}^i (M_1)^j_k (M_2)^k_j
\nonumber\\
&+ P_{TE}\, B_i \, \tilde{H}^{ji}_k (M_1)^k_l (M_2)^l_j
+ P_{AS}\, B_i \, \tilde{H}^{ji}_l (M_1)^l_j (M_2)^k_k
+ P_{SS}\, B_i \, \tilde{H}^i (M_1)^j_j (M_2)^k_k
\nonumber\\
&+ P_{ES}\, B_i \, \tilde{H}^{ij}_l (M_1)^l_j (M_2)^k_k + (M_1 \leftrightarrow M_2)\,.
\end{align}
Here the non-zero components of $\bar{H}$ and $\tilde{H}$ are given, respectively, by
\begin{equation}
\begin{aligned}
\bar{H}_{1}^{12} &= \lambda_{u}^{(d)}, &
\bar{H}_{1}^{13} &= \lambda_{u}^{(s)}, \\
\tilde{H}_{1}^{12} & = -\lambda_{t}^{(d)}, &
\tilde{H}_{1}^{13} & = -\lambda_{t}^{(s)},
\end{aligned}
\qquad\qquad
\begin{aligned}
\bar{H}^{2} &= \lambda_{u}^{(d)}, &
\bar{H}^{3} &= \lambda_{u}^{(s)}, \\
\tilde{H}^{2} & = -\lambda_{t}^{(d)}, &
\tilde{H}^{3} & = -\lambda_{t}^{(s)},
\end{aligned}
\end{equation}
with the definitions
\begin{equation}
  \lambda_{u}^{(D)} = V_{ub}\, V_{uD}^{*}, \qquad \lambda_{t}^{(D)} = V_{tb}\, V_{tD}^{*}
\end{equation}
for $D=d, s$. All the decay dynamics is then encoded by the ten tree-like ($T$, $C$, $A$, $E$, $T_{ES}$, $T_{AS}$, $T_S$, $T_{PA}$, $T_P$, and $T_{SS}$) and the ten penguin-like topological amplitudes ($P$, $P_T$, $S$, $P_C$, $P_{TA}$, $P_{A}$, $P_{TE}$, $P_{AS}$, $P_{SS}$, and $P_{ES}$), all of which have to be fitted to the available experimental data. Note that in the exact $SU(3)$ flavour symmetry limit, these topological amplitudes are symmetric under the permutation of the final-state mesons. As the symmetry is already broken by the splitting between the up/down and strange quark masses, here we do not treat them as symmetric but rather consider them to be dependent on the ordering of the final-state mesons $M_1$ and $M_2$, which provides us an easy way to quantify the amount of flavour-$SU(3)$ breaking effects in terms of transition form factors and decay constants. As will be argued in section~\ref{sec:SU(3)-breaking-effects}, we think that this captures the dominant $SU(3)$ breaking effects in these charmless $B\to PP$ decays.

\subsection{Connections to the QCD factorization results}

In the QCDF framework, we can write the amplitudes for two-body charmless $\bar{B} \to M_1 M_2$ decays as~\cite{Beneke:2003zv}
\begin{equation} \label{eq:amp_QCDF}
\begin{aligned}
\mathcal{A}^{\mathrm{QCDF}} = & i \frac{G_F}{\sqrt{2}} \sum_{p=u, c} A_{M_1 M_2}\biggl\{B M_1\left(\alpha_1 \delta_{p u} \hat{U}_p+\alpha_4^p \hat{I}+\alpha_{4, \mathrm{EW}}^p \hat{Q}\right) M_2 \Lambda_p \\
& +B M_1 \Lambda_p \cdot \operatorname{Tr}\left[\left(\alpha_2 \delta_{p u} \hat{U}_p+\alpha_3^p \hat{I}+\alpha_{3, \mathrm{EW}}^p \hat{Q}\right) M_2\right] \\
& +B\left(\beta_2 \delta_{p u} \hat{U}_p+\beta_3^p \hat{I}+\beta_{3, \mathrm{EW}}^p \hat{Q}\right) M_1 M_2 \Lambda_p \\
& +B \Lambda_p \cdot \operatorname{Tr}\left[\left(\beta_1 \delta_{p u} \hat{U}_p+\beta_4^p \hat{I}+\beta_{4, \mathrm{EW}}^p \hat{Q}\right) M_1 M_2\right] \\
& +B\left(\beta_{S 2} \delta_{p u} \hat{U}_p+\beta_{S 3}^p \hat{I}+\beta_{S3, \mathrm{EW}}^p \hat{Q}\right) M_1 \Lambda_p \cdot \operatorname{Tr} M_2 \\
& +B \Lambda_p \cdot \operatorname{Tr}\left[\left(\beta_{S 1} \delta_{p u} \hat{U}_p+\beta_{S 4}^p \hat{I}+\beta_{S 4, \mathrm{EW}}^p \hat{Q}\right) M_1\right] \cdot \operatorname{Tr} M_2\biggr\} + (M_1 \leftrightarrow M_2)\,,
\end{aligned}
\end{equation}
where terms proportional to the $3\times 3$ matrices $\hat{U}_p$, $\hat{I}$, and $\hat{Q}$ are from the current–current, the QCD penguin, and the EWP operators, respectively. Here, $\hat{I}$ is the unit matrix, $\hat{Q}=\mathrm{diag}(1,-1/2,-1/2)$, $\hat{U}_p$ has entries $(\hat{U}_p)_{ij} = \delta_{pu} \delta_{i1} \delta_{j1}$, and $\Lambda_p=(0,\lambda_p^{(d)},\lambda_p^{(s)})^T$~\cite{Beneke:2003zv}. 

In full generality, the QCDF parameters $\alpha_i$ and $\beta_i$ depend on the initial- and final-state mesons through \textit{e.g.} the light-cone distribution amplitudes of these mesons~\cite{Beneke:2001ev,Beneke:2003zv}. For simplicity, however, we will ignore this dependence and assume that $\alpha_i$ and $b_i$ (see \eqref{eq:bibetai} below) are all universal for different decay modes. It is also seen from eq.~\eqref{eq:amp_QCDF} that the parameters $\alpha_4^p$, $\beta_3^p$ as well as $\alpha_3^p$, $\beta_{S3}^p$ can always be grouped together due to their identical matrix structures. Furthermore, as the charge matrix $\hat{Q}$ can be further decomposed as
\begin{equation}
\hat{Q}=\frac{3}{2} \hat{U}_u-\frac{1}{2} \hat{I}\,,
\end{equation}
we can group the electroweak parameters associated with the CKM factor $\lambda_u^{(D)}$ into the tree parameters, while keeping their $\lambda_c^{(D)}$ counterparts as free parameters. In this way, we can see that there are a total of twenty independent complex parameters in the QCDF parametrization, same as what is observed in the TDA parametrization. This in turn demonstrates the equivalence between the two parametrizations~\cite{beneketalk,He:2018joe,Huber:2021cgk}.

The parameters $\beta_i$ in eq.~\eqref{eq:amp_QCDF} represent contributions from the exchange and annihilation topologies. As the prefactor $A_{M_1 M_2}=(m_B^2-m_{M_1}^2) F_0^{B\to M_1}(m_{M_2}^2)f_{M_2}$ cannot be defined for pure annihilation decay modes like $\bar{B}_s \to \pi^+ \pi^-$ due to the absence of the transition form factors $F_0^{B_s \to \pi}$, it is more convenient to use the original parameters $b_i$~\cite{Beneke:2001ev}, which are related to $\beta_i$ through~\cite{Beneke:2003zv}
\begin{equation}\label{eq:bibetai}
A_{M_1 M_2} \beta_i^p = B_{M_1 M_2} b_i^p\,,
\end{equation}
where $B_{M_1 M_2}=f_B f_{M_1} f_{M_2}$, with $f_M$ the decay constant of the meson $M$. Notice that in case the transition form factor $F_0^{B\to M_1}$ does not exist, we have to set $A_{M_1M_2}$ to unity.

Equating eq.~\eqref{eq:amp_QCDF} to eq.~\eqref{eq:AmpTDA} and making use of the unitarity relations of the CKM matrix element, $\lambda_{c}^{(D)}=-\lambda_{u}^{(D)}-\lambda_{t}^{(D)}$, we can obtain the connections between the TDA and QCDF parameterizations, which are given explicitly by~\cite{Huber:2021cgk}
\begin{equation}
\begin{aligned}
T &= A_{M_1 M_2}
\left[
\alpha_1
+ \frac{3}{2}\alpha_{4,\mathrm{EW}}^{u}
- \frac{3}{2}\alpha_{4,\mathrm{EW}}^{c}
\right],
& \hspace{-7cm}
C &= A_{M_1 M_2}
\left[
\alpha_2
+ \frac{3}{2}\alpha_{3,\mathrm{EW}}^{u}
- \frac{3}{2}\alpha_{3,\mathrm{EW}}^{c}
\right],
\\[1ex]
E &= A_{M_1 M_2}
\left[
\beta_1
+ \frac{3}{2} \beta_{4,\mathrm{EW}}^{u}
- \frac{3}{2} \beta_{4,\mathrm{EW}}^{c}
\right],
& \hspace{-7cm}
A &= A_{M_1 M_2}
\left[
\beta_2
+ \frac{3}{2}\beta_{3,\mathrm{EW}}^{u}
- \frac{3}{2}\beta_{3,\mathrm{EW}}^{c}
\right],
\\[1ex]
T_{AS} &= A_{M_1 M_2}
\left[
\beta_{S1}
+ \frac{3}{2} \beta_{S4,\mathrm{EW}}^{u}
- \frac{3}{2} \beta_{S4,\mathrm{EW}}^{c}
\right],
\\[1ex]
T_{ES} &= A_{M_1 M_2}
\left[
\beta_{S2}
+ \frac{3}{2}\beta_{S3,\mathrm{EW}}^{u}
- \frac{3}{2}\beta_{S3,\mathrm{EW}}^{c}
\right],
\\[1ex]
T_{PA} &= A_{M_1 M_2}
\left[
\beta_4^{u}
- \beta_4^{c}
- \left(
\frac{\beta_{4,\mathrm{EW}}^{u}}{2}
- \frac{\beta_{4,\mathrm{EW}}^{c}}{2}
\right)
\right], 
& \hspace{-10cm} 
&
\\[1ex]
T_{SS} &= A_{M_1 M_2}
\left[
\beta_{S4}^{u}
- \beta_{S4}^{c}
- \left(
\frac{\beta_{S4,\mathrm{EW}}^{u}}{2}
- \frac{\beta_{S4,\mathrm{EW}}^{c}}{2}
\right)
\right], 
& \hspace{-10cm} 
&
\\[1ex]
T_{S} &= A_{M_1 M_2}
\Bigg[
\alpha_3^{u}
- \alpha_3^{c}
- \left(
\frac{\alpha_{3,\mathrm{EW}}^{u}}{2}
- \frac{\alpha_{3,\mathrm{EW}}^{c}}{2}
\right)
+ \left(
\beta_{S3}^{u}
- \beta_{S3}^{c}
\right)
- \left(
\frac{\beta_{S3,\mathrm{EW}}^{u}}{2}
- \frac{\beta_{S3,\mathrm{EW}}^{c}}{2}
\right)
\Bigg], 
& \hspace{-10cm} 
&
\\[1ex]
T_{P}&= A_{M_1 M_2}
\Bigg[
\alpha_4^{u}
- \alpha_4^{c}
- \left(
\frac{\alpha_{4,\mathrm{EW}}^{u}}{2}
- \frac{\alpha_{4,\mathrm{EW}}^{c}}{2}
\right)
+ \left(
\beta_3^{u}
- \beta_3^{c}
\right)
- \left(
\frac{\beta_{3,\mathrm{EW}}^{u}}{2}
- \frac{\beta_{3,\mathrm{EW}}^{c}}{2}
\right)
\Bigg],
\end{aligned}
\label{eq:Tree_dict}
\end{equation}
for the tree, and
\begin{equation}
\begin{aligned}
S &= A_{M_1 M_2}
\left[
\alpha_3^{c}
+ \beta_{S3}^{c}
- \frac{\alpha_{3,\mathrm{EW}}^{c}}{2}
- \frac{\beta_{S3,\mathrm{EW}}^{c}}{2}
\right],
\\[1ex]
P & = A_{M_1 M_2}
\left[
\alpha_4^{c}
+ \beta_3^{c}
- \frac{\alpha_{4,\mathrm{EW}}^{c}}{2}
- \frac{\beta_{3,\mathrm{EW}}^{c}}{2}
\right],
\\[1ex]
P_A &= A_{M_1 M_2}
\left[
\beta_4^{c}
- \frac{\beta_{4,\mathrm{EW}}^{c}}{2}
\right],
& \hspace{-10cm}
P_{SS} &= A_{M_1 M_2}
\left[
\beta_{S4}^{c}
- \frac{\beta_{S4,\mathrm{EW}}^{c}}{2}
\right],
\\[1.2ex]
P_C &= \frac{3}{2} A_{M_1 M_2} \alpha_{3,\mathrm{EW}}^{c},
& \hspace{-10cm}
P_T &= \frac{3}{2} A_{M_1 M_2} \alpha_{4,\mathrm{EW}}^{c},
\\[1.2ex]
P_{TA} &= \frac{3}{2} A_{M_1 M_2} \beta_{3,\mathrm{EW}}^{c},
& \hspace{-10cm}
P_{TE} &= \frac{3}{2} A_{M_1 M_2} \beta_{4,\mathrm{EW}}^{c},
\\[1.2ex]
P_{AS} &= \frac{3}{2} A_{M_1 M_2} \beta_{S4,\mathrm{EW}}^{c},
& \hspace{-10cm}
P_{ES} &= \frac{3}{2} A_{M_1 M_2} \beta_{S3,\mathrm{EW}}^{c},
\end{aligned}
\label{eq:Peng_dict}
\end{equation}
for the penguin sector.

Inspired by the NLO results for the QCDF sub-amplitudes $\alpha_i$ and $\beta_i$~\cite{Beneke:2003zv} as well as the next-to-next-to-leading-order (NNLO) results for $\alpha_{1,2}$ and the leading QCD penguin amplitudes $a_4^p$~\cite{Beneke:2009ek,Bell:2009fm,Bell:2015koa,Bell:2020qus}, we are motivated to consider the simplified expressions
\begin{equation}
\begin{aligned}
T &= A_{M_1 M_2}\,\alpha_1,
&\qquad
C &= A_{M_1 M_2}\,\alpha_2,
&\qquad
E &= B_{M_1 M_2}\,b_1, 
\\[1ex]
A &= B_{M_1 M_2}\,b_2,
&\qquad
T_{AS} &= B_{M_1 M_2}\, b_{S1},
&\qquad
T_{ES} &= B_{M_1 M_2}\, b_{S2}, \nonumber
\end{aligned}
\end{equation}
\begin{equation}
\begin{aligned}
T_{PA} = 0,
\qquad
T_{SS} = 0,
\qquad
T_{S} = 0,
\qquad
\lvert T_{P} \rvert /A_{M_1 M_2} < 0.02,
\label{eq:simplifications}
\end{aligned}
\end{equation}
for the tree, and
\begin{equation}
\begin{aligned}
S &= A_{M_1 M_2}
\left[
\alpha_3
+ \beta_{S3}
- \frac{\alpha_{3,\mathrm{EW}}}{2}
- \frac{\beta_{S3,\mathrm{EW}}}{2}
\right], &
P &= A_{M_1 M_2}
\left[
\alpha_4^{c}
+ \beta_3
- \frac{\alpha_{4,\mathrm{EW}}^{c}}{2}
- \frac{\beta_{3,\mathrm{EW}}}{2}
\right],\nonumber
\end{aligned}
\end{equation}
\begin{equation}
\begin{aligned}
P_A &= B_{M_1 M_2}
\left(
b_4
- \frac{b_{4,\mathrm{EW}}}{2}
\right),
&\qquad
P_{SS} &= B_{M_1 M_2}
\left(
b_{S4}
- \frac{b_{S4,\mathrm{EW}}}{2}
\right),
\\[1.2ex]
P_C &= \frac{3}{2} A_{M_1 M_2}\,\alpha_{3,\mathrm{EW}},
&\qquad
P_T &= \frac{3}{2} A_{M_1 M_2}\,\alpha_{4,\mathrm{EW}}^{c},
\\[1.2ex]
P_{TA} &= \frac{3}{2} B_{M_1 M_2}\,b_{3,\mathrm{EW}},
&\qquad
P_{TE} &= \frac{3}{2} B_{M_1 M_2}\,b_{4,\mathrm{EW}},
\\[1.2ex]
P_{AS} &= \frac{3}{2} B_{M_1 M_2}\,b_{S4,\mathrm{EW}},
&\qquad
P_{ES} &= \frac{3}{2} B_{M_1 M_2}\,b_{S3,\mathrm{EW}}, \label{eq:simplifications_penguin}
\end{aligned}
\end{equation}
for the penguin sector. This is based on the observations that 
\begin{equation} \label{eq:NNLO_conditions}
\begin{aligned}
    & \alpha_3^u=\alpha_3^c=\alpha_3, \qquad & &\alpha_{3,\mathrm{EW}}^u=\alpha_{3,\mathrm{EW}}^c=\alpha_{3,\mathrm{EW}}, \\
    & \beta_i^u=\beta_i^c=\beta_i, \qquad 
    & & b_i^u=b_i^c=b_i,
\end{aligned}    
\end{equation}
and the difference $|\alpha_{4,\mathrm{EW}}^c-\alpha_{4,\mathrm{EW}}^u|$ is at most of $\mathcal{O}(10^{-3})$ at the NLO~\cite{Beneke:2003zv}, while the difference $|a_4^c-a_4^u|=0.013^{+0.008}_{-0.007}$ at the NNLO~\cite{Beneke:2009ek,Bell:2015koa,Bell:2020qus}. This exercise also demonstrates the advantage of the ``$\lambda_u-\lambda_t$'' convention when selecting the independent linear combinations of the QCDF sub-amplitudes.

\subsection{\texorpdfstring{Inclusion of leading factorizable $SU(3)$-breaking effects}{Inclusion of leading factorizable SU(3)-breaking effects}}
\label{sec:SU(3)-breaking-effects}

There exist several sources of $SU(3)$-breaking effects in two-body charmless $B$-meson decays, such as the transition form factors, the decay constants, the light-quark masses, and the Gegenbauer moments. It is not realistic to take into account all these effects during the fit. In fact, relaxing completely the $SU(3)$ symmetry would heavily increase the number of topological parameters beyond the number of observables, making any reasonable fit infeasible. Here we will instead assume that the dominant $SU(3)$-breaking effects arise from the following channel-dependent factors $A_{M_1 M_2}$ and $B_{M_1 M_2}$ defined, respectively, by~\cite{Huber:2021cgk,BurgosMarcos:2025xja}
\begin{align} \label{eq:AB_definitions}
A_{M_1 M_2} = (m_B^2-m_{M_1}^2) F^{B\rightarrow M_1}_0 (m^2_{M_2}) f_{M_2}\,, \qquad 
B_{M_1 M_2} = f_{B_q}f_{M_1}f_{M_2}\,,
\end{align}
where $F^{B\rightarrow M_1}_0 (m^2_{M_2})$ is the $B\rightarrow M_1$ transition form factor evaluated at $q^2=m^2_{M_2}$, while $f_{M}$ and $m_M$ are the decay constant and mass of the meson $M$. Their definitions could be found, \textit{e.g.}, in refs.~\cite{Beneke:2003zv,Beneke:2000wa}. This observation is motivated by the assumed factorization for some decay topologies and the established QCDF formulae for the decay amplitudes of two-body charmless $B_{(s)}$ decays~\cite{Beneke:1999br,Beneke:2000ry,Beneke:2001ev,Beneke:2003zv}. Numerically, these corrections are estimated to be of $\mathcal{O}(20- 40\%)$ at the amplitude level. It should be noted that $A_{M_1 M_2}$ depends on the order of the final-state mesons, while we have $B_{M_1 M_2}=B_{M_2 M_1}$. According to the power counting presented in ref.~\cite{Beneke:2000ry}, we also know that the numerical size of $B_{M_1 M_2}$ is suppressed by roughly $4.4\times 10^{-3}$ with respect to that of $A_{M_1 M_2}$, which should be kept in mind when comparing the fitted values of the QCDF parameters $\alpha_i$ and $\beta_i$ ($b_i$). 

To perform our analysis in the topological diagram basis, we introduce the so-called bare $SU(3)_F$ quantities $\tilde{T}$, $\tilde{C}$, $\tilde{E}$, $\tilde{A}$, $\tilde{T}_{AS}$, $\tilde{T}_{ES}$, $\tilde{S}$, $\tilde{P}$, $\tilde{P}_A$, $\tilde{P}_{SS}$, $\tilde{P}_{C}$, $\tilde{P}_{T}$, $\tilde{P}_{TA}$, $\tilde{P}_{TE}$, $\tilde{P}_{AS}$ and $\tilde{P}_{ES}$, which are connected to the original topological amplitudes through
\begin{equation} \label{eq:bare_TDA}
\begin{aligned}
T      &= A_{M_1M_2}\,\tilde{T},      &\qquad
C      &= A_{M_1M_2}\,\tilde{C},      &\qquad
E      &= B_{M_1M_2}\,\tilde{E}, \\[4pt]
A      &= B_{M_1M_2}\,\tilde{A},      &
T_{AS} &= B_{M_1M_2}\,\tilde{T}_{AS}, &
T_{ES} &= B_{M_1M_2}\,\tilde{T}_{ES}.\\[4pt]
S      &= A_{M_1M_2}\,\tilde{S},      &\qquad
P      &= A_{M_1M_2}\,\tilde{P},      &\qquad
P_{A}  &= B_{M_1M_2}\,\tilde{P}_{A},\\[4pt]
P_{SS} &= B_{M_1M_2}\,\tilde{P}_{SS},&\qquad
P_{C}  &= A_{M_1M_2}\,\tilde{P}_{C},&\qquad
P_{T}  &= A_{M_1M_2}\,\tilde{P}_{T},\\[4pt]
P_{TA} &= B_{M_1M_2}\,\tilde{P}_{TA},&\qquad
P_{TE} &= B_{M_1M_2}\,\tilde{P}_{TE},&\qquad
P_{AS}  &= B_{M_1M_2}\,\tilde{P}_{AS},\\[4pt]
P_{ES} &= B_{M_1M_2}\,\tilde{P}_{ES}.&\qquad
&\qquad
\end{aligned}
\end{equation}
A $\chi^2$ fit is then performed over these bare parameters, with the dominant $SU(3)$-breaking effects arising from the transition form factors and the decay constants taken into account through the factors $A_{M_1 M_2}$ and $B_{M_1 M_2}$. 

From eqs.~\eqref{eq:simplifications}, \eqref{eq:simplifications_penguin} and \eqref{eq:bare_TDA}, we can also obtain the connections between the bare $SU(3)_F$ quantities and the QCDF parameters, with
\begin{equation}
\begin{alignedat}{3}
\alpha_1 & = \tilde{T},
& \alpha_2 & = \tilde{C},
& \beta_1 & = \frac{B_{M_1 M_2}}{A_{M_1 M_2}}\,\tilde{E}, \\[6pt]
\beta_2 & = \frac{B_{M_1 M_2}}{A_{M_1 M_2}}\,\tilde{A},
& \beta_{S1} & = \frac{B_{M_1 M_2}}{A_{M_1 M_2}}\,\tilde{T}_{AS},
& \beta_{S2} & = \frac{B_{M_1 M_2}}{A_{M_1 M_2}}\,\tilde{T}_{ES}, \\[10pt]
\beta_4-\frac{\beta_{4,\mathrm{EW}}}{2}
& = \frac{B_{M_1 M_2}}{A_{M_1 M_2}}\,\tilde{P}_{A},
& \beta_{S4}-\frac{\beta_{S4,\mathrm{EW}}}{2}
& = \frac{B_{M_1 M_2}}{A_{M_1 M_2}}\,\tilde{P}_{SS},
& \alpha_{3,\mathrm{EW}}
& = \frac{2}{3}\,\tilde{P}_{C}, \\[10pt]
\alpha^{c}_{4,\mathrm{EW}}
& = \frac{2}{3}\,\tilde{P}_{T},
& \beta_{3,\mathrm{EW}}
& = \frac{2}{3}\,\frac{B_{M_1 M_2}}{A_{M_1 M_2}}\,\tilde{P}_{TA},
& \quad \beta_{4,\mathrm{EW}}
& = \frac{2}{3}\,\frac{B_{M_1 M_2}}{A_{M_1 M_2}}\,\tilde{P}_{TE}, \\[10pt]
\beta_{S4,\mathrm{EW}}
& =  \frac{2}{3}\,\frac{B_{M_1 M_2}}{A_{M_1 M_2}}\,\tilde{P}_{AS},
& 
& &\nonumber
\beta_{S3,\mathrm{EW}}
& = \frac{2}{3}\,\frac{B_{M_1 M_2}}{A_{M_1 M_2}}\,\tilde{P}_{ES},\\
\end{alignedat}
\end{equation}
\begin{equation}
\alpha_3
+ \beta_{S3}
- \frac{\alpha_{3,\mathrm{EW}}}{2}
- \frac{\beta_{S3,\mathrm{EW}}}{2}
= \tilde{S},
\qquad \qquad 
\alpha_4^{c}
+ \beta_{3}
- \frac{\alpha_{4,\mathrm{EW}}^{c}}{2}
- \frac{\beta_{3,\mathrm{EW}}}{2} = \tilde{P} .
\label{eq:QCDF2TDA}
\end{equation}
It must be noted that, while the QCDF coefficients $\alpha_i$, $\alpha_{i,\mathrm{EW}}$, $b_i$ and $b_{i,\mathrm{EW}}$ are assumed to be channel independent, the coefficients $\beta_i$ and $\beta_{i,\mathrm{EW}}$ depend on the channels considered through the factors $A_{M_1 M_2}$ and $B_{M_1 M_2}$ (see also the comment below eq.~\eqref{eq:bibetai}). 

\section{Physical amplitudes and observables}
\label{sec:physampsandobservables}

\subsection{Physical amplitudes}
\label{sec:physamps}

The TDA amplitudes, including the dominant $SU(3)$-breaking effects, are expressed in terms of the TDA parameters as
\begin{equation} \label{eq:AmpTDA-2}
  \mathcal{A}^{\mathrm{TDA}} = i\,\frac{G_F}{\sqrt{2}} \sum_{i=1}^{10} \left[\lambda_u^{(D)}\, t_i\, \mathcal{T}_i^{\mathrm{TDA}} + \lambda_t^{(D)}\, p_i\, \mathcal{P}_i^{\mathrm{TDA}}\right]\,,
\end{equation}
with
\begin{align}
\mathcal{T}_i^{\mathrm{TDA}} &= \left\{T, C, A, E, T_{PA}, T_{ES}, T_{AS}, T_{SS}, T_P, T_S\right\}\,,\nonumber \\
\mathcal{P}_i^{\mathrm{TDA}} &= \left\{P_T, P_C, P_{TA}, P, P_{TE}, P_{A}, P_{AS}, P_{ES}, P_{SS}, S\right\}\,,
\end{align}
which can be further reduced to the bare TDA quantities as introduced in eq.~\eqref{eq:bare_TDA}, while including the dominant $SU(3)$-breaking effects. The tree ($t_i$) and penguin ($p_i$) coefficients are collected in tables~\ref{tab:tree-without eta}~--~\ref{tab:tree-with eta and eta'} and \ref{tab:penguin-without eta}~--~\ref{tab:penguin-with eta and eta'}, respectively, which we relegate to appendix~\ref{app:ampcoeffs} for better readability of the paper.

In the Feldmann–Kroll–Stech (FKS) scheme~\cite{Feldmann:1998vh} for $\eta$ and $\eta^\prime$ mixing, the physical amplitudes with a single $\eta^{(\prime)}$ meson in the final state are given, respectively, by
\begin{align}
\mathcal{A}(\bar{B} \rightarrow M \eta) &= \mathcal{A}(\bar{B} \rightarrow M \eta_q) + \mathcal{A}(\bar{B} \rightarrow M \eta_s)\,, \nonumber\\
\mathcal{A}(\bar{B} \rightarrow M \eta^\prime) &= \mathcal{A}(\bar{B} \rightarrow M \eta'_q) + \mathcal{A}(\bar{B} \rightarrow M \eta'_s)\,,
\end{align}
while those with a pair of $\eta^{(\prime)}$ mesons in the final state read
\begin{align}
\mathcal{A}(\bar{B}\rightarrow \eta \eta) &= \mathcal{A}(\bar{B}\rightarrow \eta_q \eta_q) + \mathcal{A}(\bar{B}\rightarrow \eta_q \eta_s) + \mathcal{A}(\bar{B}\rightarrow \eta_s \eta_q) + \mathcal{A}(\bar{B}\rightarrow \eta_s \eta_s)\,, \nonumber\\
\mathcal{A}(\bar{B}\rightarrow \eta' \eta') &= \mathcal{A}(\bar{B}\rightarrow \eta'_q \eta'_q) + \mathcal{A}(\bar{B}\rightarrow \eta'_q \eta'_s) + \mathcal{A}(\bar{B}\rightarrow \eta'_s \eta'_q) + \mathcal{A}(\bar{B}\rightarrow \eta'_s \eta'_q)\,, \nonumber\\
\mathcal{A}(\bar{B}\rightarrow \eta \eta') &= \mathcal{A}(\bar{B}\rightarrow \eta_q \eta'_q)  +  \mathcal{A}(\bar{B}\rightarrow \eta_q \eta'_s) + \mathcal{A}(\bar{B}\rightarrow \eta_s \eta'_q) + \mathcal{A}(\bar{B}\rightarrow \eta_s \eta'_s) \,.
\end{align}

\subsection{Observable definitions}
\label{sec:observabledefs}

With the amplitude at hand, the branching ratio of a two-body charmless $B$-meson decay $\bar{B} \to M_1 M_2$ is then given by
\begin{align}
    \mathrm{Br}(\bar{B} \to M_1 M_2) = S\,\frac{\tau_B |\boldsymbol{p}|}{8\pi m_B^2} \left|\mathcal{A}(\bar{B} \to M_1 M_2)\right|^2\,,
\end{align}
where $\tau_B$ is the $B$-meson lifetime, and
\begin{align}
|\boldsymbol{p}|=\frac{\sqrt{\left[m_B^2-(m_{M_1}+m_{M_2})^2\right]\left[m_B^2-(m_{M_1}-m_{M_2})^2\right]}}{2 m_B}\,,
\end{align}
is the magnitude of the three-momentum of either meson in the $B$-meson center-of-mass frame, which accounts for the $SU(3)$-breaking effect arising from the phase space. The factor $S=1/2$ if $M_1$ and $M_2$ are identical, and $S=1$ otherwise. Here we always consider the CP-averaged branching ratio, which is defined for a given decay mode $\bar{B} \to \bar{f}$ by
\begin{align}
    \frac{1}{2}\left[\mathrm{Br}(\bar{B} \to \bar{f}) + \mathrm{Br}(B \to f)\right]\,.
\end{align}

The direct CP asymmetry (also called the partial rate asymmetry) is defined as the difference between the rate involving a $b$ quark and that involving a $\bar{b}$ quark, divided by the sum. Following this standard ``$\bar{B}$ minus $B$'' convention, we have 
\begin{align} \label{eq:ACP}
    A_{CP}(\bar{B} \to \bar{f}) = \frac{\mathrm{Br}(\bar{B} \to \bar{f}) - \mathrm{Br}(B \to f)}{\mathrm{Br}(\bar{B} \to \bar{f}) + \mathrm{Br}(B \to f)}\,,
\end{align}
which holds for the charged $B^-$ decays as well as the neutral $\bar{B}^0$ decays into flavour-specific final states. For the decays into final states that are common to the neutral $B^0$ mesons, we can also define the time-dependent CP asymmetries~\cite{HFLAV:2005zyi,Fleischer:2002ys}.\footnote{Here the neutral $B^0$ meson refers to either $B^0$ or $B_s^0$. While the decay width difference of the $B^0$ system can be safely neglected, we must take into account this effect for the $B_s^0$ system in practice, for which we refer the readers to refs.~\cite{DeBruyn:2012wj,Fleischer:2016ofb} for further details.} To this end, let us firstly introduce our phase convention for the CP transformation laws of the neutral $B^0$ mesons and the final state $f$
\begin{align}
    \mathrm{CP}|B^0\rangle &=\omega_B |\bar{B}^0\rangle, & 
    \mathrm{CP}|\bar{B}^0\rangle &=\omega_B^* |B^0\rangle\,, \nonumber\\
    \mathrm{CP}|f\rangle &=\omega_f|\bar{f}\rangle, & 
    \mathrm{CP}|\bar{f}\rangle &=\omega_f^*|f\rangle\,,
\end{align}
with $|\omega_B|=1$ and $|\omega_f|=1$. When the final state is a CP eigenstate, $f=\bar{f}=f_{CP}$, the phase factor $\omega_f$ will be replaced by the CP eigenvalue $\eta_f=\pm1$.\footnote{For the two-body charmless $B$-meson decays into pseudoscalar final states, we have $\eta_{\pi^0 K_S}=-1$ and $\eta_{\eta^{(\prime)} K_S}=-1$, while all the other final CP eigenstates have $\eta_f=+1$.} We then define the decay amplitudes $\bar{\mathcal{A}}_f$ and $\mathcal{A}_f$ according to
\begin{align}
    \bar{\mathcal{A}}_f = \langle f|\mathcal{H}_{\rm eff}| \bar{B}^0\rangle\,, \qquad
    \mathcal{A}_f = \langle f|\mathcal{H}_{\rm eff}| B^0\rangle\,,
\end{align}
where $\mathcal{H}_{\rm eff}$ is the effective weak Hamiltonian governing the decay processes, defined already in eq.~\eqref{eq:Heff}. To discuss CP violation in the interference of decays with and without mixing, we introduce a complex quantity $\lambda_f$ defined by
\begin{align}
    \lambda_f = \frac{q}{p}\,\frac{\bar{\mathcal{A}}_f}{\mathcal{A}_f}\,,
\end{align}
which depends on the factor $q/p$ related to $B^0-\bar{B}^0$ mixing and on the decay amplitude ratios $\bar{\mathcal{A}}_f/\mathcal{A}_f$, but is independent of the phase convention. Equipped with the above information, we can finally write the time-dependent CP asymmetry as
\begin{align}
A_{CP}(t) \equiv \frac{\Gamma_{\bar{B}^0 \rightarrow f}(t)-\Gamma_{B^0 \rightarrow f}(t)}{\Gamma_{\bar{B}^0 \rightarrow f}(t)+\Gamma_{B^0 \rightarrow f}(t)} \propto S_f \sin (\Delta M_q t) - C_f \cos (\Delta M_q t)\,,
\end{align}
where $\Delta M_q$ is the mass difference between the heavy and light mass eigenstates of the neutral $B$-meson system, and
\begin{align}
    S_f = \frac{2 \operatorname{Im}\left(\lambda_f\right)}{1+\left|\lambda_f\right|^2}\,, \qquad
    C_f = \frac{1-\left|\lambda_f\right|^2}{1+\left|\lambda_f\right|^2}\,.
\end{align}
It is noted that $C_f = - A_{CP}$, with the latter defined already by eq.~\eqref{eq:ACP}. When the CP violation in mixing is neglected and the decay amplitude contains terms with only a single weak phase, we have $q/p=\omega_B\,e^{-2i\phi_B}$ and $\bar{\mathcal{A}}_f/\mathcal{A}_f=\omega_f\, \omega_B^\ast\,e^{-2i\phi_f}$, with $\phi_B$ and $\phi_f$ being the appropriate CP-violating weak phases, which imply that
\begin{align}
    \lambda_f = \eta_f\, e^{-2i(\phi_B+\phi_f)}\,,
\end{align}
for a final CP eigenstate $f$. In this case, $S_f=-\eta_f\,\sin2(\phi_B+\phi_f)$ and $C_f=0$, showing a clean interpretation of these observables. If a decay channel receives contributions from several amplitudes with different weak phases, no clean interpretation is possible anymore for these observables.

\section{Fit to data}
\label{sec:fit}

\subsection{Experimental data and input parameters}

The experimental data on CP-averaged branching ratios, as well as the direct and mixing-induced CP asymmetries in two-body charmless $\bar{B} \to P P$ decays can be found in tables~\ref{tab:resultsBRI}~--~\ref{tab:resultsMixCPIII} (together with the fit results). Unless stated otherwise, all the experimental data is taken from the latest version of Particle Data Group (PDG)~\cite{ParticleDataGroup:2024cfk} and Heavy Flavor Averaging Group (HFLAV)~\cite{HeavyFlavorAveragingGroupHFLAV:2024ctg}. Notice that the experimental numbers are usually averages over several experimental measurements, and the experimental correlations among them are not known or not quoted.

\begin{table}[htbp]
\renewcommand{\arraystretch}{1.28} 
\setlength{\tabcolsep}{13pt} 
\centering
\begin{tabular}{c|c}
\hline
\multicolumn{2}{c}
{\textbf{\boldmath $B$-meson lifetimes~\cite{ParticleDataGroup:2024cfk}}} \\
\hline
$\tau_{B^-}$ & $1.638 \pm 0.004$~ps \\ 
$\tau_{\bar{B}^0}$ & $1.519 \pm 0.004$~ps  \\ 
$\tau_{\bar{B}_s}$ & $1.516 \pm 0.006$~ps \\
\hline
\multicolumn{2}{c}
{\textbf{\boldmath $B_{(s)} \to P$ transition form factors~\cite{Cui:2022zwm}}} \\
\hline
$F_{0}^{B \to \pi}(q^2 = 0)$ & $0.192 \pm 0.022$ \\
$F_{0}^{B \to \pi}(q^2 \approx 0.019 \text{GeV}^2)$ & $0.192 \pm 0.022$ \\ 
$F_{0}^{B \to \pi}(q^2 \approx 0.244 \text{GeV}^2)$ & $0.193 \pm 0.022$ \\
$F_{0}^{B \to \pi}(q^2 \approx 0.300 \text{GeV}^2)$ & $0.193 \pm 0.022$ \\ 
$F_{0}^{B \to \pi}(q^2 \approx 0.917 \text{GeV}^2)$ & $0.197 \pm 0.021$ \\ 
\hline
$F_{0}^{B_s \to \bar{K}}(q^2 = 0)$ & $0.203 \pm 0.014$ \\ 
$F_{0}^{B_s \to \bar{K}}(q^2 \approx 0.019 \text{GeV}^2)$ & $0.203 \pm 0.014$ \\ 
$F_{0}^{B_s \to \bar{K}}(q^2 \approx 0.244 \text{GeV}^2)$ & $0.205 \pm 0.014$ \\
$F_{0}^{B_s \to \bar{K}}(q^2 \approx 0.300 \text{GeV}^2)$ & $0.206 \pm 0.014$ \\ 
$F_{0}^{B_s \to \bar{K}}(q^2 \approx 0.917 \text{GeV}^2)$ & $0.210 \pm 0.014$ \\ 
\hline
$F_{0}^{B \to K}(q^2 = 0)$ & $0.326 \pm 0.010$ \\ 
$F_{0}^{B \to K}(q^2 \approx 0.019 \text{GeV}^2)$ & $0.326 \pm 0.010$ \\ 
$F_{0}^{B \to K}(q^2 \approx 0.244 \text{GeV}^2)$ & $0.328 \pm 0.010$ \\ 
$F_{0}^{B \to K}(q^2 \approx 0.300 \text{GeV}^2)$ & $0.328 \pm 0.010$ \\ 
$F_{0}^{B \to K}(q^2 \approx 0.917 \text{GeV}^2)$ & $0.333 \pm 0.010$ \\ 
\hline
\multicolumn{2}{c}{\textbf{Decay constants [MeV]~\cite{FlavourLatticeAveragingGroupFLAG:2024oxs,ParticleDataGroup:2024cfk}}} \\
\hline
$f_{B} = 190.0 \pm 1.3$ & $\frac{f_{B_s}}{f_B} = 1.209 \pm 0.005$ \\
$f_{\pi} = 130.2 \pm 1.2$ & $f_{K} = 155.7 \pm 0.3$ \\
\hline
\multicolumn{2}{c}{\textbf{\boldmath $\eta$ and $\eta^\prime$ mixing parameters~\cite{Feldmann:1998vh}}} \\
\hline
$f_{q} = (1.07 \pm 0.02)\,f_\pi$ & 
$f_{s} = (1.34 \pm 0.06)\,f_\pi$ \\
$\phi = 39.3^\circ \pm 1.0^\circ$ \\
\hline
\multicolumn{2}{c}{\textbf{Wolfenstein parameters~\cite{Charles:2004jd}}} \\
\hline
$A = 0.8215^{+0.0045}_{-0.0146}$ & 
$\lambda = 0.22504^{+0.00020}_{-0.00022}$ \\
$\bar{\rho} = 0.1562^{+0.0102}_{-0.0045}$ &
$\bar{\eta} = 0.3564^{+0.0061}_{-0.0065}$\\
\hline
\end{tabular}
\caption{Input parameters used throughout this paper. The $B \to \pi$, $B_s \to \bar{K}$, and $B \to K$ transition form factors are evaluated at $q^2 = 0$, $q^2 \approx 0.019$ GeV$^2$ (for pion), $q^2 \approx 0.244$ GeV$^2$ (for kaon), $q^2 \approx 0.300$ GeV$^2$ (for $\eta$), and $q^2 \approx 0.917$ GeV$^2$ (for $\eta^\prime$) based on the combined BCL fit performed in ref.~\cite{Cui:2022zwm}. For the decay constants, we assume the isospin symmetry, and hence $f_{B^-}=f_{\bar{B}^0}$, $f_{\pi^{\pm}}=f_{\pi^0}$, $f_{K^{\pm}}=f_{K^0}$.
\label{tab:formfactors_decayconstants}}
\end{table}

All the input parameters involved in this study are collected in table~\ref{tab:formfactors_decayconstants}. For the $B\to P$ transition form factors $F^{B \to P}_{0}(q^2)$, we use the formulae and numerical values provided in ref.~\cite{Cui:2022zwm}, which are obtained through a combined fit to both the LCSR and lattice simulation results based on the Bourrely-Caprini-Lellouch (BCL) $z$-series parameterizations~\cite{Lellouch:1995yv,Bourrely:2005hp,Bourrely:2008za} for the transition form factors. It must be emphasized that, compared to the BCL fitting procedure with only the lattice QCD data points, such a combined fit is especially beneficial for further improving the theory accuracy of the transition form factors in the low-$q^2$ regime. As the $q^2$ dependence of these scalar form factors is quite marginal when $q^2$ varies from $m_\pi^2$ to $m_{\eta'}^2$, with the resulting central values being always within the uncertainties of the corresponding transition form factors, we can consistently fix them at the $q^2=0$ point for the sake of performing the fit. It is also interesting to note that sizeable $SU(3)$-flavour symmetry violating effects (numerically of $\mathcal{O}(30\%)$) are predicted between the $B \to \pi$ and $B \to K$ transition form factors in the large recoil region~\cite{Cui:2022zwm}, which is helpful to explain the current experimental data on charmless $\bar{B} \to P P$ decays. 

The parameters related to $\eta$ and $\eta^\prime$ mesons are taken from refs.~\cite{Feldmann:1998vh,Beneke:2002jn}. In the FKS mixing scheme~\cite{Feldmann:1998vh}, we need only a single mixing angle $\phi$ in the quark-flavour basis, and the physical states are expressed in terms of the flavour states as
\begin{equation}
\binom{|\eta\rangle}{\left|\eta^{\prime}\right\rangle}=\left(\begin{array}{rr}
\cos \phi & -\sin \phi \\
\sin \phi & \cos \phi
\end{array}\right)\binom{\left|\eta_q\right\rangle}{\left|\eta_s\right\rangle}\,,
\end{equation}
where $\left|\eta_q\right\rangle=(|u \bar{u}\rangle+|d \bar{d}\rangle)/\sqrt{2}$ and $\left|\eta_s\right\rangle=|s \bar{s}\rangle$. The physical decay constants $f_{\eta}^q$, $f_{\eta}^s$, $f_{\eta'}^q$, and $f_{\eta'}^s$ are then given, respectively, by
\begin{equation}
\begin{aligned}
f_\eta^q &=f_q \cos \phi, & f_\eta^s &=-f_s \sin \phi, \\
f_{\eta^{\prime}}^q &=f_q \sin \phi, & f_{\eta^{\prime}}^s &=f_s \cos \phi,
\end{aligned}
\end{equation}
while the $B\to \eta^{(\prime)}$ and $B_s\to \eta^{(\prime)}$ transition form factors can be parameterized, respectively, by~\cite{Beneke:2002jn}\footnote{These transition form factors can also be directly calculated in the LCSR and lattice QCD approaches. We refer the readers to refs.~\cite{Melic:2025uha,Zhang:2025yeu,Mandal:2024pwz} and references therein for the recent LCSR calculations, and to refs.~\cite{Parrott:2022rgu,Parrott:2020vbe} for the first lattice QCD calculation of the unphysical $B_s \to \eta_s$ transition form factors. It must be noted that these results are still plagued by large uncertainties, and are also consistent with our numerical results obtained through eq.~\eqref{eq:FFB2etaandetaprime} within uncertainties for $F_0^{B \rightarrow \eta^{(\prime)}}(0)$. For $F_0^{B_s \rightarrow \eta^{(\prime)}}(0)$, on the other hand, our numerical results are generally lower than what those references obtained. One has to admit that the dynamics behind these transition form factors, especially the two-gluon contributions, are still not well known~\cite{Beneke:2002jn}. Hence, we prefer to use the parametrizations given by eq.~\eqref{eq:FFB2etaandetaprime}, because we  consider the LCSR and lattice QCD calculations as well as the data information on $F_0^{B \rightarrow \pi}$ and $F_0^{B_s \rightarrow K}$ more solid and accurate. The numerical results given in ref.~\cite{Cui:2022zwm} also enable us to properly take into account the correlations among these transition form factors.}
\begin{equation} \label{eq:FFB2etaandetaprime}
\begin{aligned}
F_0^{B \rightarrow \eta^{(\prime)}} &= F_0^{B \rightarrow \pi} \frac{f_{\eta^{(\prime)}}^q}{f_\pi}+F_2 \frac{\sqrt{2} f_{\eta^{(\prime)}}^q+f_{\eta^{(\prime)}}^s}{\sqrt{3} f_\pi},\\
F_0^{B_s \rightarrow \eta^{(\prime)}} &= F_0^{B_s \rightarrow K} \frac{f_{\eta^{(\prime)}}^s}{f_\pi}+F_2 \frac{\sqrt{2} f_{\eta^{(\prime)}}^q+f_{\eta^{(\prime)}}^s}{\sqrt{3} f_\pi},
\end{aligned}
\end{equation}
where the parameters $f_q$ and $f_s$, together with the mixing angle $\phi$ can be determined from a fit to experimental data~\cite{Feldmann:1998vh}. The parameter $F_2$ accounts for the unknown two-gluon contribution to the $B\to \eta^{(\prime)}$ and $B_s\to \eta^{(\prime)}$ transition form factors and, for the time being, we will take the value $F_2=0$ due to lack of better knowledge~\cite{Beneke:2002jn}.

For the CKM matrix elements, we will adopt the Wolfenstein parametrization~\cite{Wolfenstein:1983yz} and keep the expansion up to $\mathcal{O}(\lambda^8)$, with the four Wolfenstein parameters $A$, $\lambda$, $\bar{\rho}$ and $\bar{\eta}$ extracted from a global fit to experimental data~\cite{Charles:2004jd},\footnote{Here we use the results provided by the CKMfitter group~\cite{Charles:2004jd}, which are consistent with that of the UTfit group~\cite{UTfit:2022hsi} within the quoted uncertainties.} where the exact relation to the original parameters $\rho$ and $\eta$ is given by~\cite{Antonelli:2009ws}
\begin{equation}
\rho + i \eta=\sqrt{\frac{1-A^2 \lambda^4}{1-\lambda^2}} \frac{\bar{\rho}+i \bar{\eta}}{1-A^2 \lambda^4(\bar{\rho}+i \bar{\eta})} \simeq\left(1+\frac{\lambda^2}{2}\right)(\bar{\varrho}+i \bar{\eta})+\mathcal{O}\left(\lambda^4\right)\,.
\end{equation}
All the other input parameters, such as the $B$-meson lifetimes and masses, the pseudoscalar masses, and some basic electroweak parameters, are taken from ref.~\cite{ParticleDataGroup:2024cfk}.

\subsection{Fitting procedure}

Here we follow two independent approaches to perform our fitting. The first approach proceeds along the lines of ref.~\cite{Huber:2021cgk}. For a given set of observables $\mathcal{O}_i$, the comparison between the corresponding theoretical determination $\mathcal{O}^{\rm Theo}_i$ and experimental result $\mathcal{O}^{\rm Exp}_i$ is done in terms of the function
\begin{eqnarray}
    \chi^2_{\mathcal{O}}&=&\sum_i\Bigl(\frac{\mathcal{O}^{\rm Theo}_i - \mathcal{O}^{\rm Exp}_i}{\sigma^{\rm Exp}_i}\Bigl)^2\,,     
\end{eqnarray}
where $\sigma^{\rm Exp}_i$ is the experimental uncertainty and the sum runs over all the branching fractions and CP asymmetries included in this study. During our statistical analysis, the transition form factors, decay constants, $\eta-\eta^\prime$ mixing angle, and Wolfenstein parameters are considered as nuisance parameters, and they are therefore allowed to take values based on Gaussian distributions with central values $p^{\rm input}_j$ and standard deviations $\sigma^{\rm input}_j$ as given in table~\ref{tab:formfactors_decayconstants}. This leads us to consider a second sector in our $\chi^2$ function
\begin{eqnarray}
    \chi^2_{np}&=&\sum_j\Bigl(\frac{p^{\rm fit}_j - p^{\rm input}_j}{\sigma^{\rm input}_j}\Bigl)^2.  
\end{eqnarray}
Our full $\chi^2$ function is then defined by 
\begin{eqnarray}
\chi^2&=& \chi^2_{\mathcal{O}} + \chi^2_{np}.
\end{eqnarray}

As a first step, we aim at finding suitable starting points for minimizing the $\chi^2$ function. To this end, we generate random values for the different bare sub-amplitudes defined in eq.~\eqref{eq:bare_TDA}. Based on the available dynamical information~\cite{Beneke:2003zv,Beneke:2009ek,Bell:2015koa,Bell:2020qus}, we expect that the values of the QCDF sub-amplitudes $\alpha_i$ and $\beta_i$ obey $|\alpha_1|\leq 1.5$, $|\alpha_{i\neq1}|\leq 1$ and $|\beta_i|\leq 1$. Thus, considering the TDA-QCDF relations given by eqs.~\eqref{eq:Tree_dict}~--~\eqref{eq:simplifications_penguin} together with \eqref{eq:bare_TDA}, we have the following conservative bounds for the bare topological sub-amplitudes that are directly related to $\alpha_i$:
\begin{equation} \label{eq:Xabound}
-1.5 \leq \mathrm{Re}(X_{\alpha}), \, \mathrm{Im}(X_{\alpha}) \leq 1.5\,,
\end{equation}
with
\begin{equation}
X_{\alpha}\in \bigl\{\tilde{T}, \, \tilde{C}, \, \tilde{P}_T, \, \tilde{P}_C, \, \tilde{P}, \, \tilde{S} \bigl\}.
\end{equation}
In view of the bound for $T_P$ in eq.~\eqref{eq:simplifications}, we consider
\begin{eqnarray} \label{eq:TPbound}
-0.02\leq {\rm Re}(\tilde{T}_P), \, {\rm Im}(\tilde{T}_P) \leq 0.02.  
\end{eqnarray}
Keeping in mind that $|\beta_i|\leq 1$ and $|B_{M_1 M_2}/A_{M_1 M_2}|\leq 4.4\times 10^{-3}$, we set the following bounds for the bare sub-amplitudes related to the annihilation topologies:
\begin{equation} \label{eq:Xbbound}
-100 \leq \mathrm{Re}(X_{b}), \, \mathrm{Im}(X_{b}) \leq 100\,,
\end{equation}
with
\begin{equation}
X_{b}\in \bigl\{ \tilde{A}, \, \tilde{E}, \, \tilde{T}_{AS}, \, \tilde{P}_{TA}, \, \tilde{P}_{TE}, \, \tilde{P}_{TE}, \, \tilde{P}_{A}, \, \tilde{P}_{AS}, \, \tilde{P}_{ES}, \, \tilde{P}_{SS} \bigl\}.
\end{equation}
These bounds are rather conservative compared to the true values of $b_i$ obtained in the QCDF approach~\cite{Beneke:1999br,Beneke:2000ry,Beneke:2001ev,Beneke:2003zv}. In practice, we scan the relevant real and imaginary parts of these bare sub-amplitudes using a computer cluster by sending $2000$ instances in parallel. In each instance, we produce $10^{6}$ random points and evaluate the corresponding $\chi^2$ function in each of these points, while selecting only the ones with $\chi^2<10^6$. We use each point satisfying the previous condition as the starting point for the python minimization function ``\texttt{minimize}''; in  particular, we select the algorithm  ``\texttt{sequential-least squares (SQLS)}'' that is included in the library \texttt{scipy.optimize}, because it allows us to include the following observables with only upper bounds during the minimization procedure (see tables~\ref{tab:resultsBRI}~--~\ref{tab:resultsBRIII} and~\ref{tab:resultsACPII}):
\begin{align} \label{eq:exp_bound}
{\rm Br}(\bar{B}_s\rightarrow \pi^0\pi^0) &< 7.7\times 10^{-6},\, &
{\rm Br}(\bar{B}_s\rightarrow \eta\pi^0) &< 1.0 \times 10^{-3},\, \nonumber\\
{\rm Br}(\bar{B}^0 \rightarrow \eta\eta) &< 1.0\times 10^{-6}, &
{\rm Br}(\bar{B}^0\rightarrow \eta'\eta') &< 1.7 \times  10^{-6}, \nonumber\\
{\rm Br}(\bar{B}^0\rightarrow \eta'\eta) &< 1.2\times 10^{-6},\, &
{\rm Br}(\bar{B}_s \rightarrow \eta'\eta) &< 6.5 \times 10^{-5},\, \nonumber \\
{\rm Br}(\bar{B}_s\rightarrow \eta' K^0) &< 8.16\times  10^{-6},\, &
{\rm Br}(\bar{B}_s\rightarrow \eta\eta) &< 1.43 \times 10^{-4},\, \nonumber\\
{A}_{CP}(\bar{B}_s \rightarrow \eta K^0) &< 0.001. 
\end{align}

Unlike in ref.~\cite{Huber:2021cgk}, in this analysis we do not impose the exact constraint on $\tilde{T}$ derived from our current knowledge of $\alpha_1$ in QCDF~\cite{Beneke:2009ek}, but assume only the conservative bound specified by eq.~\eqref{eq:Xabound} at the random sampling stage to find a suitable starting point for the minimization itself. Nevertheless, we account for the NLO and NNLO results obtained in QCDF by enforcing the relationships indicated in eq.~\eqref{eq:NNLO_conditions}, which imply
\begin{eqnarray}
\tilde{T}_{PA}=0,\qquad
\tilde{T}_{SS}=0,\qquad
\tilde{T}_{S}=0,\qquad
|\tilde{T}_{P}|<0.02,
\label{eq:tilde_conditions}
\end{eqnarray}
and also reduce the number of degrees of freedom to be determined from the $\chi^2$-fit to experimental data. 

Our second approach employs a Bayesian inference framework, whose principal advantage lies in its ability to integrate prior physical knowledge with experimental observations in a probabilistic manner, yielding directly the posterior distributions of the parameters and thereby providing a parameter estimation with full uncertainty quantification~\cite{Albert:2024zsh,Caldwell:2008fw,Gregory_2005}. To this end, we perform the Markov-Chain-Monte-Carlo (MCMC) sampling using the \texttt{Stan} probabilistic programming platform~\cite{Carpenter2017}. Internally, \texttt{Stan} utilizes the state-of-the-art No-U-Turn Sampler (NUTS), an adaptive variant of the Hamiltonian Monte Carlo (HMC) algorithm that exploits gradient information of the log-posterior density to automatically adjust the step size and trajectory length~\cite{Hoffman2014}. This enables efficient exploration of the high-dimensional parameter space considered in this analysis. To ensure adequate chain mixing and robust convergence, we set the target acceptance rate parameter \texttt{adapt\_delta} to $0.99$ and increase the maximum tree depth \texttt{max\_treedepth} to $20$ in the \texttt{Stan} configuration.

During such a Bayesian analysis, the specification of prior distributions can have a substantial impact on the fitting results. As the preliminary constraints on the TDA parameters have already been obtained in our first approach, we do not resort to uninformative priors; rather, we adopt informative priors centered at the previously determined best-fit values~\cite{Trotta:2008bp}. Concretely, the real and imaginary parts of each TDA amplitude are assigned as independent normal distributions $\mathcal{N}(\mu_i, \sigma_i)$, where the mean $\mu_i$ is set to the corresponding earlier central value, while the standard deviation $\sigma_i$ is moderately relaxed to preserve sufficient freedom for the parameter space to be explored by the data. This strategy strikes an appropriate balance between prior guidance and data-driven inference, mitigating the sampling inefficiency often associated with purely uninformative priors while avoiding the risk of over-constraining the parameters and thereby obscuring the information contained in the data. Regarding the convergence diagnostics, the $\widehat{R}$ statistic for all parameters remains well below $1.05$, and the effective sample sizes substantially exceed the commonly recommended thresholds, indicating that the MCMC chains have mixed satisfactorily and that the posterior estimates are numerically stable and reliable~\cite{Vehtari2021}.

As an important cross-check of the two analysis frameworks, we have performed a systematic comparison between the central values and credible intervals of the posterior distributions obtained from this Bayesian fit and the corresponding results derived earlier by using an independent fitting methodology. The two sets of results are found to be in good agreement within their statistical uncertainties, without any significant discrepancies observed. Consequently, this consistency reinforces the reliability and robustness of the conclusions presented in this work.

\section{Results}
\label{sec:results}

\subsection{Fit results}
\label{sec:Fit_results}

\begin{table}[t]
\setlength{\tabcolsep}{13pt}
\renewcommand{\arraystretch}{1.45}
\centering
\begin{tabular}{|c|c|}
\hline
Amplitude & Value \\
\hline
\rule{0pt}{16.8pt}
$\tilde{T}$   & $(1.073^{+0.024}_{-0.024}) + (0.044^{+0.017}_{-0.017})\,i$ \\[0.2em]
$\tilde{C}$   & $(0.334^{+0.046}_{-0.047}) + (-0.689^{+0.048}_{-0.047})\,i$ \\[0.2em]
$\tilde{A}$   & $(4.782^{+8.455}_{-8.745}) + (8.524^{+7.830}_{-7.725})\,i$ \\[0.2em]
$\tilde{E}$   & $(-7.655^{+7.551}_{-6.961}) + (12.782^{+5.218}_{-10.542})\,i$ \\[0.2em]
$\tilde{T}_{ES}$ & $(-80.703^{+9.453}_{-9.697}) + (8.656^{+28.430}_{-28.362})\,i$ \\[0.2em]
$\tilde{T}_{AS}$ & $(-8.747^{+92.947}_{-66.053}) + (20.809^{+\phantom{0}49.991}_{-107.809})\,i$ \\[0.2em]
$\tilde{T}_{P}$  & $\left(-7.421^{+7.411}_{-2.579}\right)\times10^{-5}
 + (-0.020^{+0.012}_{-0.018})\,i$ \\[0.2em]
\hline
\rule{0pt}{16.8pt}
$\tilde{P}_{T}$  & $(-0.245^{+0.002}_{-0.002}) + (-0.133^{+0.003}_{-0.003})\,i$ \\[0.2em]
$\tilde{P}_{C}$  & $(0.179^{+0.002}_{-0.002}) + (0.154^{+0.004}_{-0.004})\,i$ \\[0.2em]
$\tilde{P}_{TA}$ & $(-57.273^{+0.629}_{-0.619}) + (-40.248^{+1.011}_{-1.020})\,i$ \\[0.2em]
$\tilde{P}$      & $(0.081^{+0.002}_{-0.002}) + (0.186^{+0.002}_{-0.002})\,i$ \\[0.2em]
$\tilde{P}_{TE}$ & $(35.058^{+2.022}_{-2.088}) + (28.547^{+0.721}_{-1.047})\,i$ \\[0.2em]
$\tilde{P}_{A}$  & $(-18.150^{+0.885}_{-0.910}) + (-10.960^{+0.354}_{-0.474})\,i$ \\[0.2em]
$\tilde{P}_{AS}$ & $(-24.555^{+14.835}_{-28.005}) + (-41.077^{+\phantom{0}6.357}_{-24.763})\,i$ \\[0.2em]
$\tilde{P}_{ES}$ & $(77.683^{+4.093}_{-4.331}) + (54.308^{+1.296}_{-1.736})\,i$ \\[0.2em]
$\tilde{P}_{SS}^{\quad\dagger}$ & $19.785 + 18.715\,i$ \\[0.2em]
$\tilde{S}$      & $(-0.199^{+0.006}_{-0.006}) + (-0.112^{+0.004}_{-0.004})\,i$ \\[0.2em]
\hline
\end{tabular}
\caption{Our best-fit values and the associated uncertainties for the bare topological sub-amplitudes. ${}^{\dagger}$The $68.27\%$ confidence level (C.L.) region allowed by $\tilde{P}_{SS}$ obeys the rule $7.0 \leq\sqrt{|{\rm Re}(\tilde{P}_{SS})-22.00|^2 +  |{\rm Im}(\tilde{P}_{SS})-10.42|^2} \leq 10.0$. 
\label{tab:TDA_bestfitpoint}}
\end{table}

\begin{table}[t]
\renewcommand{\arraystretch}{1.4} 
\setlength{\tabcolsep}{10pt} 
\centering
\begin{tabular}{|c|c|c|c|}
\hline
Parameter & Fit & Input & Pull \\
\hline
$\lambda$ & 0.225041 & $0.22504 \pm 0.00022$ & $\phantom{+}0.005$ \\
$\bar{\rho}$ & 0.158946 & $0.1562 \pm 0.0102$ & $\phantom{+}0.269$ \\
$\bar{\eta}$ & 0.352491 & $0.3564 \pm 0.0065$ & $-0.601$ \\
$A$ & 0.821698 & $0.8215 \pm 0.0146$ & $\phantom{+}0.014$ \\
\hline
$F_0^{B\to \pi}$ & 0.200490 & $0.192 \pm 0.022$ & $\phantom{+}0.386$ \\
$F_0^{B\to K}$ & 0.325403 & $0.326 \pm 0.010$ & $-0.060$ \\
$F_0^{B_s\to \bar K}$ & 0.202150 & $0.203 \pm 0.014$ & $-0.061$ \\
\hline
$f_\pi$ & 0.130345 & $0.1302 \pm 0.0012$ & $\phantom{+}0.121$ \\
$f_K$ & 0.155693 & $0.1557 \pm 0.0003$ & $-0.024$ \\
$f_q$ & 1.071508 & $(1.07 \pm 0.02)\,f_\pi$ & $\phantom{+}0.075$ \\
$f_s$ & 1.351514 & $(1.34 \pm 0.06)\,f_\pi$ & $\phantom{+}0.192$ \\
$\phi$ & $39.384546^\circ$ & $39.3^\circ \pm 1.0^\circ$ & $\phantom{+}0.085$ \\
\hline
$\frac{f_{B_s}}{f_B}$ & 1.209075 & $1.209 \pm 0.005$ & $\phantom{+}0.015$ \\
$f_B$ & 0.190000 & $0.190 \pm 0.0013$ & $\phantom{+}0.000$ \\
$f_{B^+}$ & 0.189999 & $0.190 \pm 0.0013$ & $\phantom{+}0.000$ \\
\hline
\end{tabular}
\caption{Best-fit values of the nuisance parameters written in the notation consistent with table~\ref{tab:formfactors_decayconstants}. Pulls are computed as $(p^{\rm fit}_j - p^{\rm input}_j)/\sigma^{\rm input}_j$ for a given parameter $p_j$.
\label{tab:bestfitpoints_input}}
\end{table}

The application of the minimization procedure described above delivers several sets of minima based on the output of the \texttt{SQLS} algorithm. In particular, we find that the minimum of the $\chi^2$ function is reached at multiple values for the topological sub-amplitudes which are not necessarily adjacent in parameter space. For these minima, the $\chi^2$ function experiences only marginal fluctuations with $\Delta \chi^2_{\min}/\chi^2_{\min}<2\times 10^{-5}$. We find, however, that among all these possibilities, there is a solution that obeys the expected behaviour from the NNLO results for the real and imaginary parts of $\tilde{T}$ and the real part of $\tilde{C}$ based on the QCDF calculation~\cite{Beneke:2009ek}. Thus, we pick this set of parameters as our best-fit point, which is shown in table~\ref{tab:TDA_bestfitpoint}. We would like to stress that this emerges naturally from the minimization procedure, while the NNLO constraints are only used in eq.~\eqref{eq:tilde_conditions}. Based on our best-fit point, we have performed a conservative estimation of the $\chi^2$ per degree of freedom and the corresponding $p$-value
\begin{eqnarray}
\chi^2/d.o.f.=1.67, \quad\quad p=0.072.
\end{eqnarray}
Our estimation is conservative in the sense that in the calculation of the number of degrees of freedom, we do not include the experimental bounds given in eq.~\eqref{eq:exp_bound}. For convenience, we have also shown in table~\ref{tab:bestfitpoints_input} the best-fit values of the nuisance parameters, together with the corresponding pulls. One can see that all the fitted values are well within the corresponding ranges of the input parameters.

\begin{figure}[htbp]
\centering
\includegraphics[width=0.49\textwidth]{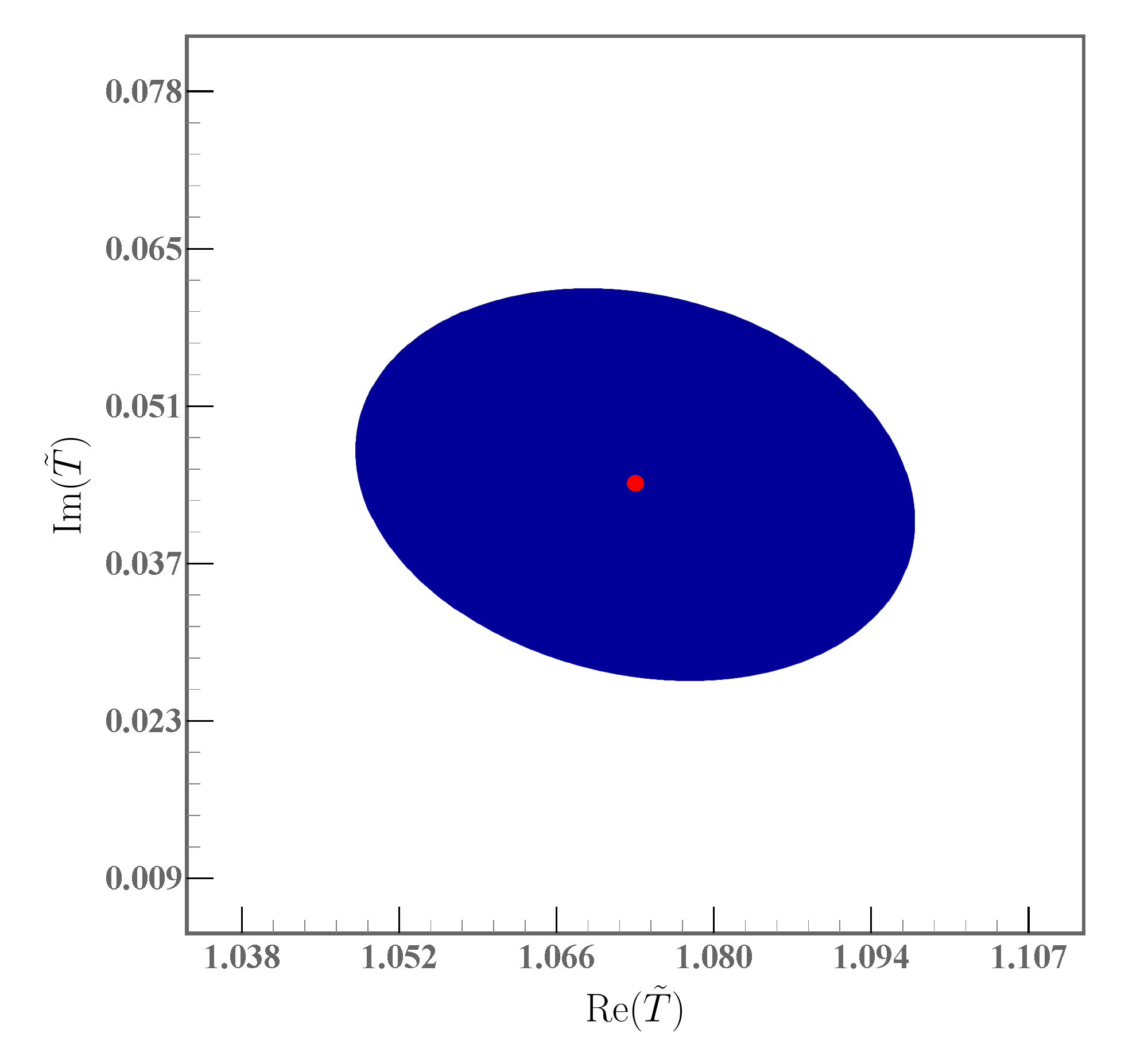}\hfill
\includegraphics[width=0.49\textwidth]{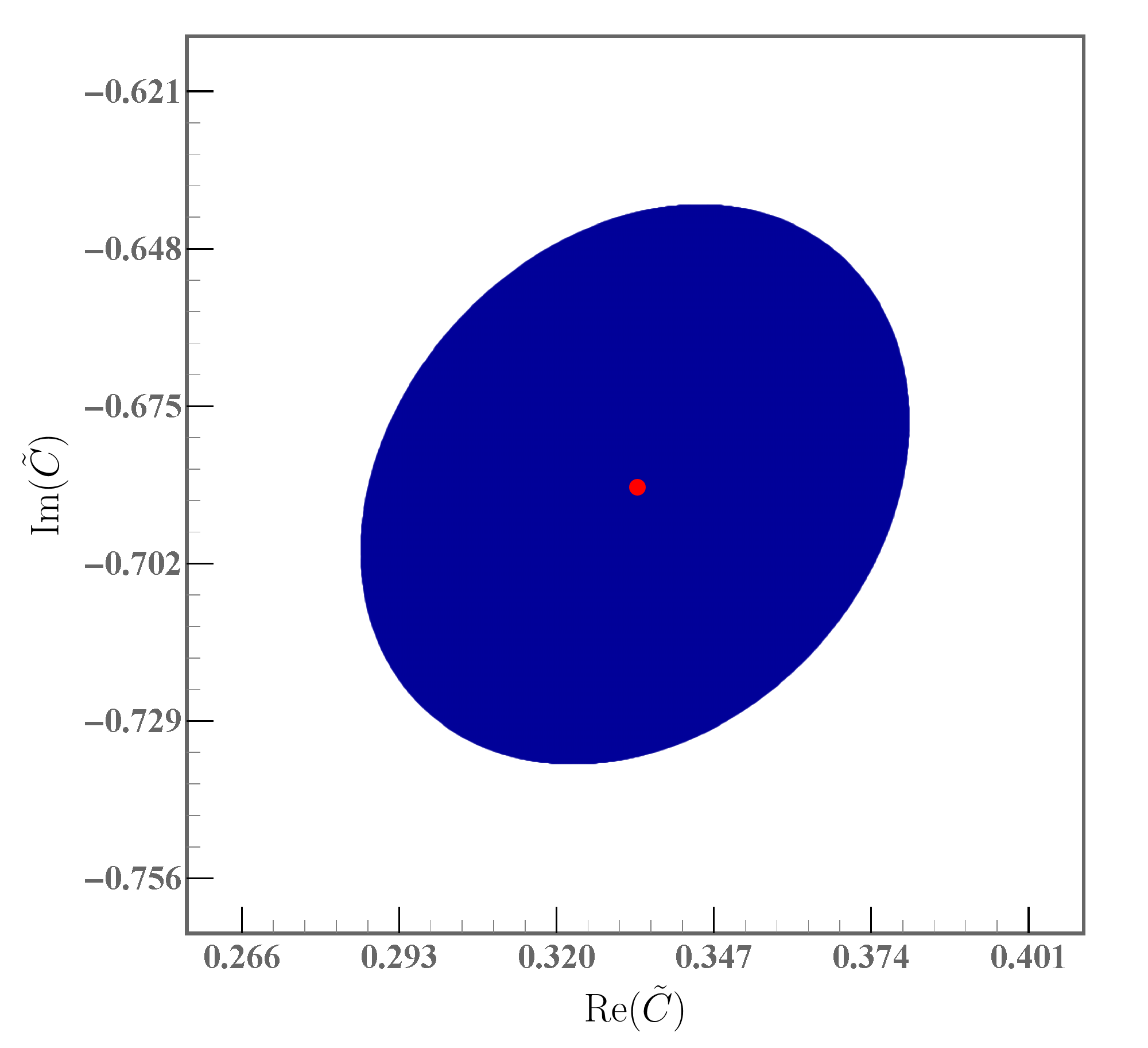}

\vspace{0.7em}

\includegraphics[width=0.49\textwidth]{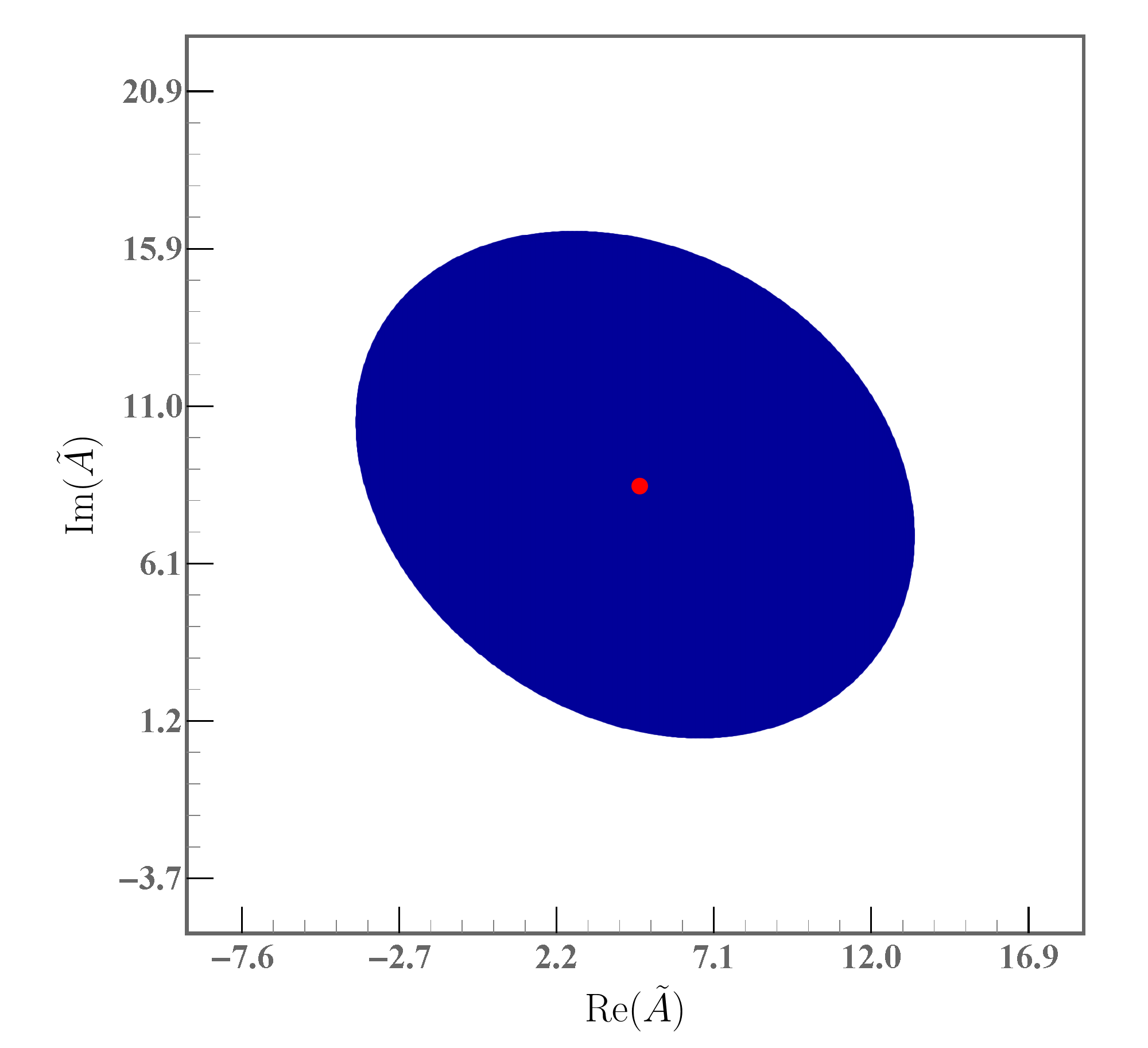}\hfill
\includegraphics[width=0.49\textwidth]{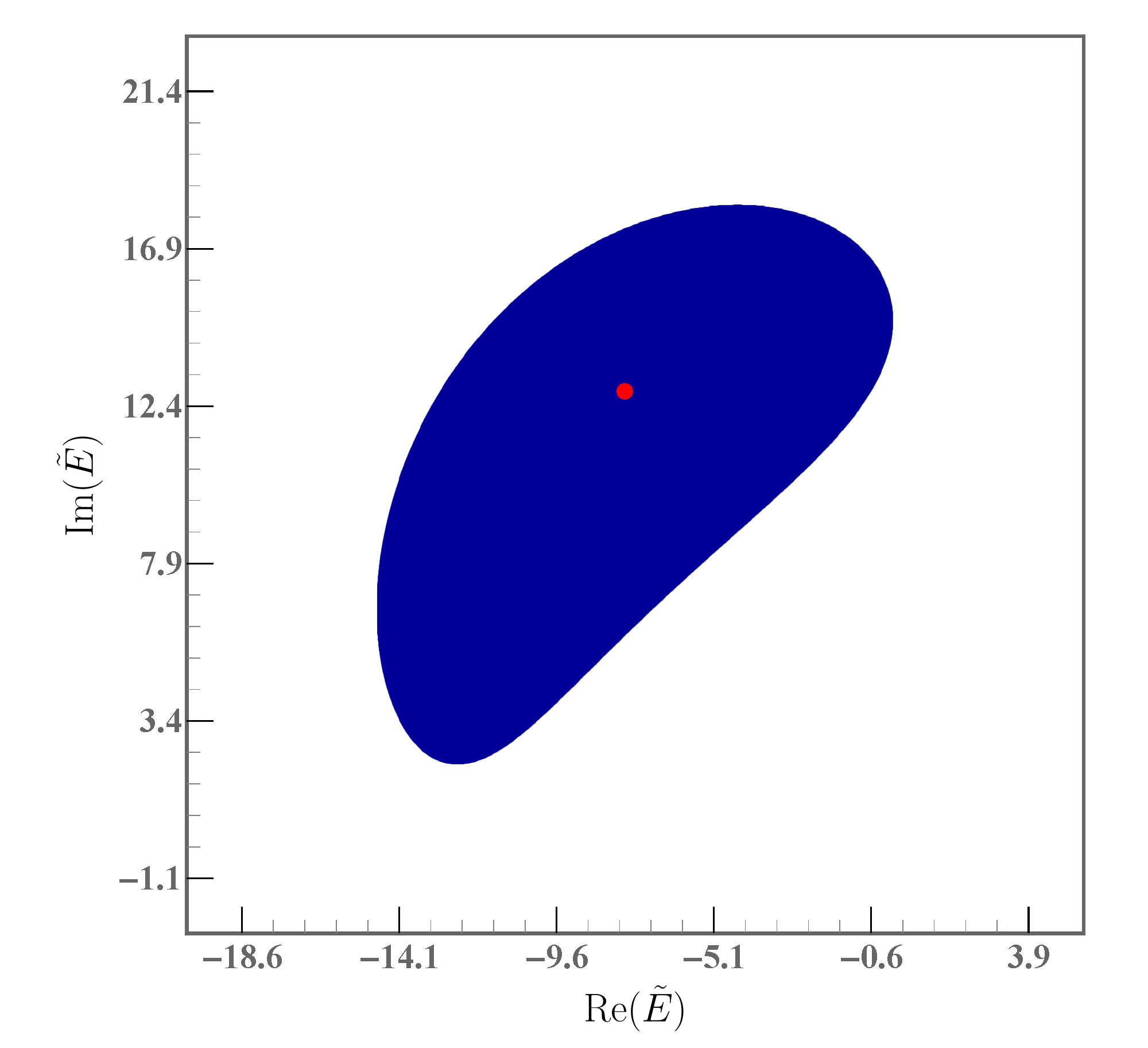}

\vspace{0.7em}

\includegraphics[width=0.49\textwidth]{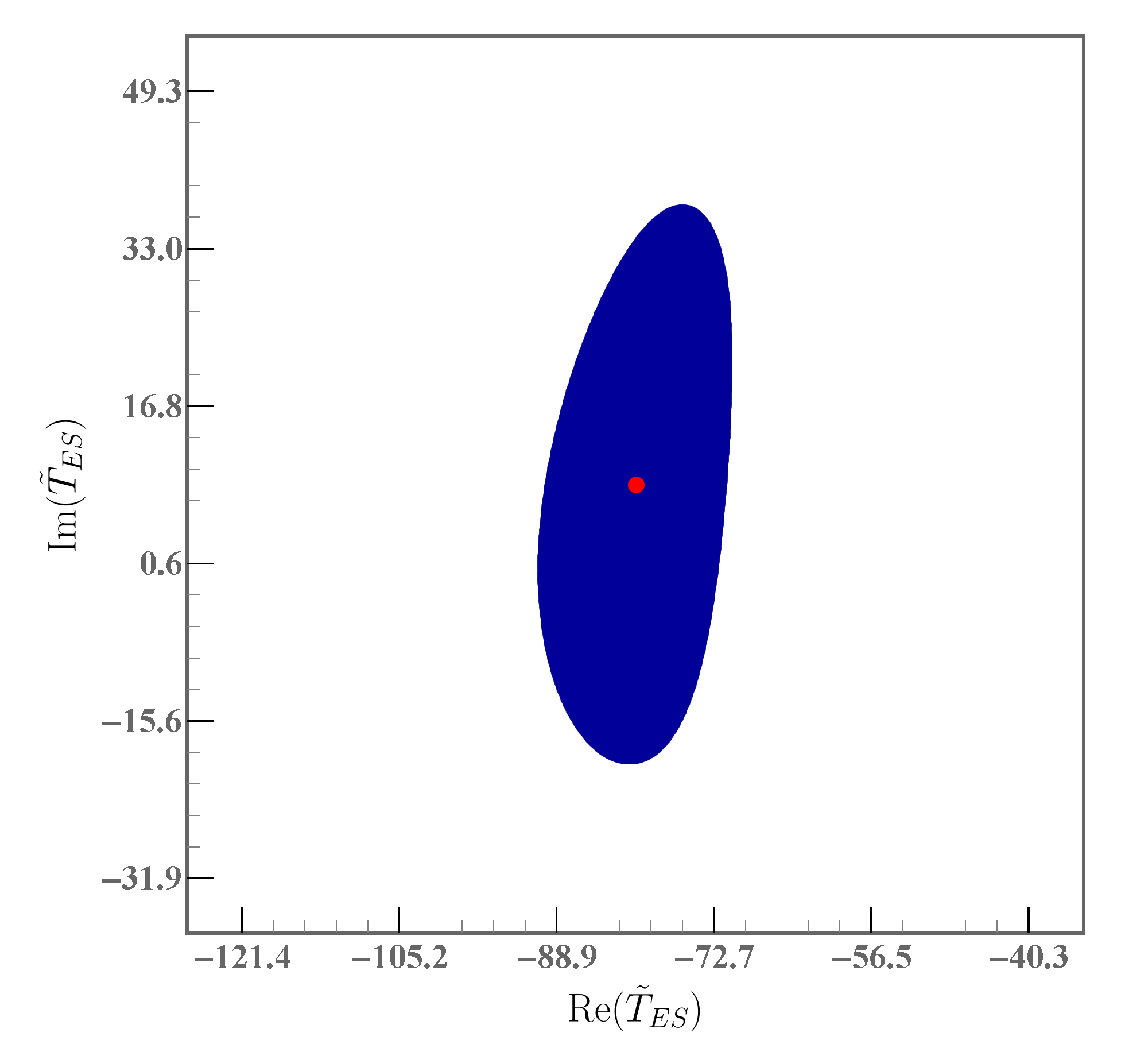}\hfill
\includegraphics[width=0.49\textwidth]{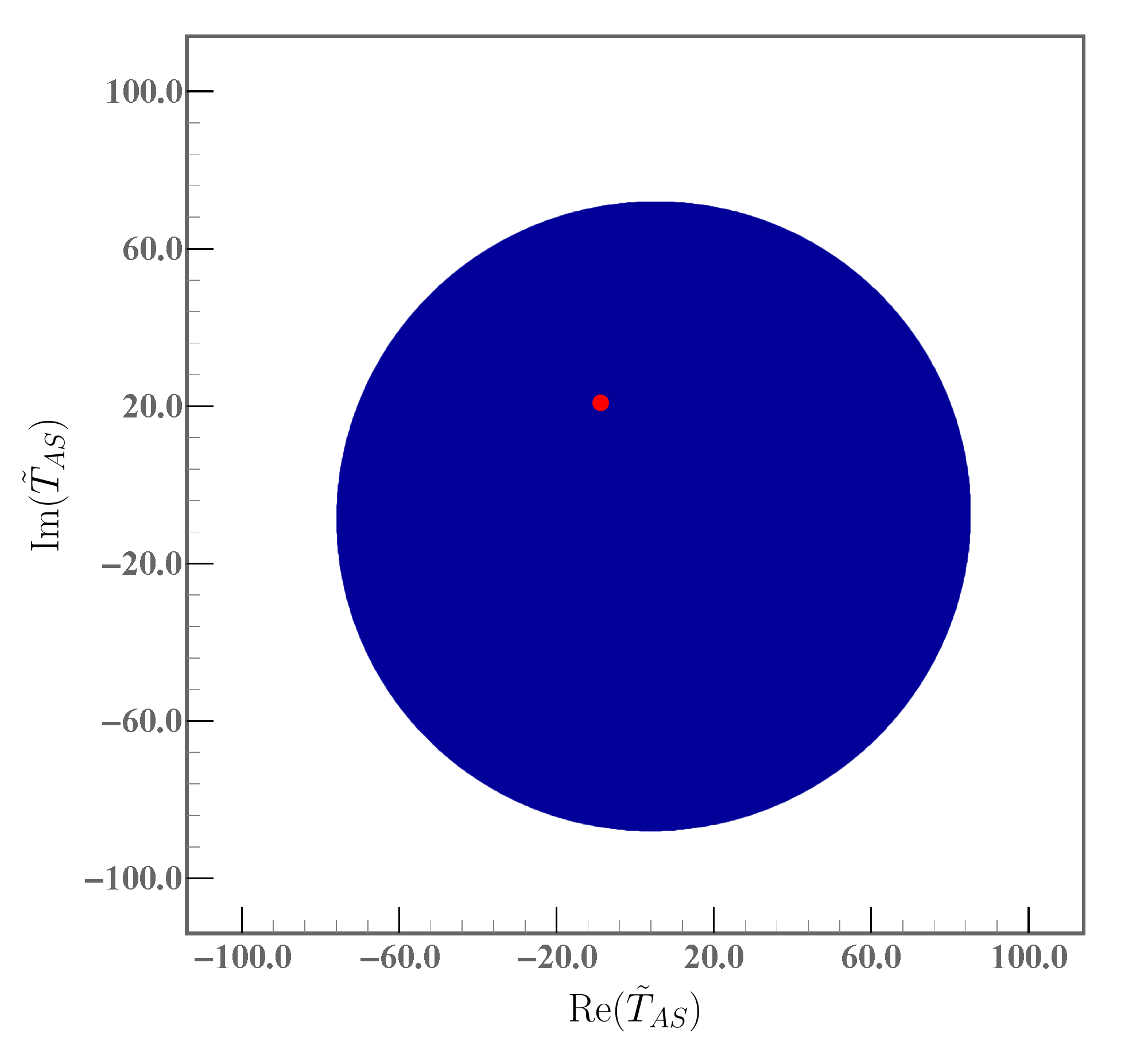}

\caption{Confidence regions for the topological sub-amplitudes in the tree sector. The best-fit point for each sub-amplitude is marked in red.}
\label{Fig:T_tilde_regions}
\end{figure}

\begin{figure}[htbp]
\centering
\includegraphics[width=0.49\textwidth]{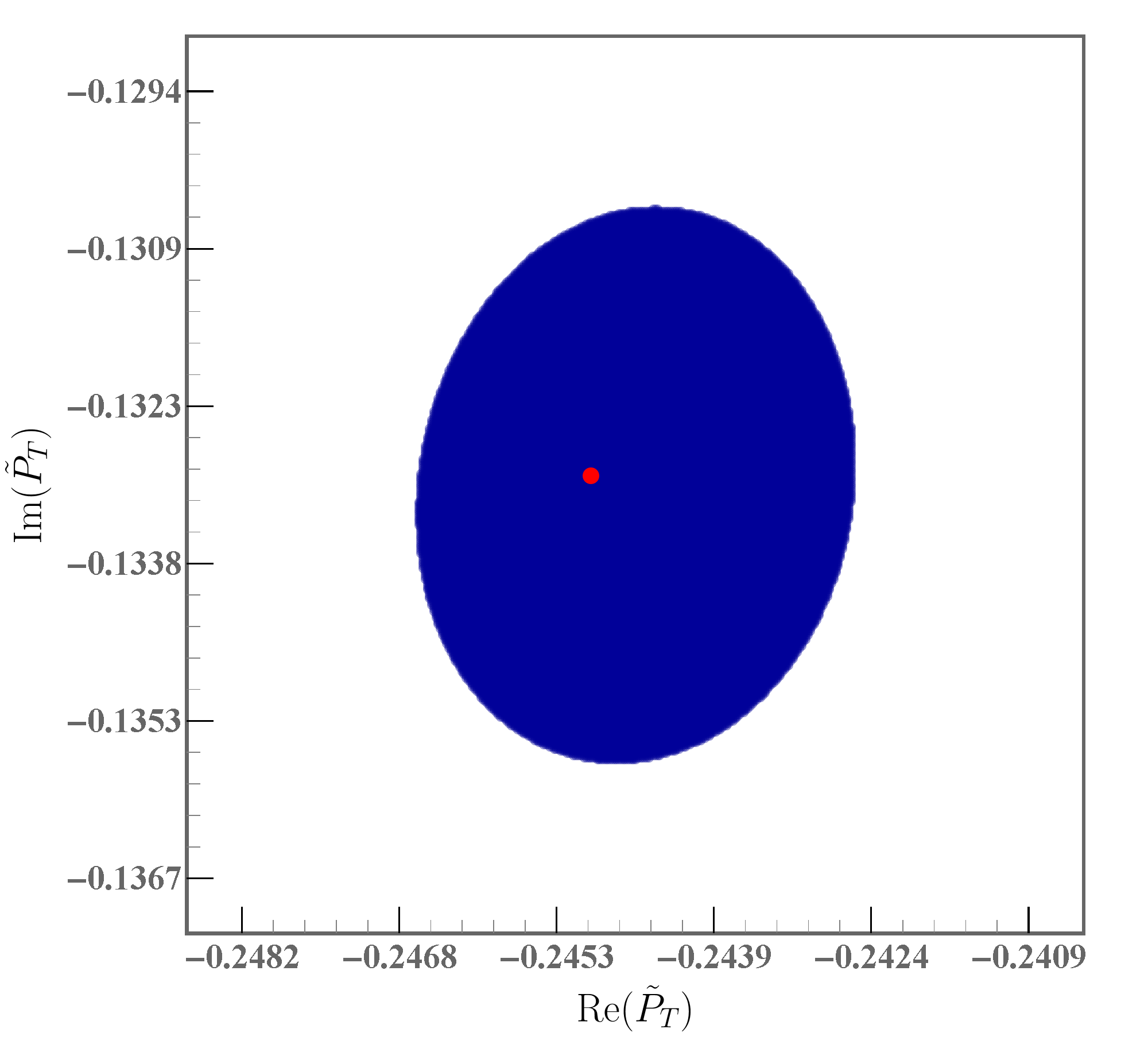}\hfill
\includegraphics[width=0.49\textwidth]{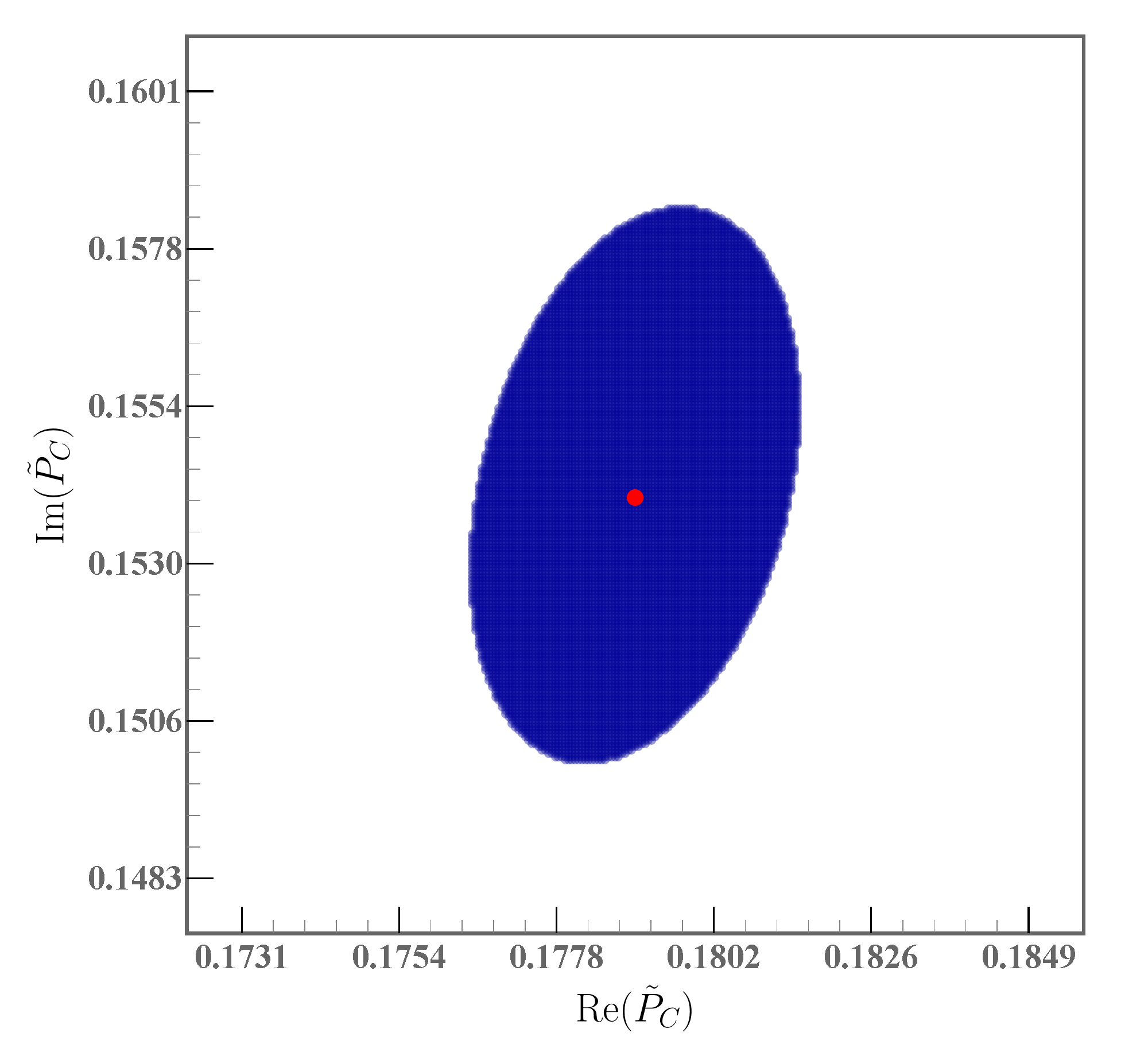}

\vspace{0.7em}

\includegraphics[width=0.49\textwidth]{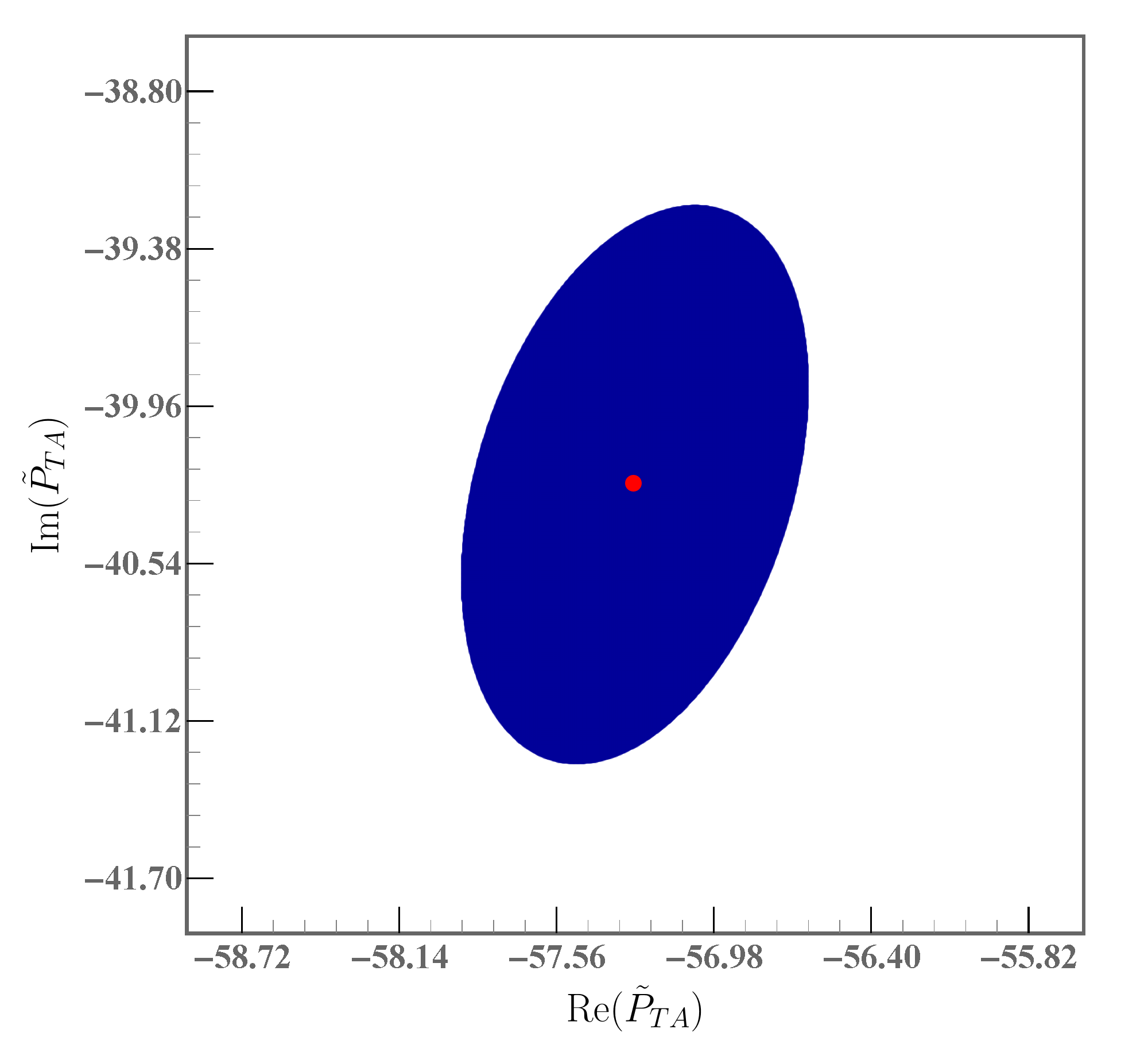}\hfill
\includegraphics[width=0.49\textwidth]{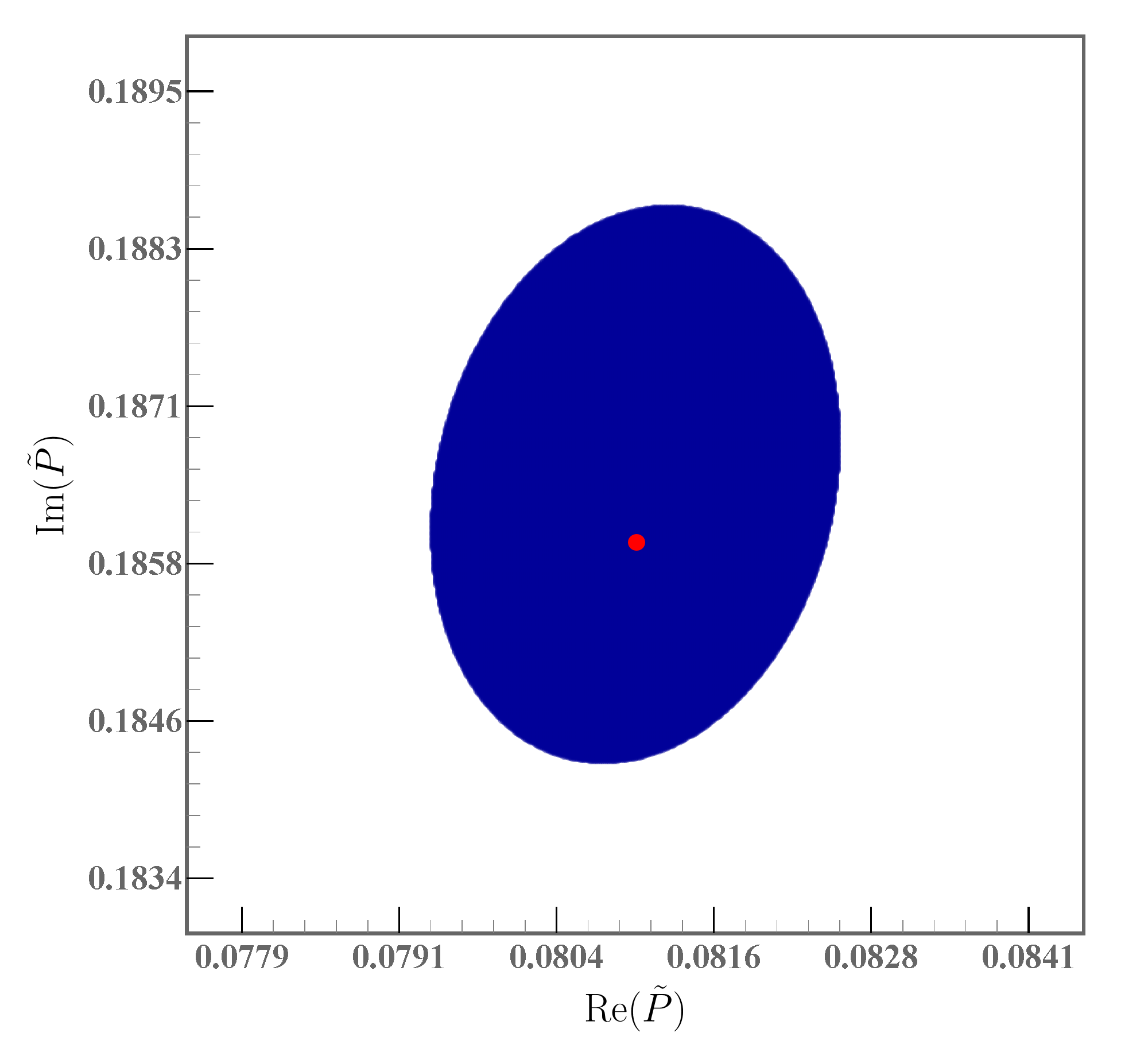}

\vspace{0.7em}

\includegraphics[width=0.48\textwidth]{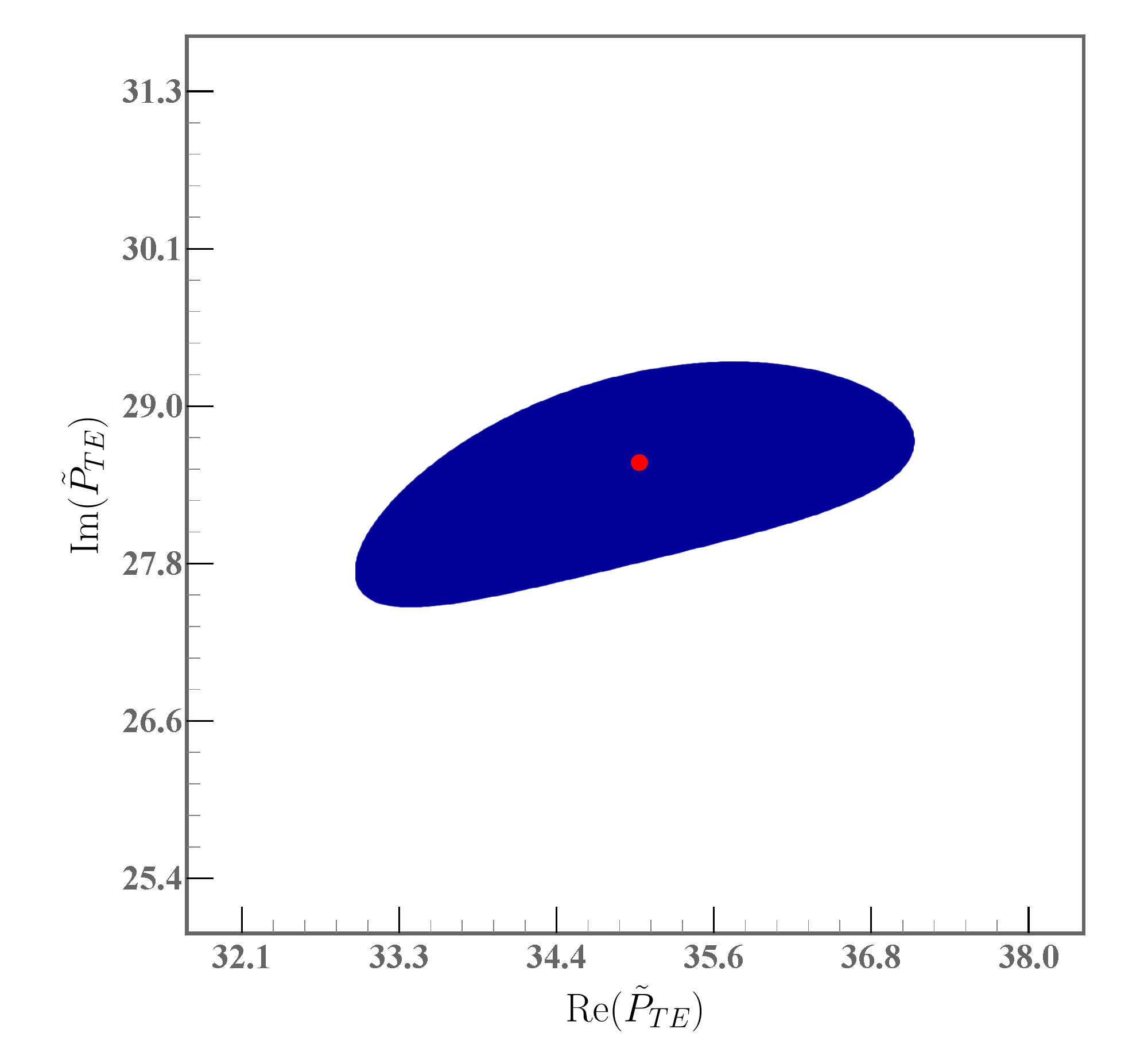}\hfill
\includegraphics[width=0.48\textwidth]{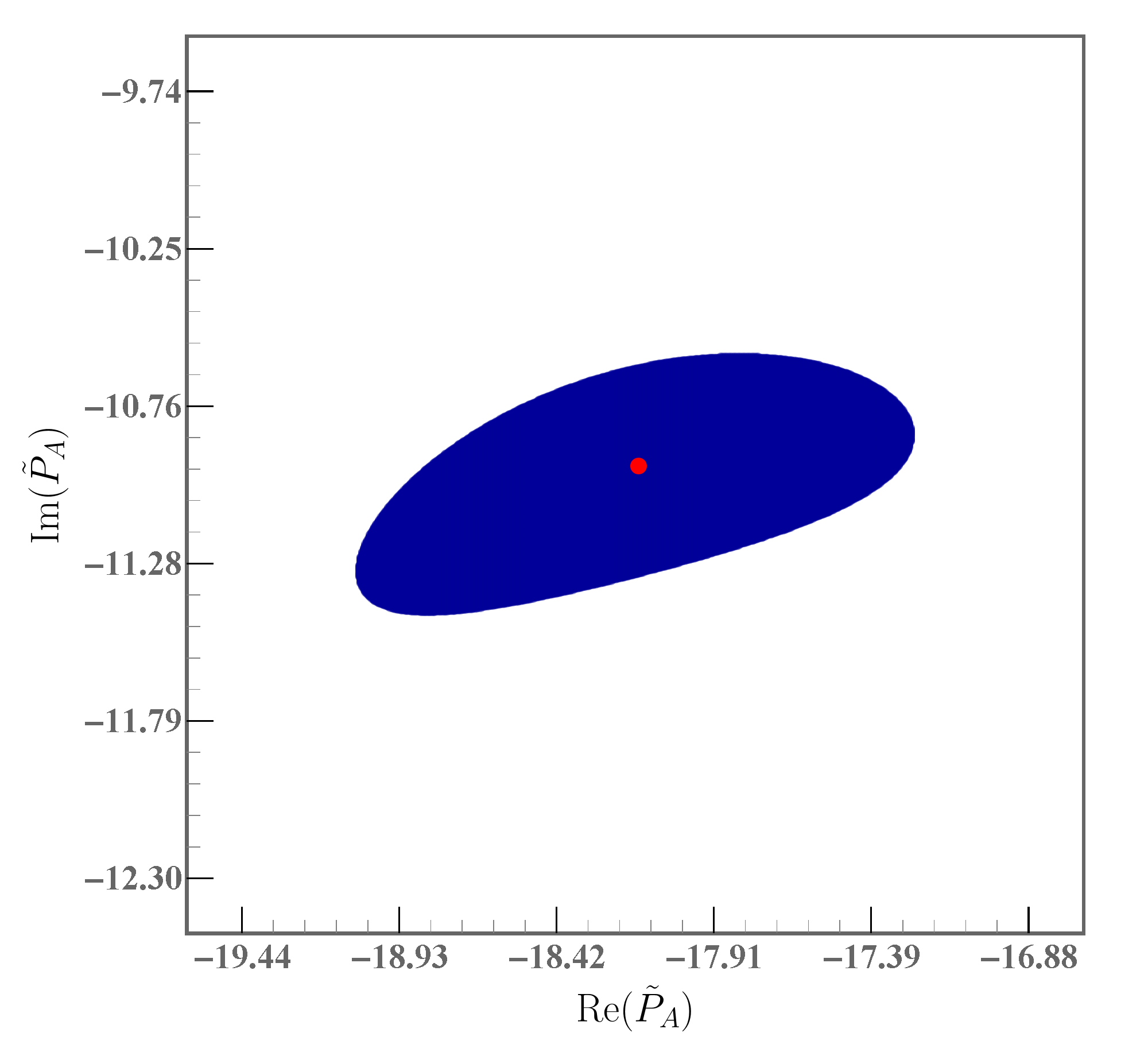}

\caption{Confidence regions for the topological sub-amplitudes in the penguin sector (part I). The best-fit point for each sub-amplitude is marked in red.
\label{Fig:P_tilde_regions}}
\end{figure}

\begin{figure}[ht]
\centering
\includegraphics[width=0.49\textwidth]{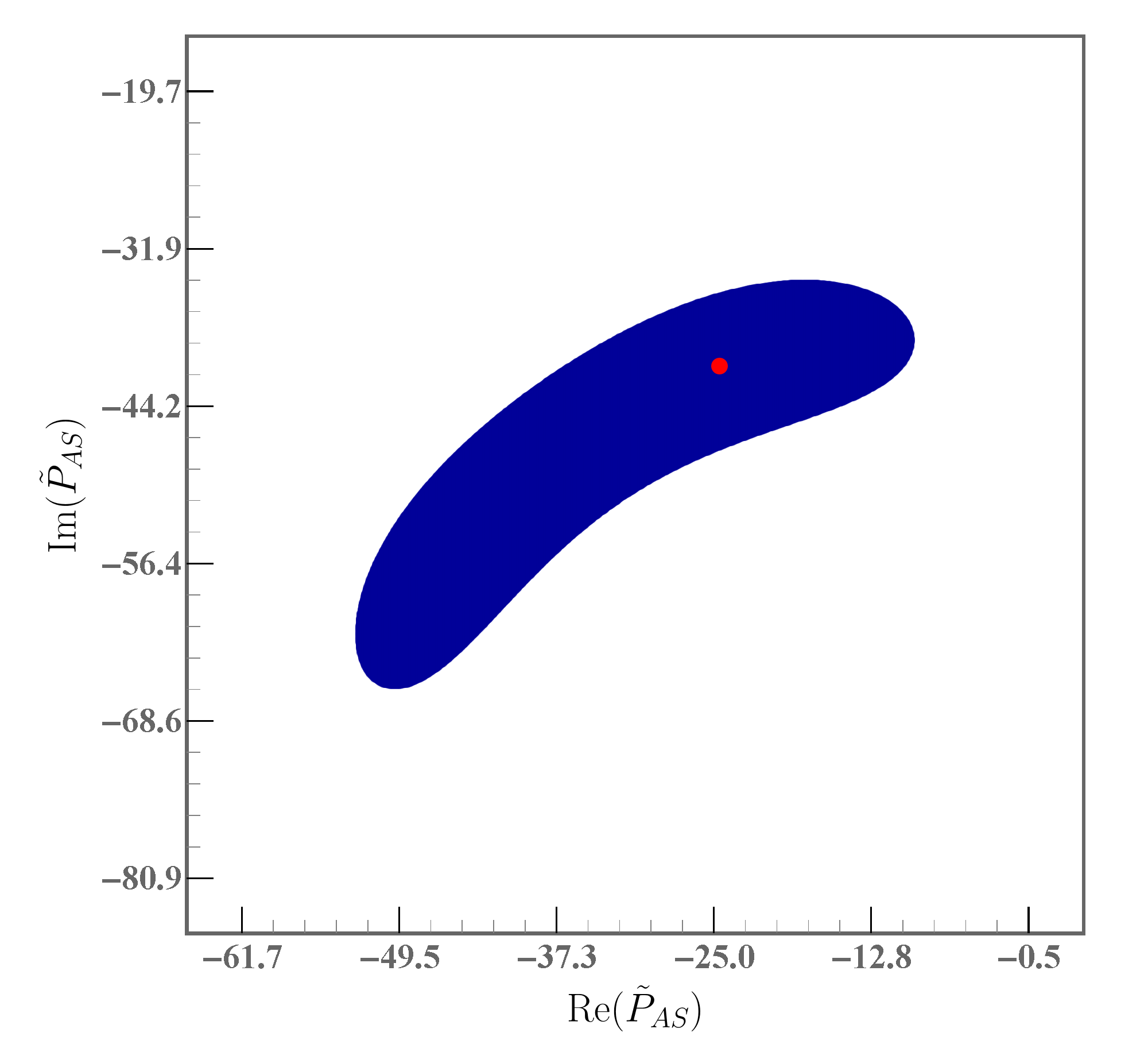}\hfill
\includegraphics[width=0.49\textwidth]{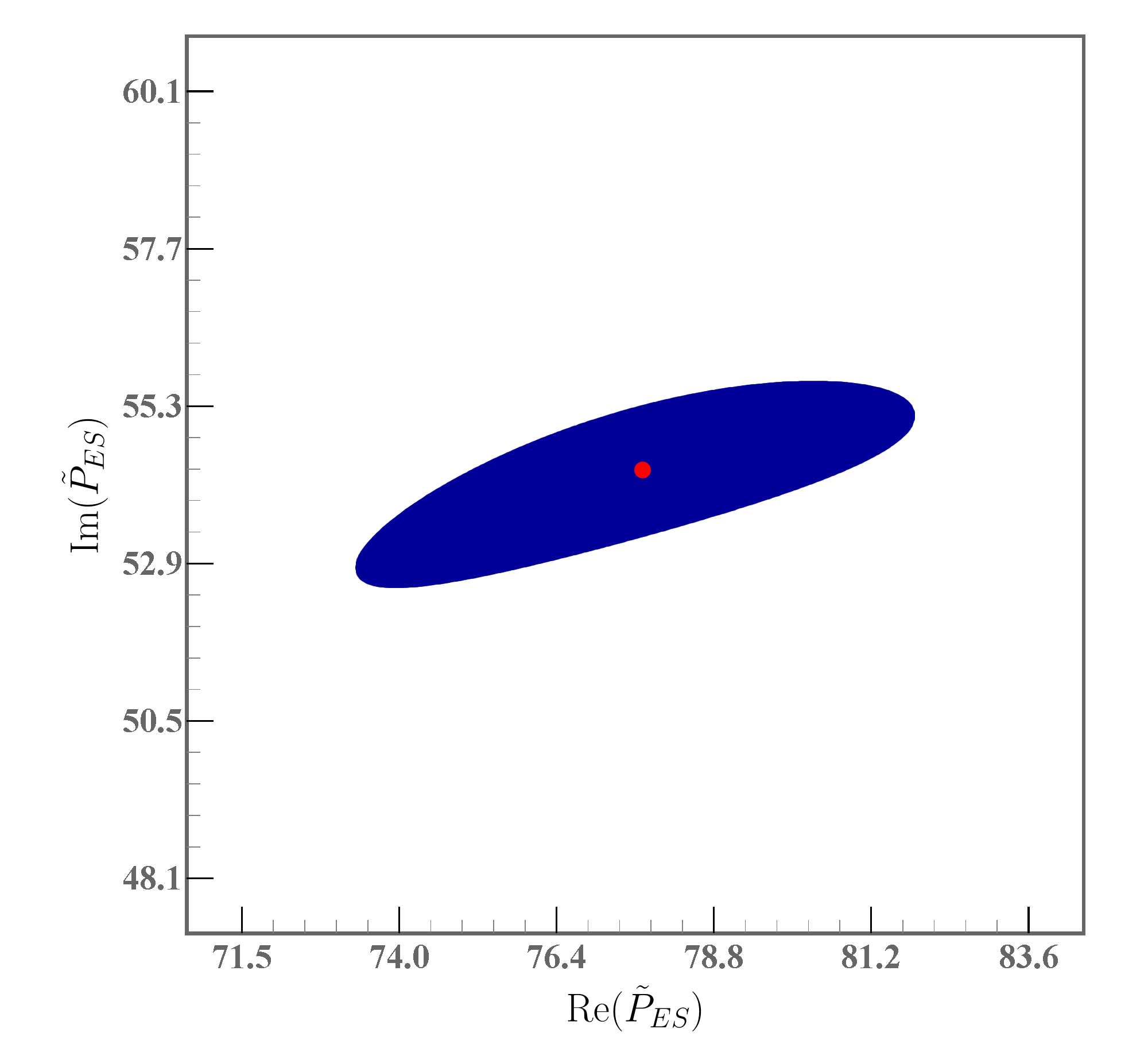}

\vspace{0.7em}

\includegraphics[width=0.49\textwidth]{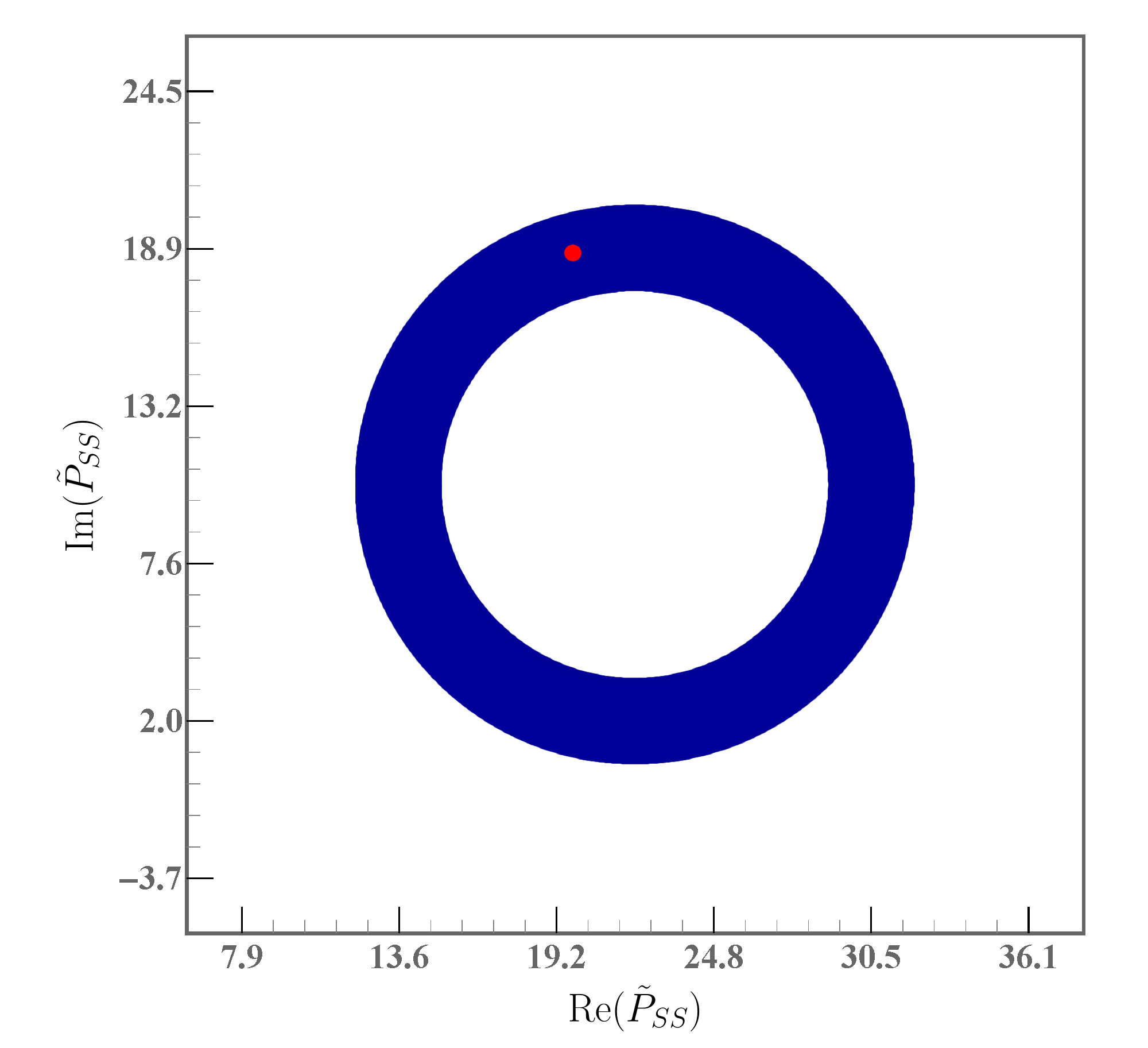}\hfill
\includegraphics[width=0.49\textwidth]{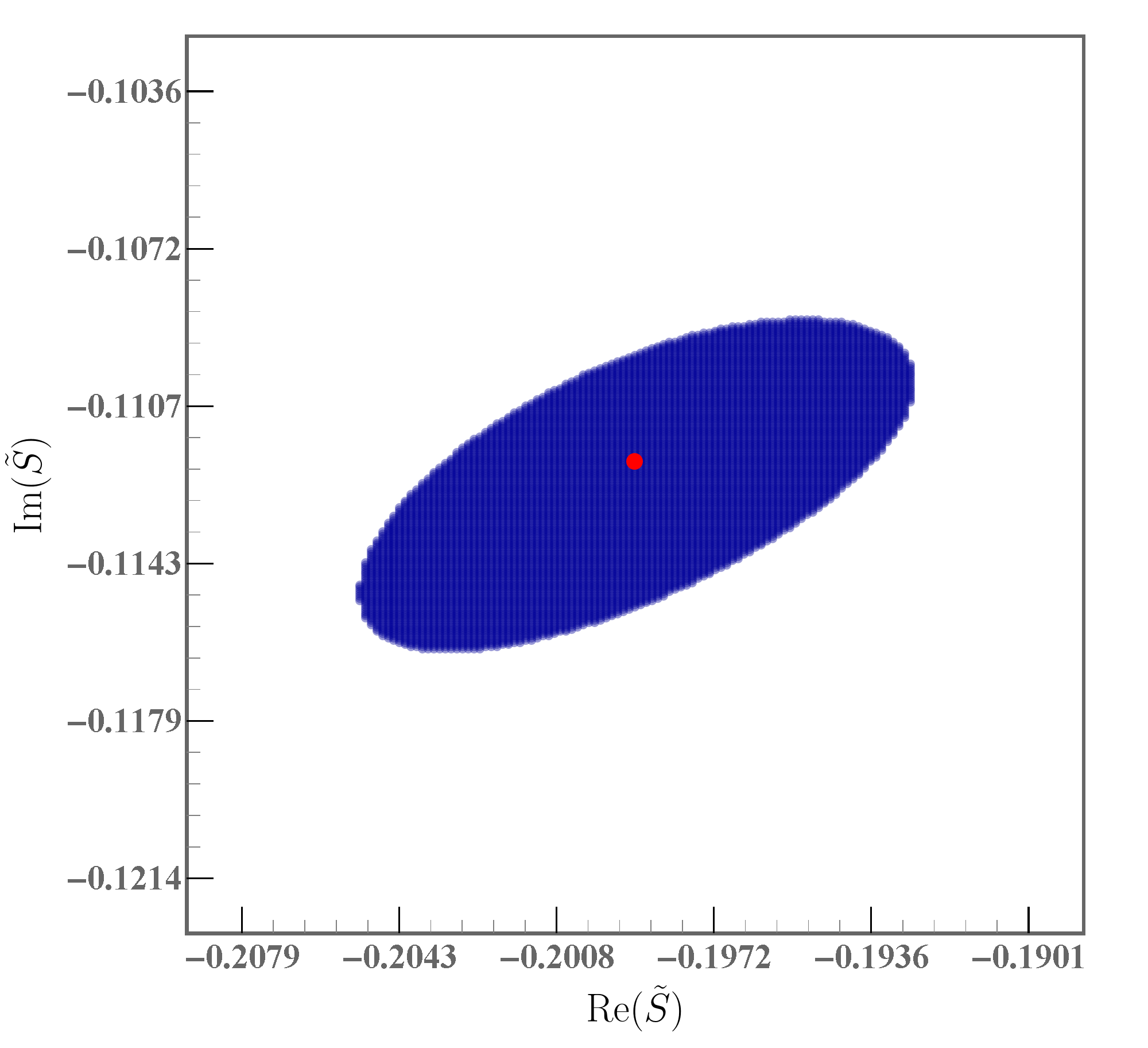}

\caption{Confidence regions for the topological sub-amplitudes in the penguin sector (part II). The best-fit point for each sub-amplitude is marked in red.
\label{Fig:P_tilde_regions_II}}
\end{figure}

The resulting two-dimensional $68\%$-confidence regions for the real and imaginary parts of the topological sub-amplitudes are shown in figures~\ref{Fig:T_tilde_regions} as well as \ref{Fig:P_tilde_regions} and \ref{Fig:P_tilde_regions_II} for the tree and penguin sectors, respectively. For convenience, we have also shown the best-fit point marked in red for each sub-amplitude. Notice that our estimations of the confidence regions are made simultaneously for the real and imaginary components of the individual sub-amplitudes, \textit{i.e.}, they are performed for two degrees of freedom. We obtain them by applying a $\Delta\chi^2$ test for two degrees of freedom at $68\%$ confidence level, which results in $\Delta\chi^2=\chi^2-\chi^2_0<2.3$, with $\chi^2_0$ being the value obtained from our best fit point and $\chi^2$ is the value obtained by assessing the null-hypothesis. Also note that we have simplified our evaluation by only allowing to float the parameters over which we are profiling the likelihood function. One can see that the confidence regions for the sub-amplitudes are all within the expected bounds given by eqs.~\eqref{eq:Xabound} and \eqref{eq:Xbbound}, and there exist strong correlations between the real and imaginary parts for some of them.

\subsection{\texorpdfstring{Individual QCDF sub-amplitudes $\beta_i$}{Individual QCDF sub-amplitudes beta}}
\label{sec:ampresults}

The fitted values of the QCDF sub-amplitudes $\beta_i$ can be inferred from eq.~\eqref{eq:QCDF2TDA}, together with the topological sub-amplitudes given in table~\ref{tab:TDA_bestfitpoint}. Our final results are collected in tables~\ref{tab:beta_amplitudes} and \ref{tab:beta_ew_channels} for the interesting decay modes like $B_{(s)}\to \pi\pi, \pi K$ and $K \bar{K}$, as well as the modes involving $\eta^{(\prime)}$. It is interesting to note that there are no strong indications that these annihilation amplitudes are numerically enhanced beyond the naive $\Lambda_{\textrm{QCD}}/m_b$ scaling, as generally expected in the QCDF and SCET frameworks~\cite{Beneke:1999br,Beneke:2000ry,Beneke:2001ev,Beneke:2003zv,Arnesen:2006vb}.\footnote{It is observed in ref.~\cite{Arnesen:2006vb} that the leading $\mathcal{O}(\alpha_s(m_b) \Lambda_{\rm QCD}/m_b)$ contributions to annihilation amplitudes are real, and a complex nonperturbative parameter from annihilation first appears at $\mathcal{O}(\alpha_s^2(\sqrt{\Lambda_{\rm QCD} m_b}) \Lambda_{\rm QCD}/m_b)$, implying that incalculable strong phases are suppressed in these amplitudes unless the $\alpha_s(\sqrt{\Lambda_{\rm QCD} m_b})$ expansion breaks down. As can be seen from tables~\ref{tab:TDA_bestfitpoint}, \ref{tab:beta_amplitudes} and \ref{tab:beta_ew_channels}, our result for the real part of each annihilation sub-amplitude is generally associated with a comparable imaginary part.} It is also noted that when both $A_{M_1 M_2}$ and $B_{M_1 M_2}$ are well-defined by eq.~\eqref{eq:AB_definitions}, the resulting QCDF sub-amplitudes $\beta_i$ would be independent of the decay constant $f_{M_2}$, as can be inferred from eqs.~\eqref{eq:bibetai} and \eqref{eq:AB_definitions}. This explains why some entries in tables~\ref{tab:beta_amplitudes} and \ref{tab:beta_ew_channels} are the same for different decay modes.

\begin{table}[htbp]
\centering
\scriptsize
\setlength{\tabcolsep}{10pt}
\renewcommand{\arraystretch}{1.9}
\begin{adjustbox}{width=0.99\textwidth,center,keepaspectratio}
\begin{tabular}{|c|c|c|c|c|}
\hline
Channel & $\beta_1$ & $\beta_2$ & $\beta_{S1}$ & $\beta_{S2}$ \\
\hline

$B \to \pi\pi$ &
$(-3.60^{+3.50}_{-3.70}) + (5.90^{+5.10}_{-4.80})\,i$ &
$(2.22^{+1.84}_{-1.84}) + (3.91^{+0.72}_{-0.71})\,i$ & & \\

$B_s \to \pi\pi$ &
$(-3.00^{+3.00}_{-3.00}) + (5.00^{+4.10}_{-4.10})\,i$ & & & \\

$B \to \pi  K$ &
& $(2.22^{+1.84}_{-1.84}) + (3.91^{+0.72}_{-0.71})\,i$ & & \\

$B_s \to KK$ &
$(-4.70^{+4.70}_{-4.80}) + (8.00^{+6.30}_{-6.40})\,i$ & & & \\

$B \to KK$ &
$(-3.50^{+3.50}_{-3.40}) + (5.90^{+4.90}_{-4.80})\,i$ &
$(1.60^{+2.90}_{-2.80}) + (2.80^{+2.60}_{-2.60})\,i$ & & \\

\hline

$B \to \pi \eta_q$ &
$(-3.60^{+3.50}_{-3.70}) + (5.90^{+5.10}_{-4.80})\,i$ &
$(2.22^{+1.84}_{-1.84}) + (3.91^{+0.72}_{-0.71})\,i$ &
$(-4.14^{+43.66}_{-43.65}) + (9.75^{+50.80}_{-50.77})\,i$ &
$(-37.73^{+5.84}_{-6.85}) + (4.02^{+13.45}_{-13.29})\,i$ \\

$B \to \pi \eta_q^\prime$ &
$(-3.60^{+3.50}_{-3.70}) + (5.90^{+5.10}_{-4.80})\,i$ &
$(2.22^{+1.84}_{-1.84}) + (3.91^{+0.72}_{-0.71})\,i$ &
$(-4.14^{+43.66}_{-43.65}) + (9.75^{+50.80}_{-50.77})\,i$ &
$(-37.73^{+5.84}_{-6.85}) + (4.02^{+13.45}_{-13.29})\,i$ \\

$B \to \eta_q \pi$ &
$(-3.58^{+3.52}_{-3.64}) + (5.98^{+4.10}_{-4.62})\,i$ &
& & \\

$B \to \eta_q^\prime \pi$ &
$(-3.56^{+3.52}_{-3.60}) + (6.02^{+5.07}_{-4.93})\,i$ &
& & \\

$B \to K \eta_q$ &
& & &
$(-26.50^{+3.27}_{-3.30}) + (2.81^{+9.33}_{-9.32})\,i$ \\

$B \to K \eta_q^\prime$ &
& & &
$(-26.50^{+3.27}_{-3.30}) + (2.81^{+9.33}_{-9.32})\,i$ \\

$B \to \eta_q K$ &
& $(2.23^{+4.15}_{-4.08}) + (3.98^{+3.77}_{-3.66})\,i$ & & \\

$B \to \eta_q^\prime K$ &
& $(2.30^{+4.24}_{-4.18}) + (4.08^{+3.85}_{-3.71})\,i$ & & \\

$B \to K \eta_s$ &
& $(1.60^{+2.90}_{-2.80}) + (2.80^{+2.60}_{-2.60})\,i$ & &
$(-26.50^{+3.27}_{-3.30}) + (2.81^{+9.33}_{-9.32})\,i$ \\

$B \to K \eta_s^\prime$ &
& $(1.60^{+2.90}_{-2.80}) + (2.80^{+2.60}_{-2.60})\,i$ & &
$(-26.50^{+3.28}_{-3.33}) + (2.83^{+9.36}_{-9.39})\,i$ \\

$B \to \pi \eta_s$ &
& & $(-4.14^{+43.66}_{-43.65}) + (9.75^{+50.80}_{-50.77})\,i$ &
$(-37.73^{+5.84}_{-6.85}) + (4.02^{+13.45}_{-13.29})\,i$ \\

$B \to \pi \eta_s^\prime$ &
& & $(-4.14^{+43.66}_{-43.65}) + (9.75^{+50.80}_{-50.77})\,i$ &
$(-37.73^{+5.84}_{-6.85}) + (4.02^{+13.45}_{-13.29})\,i$ \\

$B_s \to \pi \eta_q$ &
$(-2.47^{+2.43}_{-2.43}) + (4.10^{+3.40}_{-3.39})\,i$ &
& $(-2.84^{+29.98}_{-29.97}) + (6.89^{+34.57}_{-34.90})\,i$ & \\

$B_s \to \eta_q \pi$ &
$(-2.47^{+2.42}_{-2.42}) + (4.12^{+3.41}_{-3.41})\,i$ & & & \\

$B_s \to \pi \eta_q^\prime$ &
$(-2.01^{+2.00}_{-2.01}) + (3.37^{+2.79}_{-2.78})\,i$ &
& $(-2.30^{+24.52}_{-24.49}) + (5.53^{+28.44}_{-28.41})\,i$ & \\

$B_s \to \eta_q^\prime \pi$ &
$(-2.02^{+2.01}_{-2.02}) + (3.38^{+2.79}_{-2.78})\,i$ & & & \\

$B_s \to \pi \eta_s$ &
& &$( -4.13^{+43.77}_{-43.70}) + ( 9.68^{+50.83}_{-50.62})\,i$ & \\

$B_s \to \pi \eta_s^\prime$ &
& &$( -4.05^{+43.11}_{-43.25}) + ( 9.56^{+50.35}_{-50.14})\,i$ & \\

\hline

$B \to \eta_q \eta_q^\prime$ &
$(-3.58^{+3.52}_{-3.64}) + (5.98^{+4.10}_{-4.62})\,i$ &
& $(-4.07^{+43.63}_{-43.79}) + (9.71^{+51.10}_{-51.00})\,i$ & \\

$B \to \eta_q^\prime \eta_q$ &
$(-3.56^{+3.52}_{-3.60}) + (6.02^{+5.07}_{-4.93})\,i$ &
& $(-4.14^{+43.70}_{-43.94}) + (9.75^{+50.78}_{-50.82})\,i$ & \\

$B \to \eta_q \eta_s^\prime$ &
& &$(-4.07^{+43.63}_{-43.79}) + (9.71^{+51.10}_{-51.00})\,i$ & \\


$B \to \eta_q^\prime \eta_s$ &
& & $(-4.14^{+43.70}_{-43.94}) + (9.75^{+50.78}_{-50.82})\,i$ & \\


\hline

$B_s \to \eta_q \eta_q^\prime$ &
$(-1.67^{+1.64}_{-1.65}) + (2.79^{+2.31}_{-2.30})\,i$ &
& $(-1.87^{+20.29}_{-20.27}) + (4.51^{+23.57}_{-23.48})\,i$ & \\

$B_s \to \eta_q^\prime \eta_q$ &
$(-1.67^{+1.64}_{-1.65}) + (2.79^{+2.31}_{-2.30})\,i$ &
& $(-1.87^{+20.29}_{-20.27}) + (4.51^{+23.57}_{-23.48})\,i$ & \\

$B_s \to \eta_q \eta_s^\prime$ &
& & $( -2.65^{+28.63}_{-28.81}) + ( 6.27^{+33.25}_{-33.14})\,i$ & \\


$B_s \to \eta_q^\prime \eta_s$ &
& &$( 1.93^{+20.79}_{-20.72}) + ( -4.64^{+24.04}_{-24.12})\,i$ & \\


\hline

$B \to \eta_q \eta_q$ &
$(-3.58^{+3.52}_{-3.64}) + (5.98^{+4.10}_{-4.62})\,i$ &
& $(-4.07^{+43.63}_{-43.79}) + (9.71^{+51.10}_{-51.00})\,i$ & \\

$B \to \eta_q \eta_s$ &
$(-3.58^{+3.52}_{-3.64}) + (5.98^{+4.10}_{-4.62})\,i$ &
&$(-4.07^{+43.63}_{-43.79}) + (9.71^{+51.10}_{-51.00})\,i$ & \\

\hline

$B_s \to \eta_q \eta_q$ &
$(-3.59^{+3.54}_{-3.62}) + (5.98^{+5.11}_{-4.90})\,i$ &
& $(-2.25^{+24.78}_{-24.70}) + (5.72^{+28.67}_{-29.08})\,i$ & \\

$B_s \to \eta_q \eta_s$ &
& & $( 2.42^{+25.41}_{-25.42}) + ( -5.68^{+29.53}_{-29.41})\,i$ & \\

\hline

$B \to \eta_q^\prime \eta_q^\prime$ &
$(-3.56^{+3.52}_{-3.60}) + (6.02^{+5.07}_{-4.93})\,i$ &
& $(-4.14^{+43.70}_{-43.94}) + (9.75^{+50.78}_{-50.82})\,i$ & \\

$B \to \eta_q^\prime \eta_s^\prime$ &
$(-3.56^{+3.52}_{-3.60}) + (6.02^{+5.07}_{-4.93})\,i$ &
& $(-4.14^{+43.70}_{-43.94}) + (9.75^{+50.78}_{-50.82})\,i$ & \\

\hline

$B_s \to \eta_q^\prime \eta_q^\prime$ &
$(-1.37^{+1.35}_{-1.35}) + (2.28^{+1.89}_{-1.88})\,i$ &
& $(-1.32^{+13.68}_{-13.77}) + (3.02^{+16.01}_{-15.93})\,i$ & \\

$B_s \to \eta_q^\prime \eta_s^\prime$ &
& & $(-4.11^{+43.03}_{-43.25}) + (9.50^{+50.46}_{-50.18})\,i$ & \\


\hline
\end{tabular}
\end{adjustbox}

\caption{Fitted values of the QCDF sub-amplitudes $\beta_1$, $\beta_2$, $\beta_{S1}$ and $\beta_{S2}$. All values are given in units of $10^{-2}$.
\label{tab:beta_amplitudes}}
\end{table}

\begin{table}[htbp]
\centering
\scriptsize
\setlength{\tabcolsep}{3pt}
\renewcommand{\arraystretch}{1.6}
\begin{adjustbox}{width=0.99\textwidth,center,keepaspectratio}
\begin{tabular}{|c|c|c|c|c|}
\hline
 & $\beta_{3,\mathrm{EW}}$ & $\beta_{4,\mathrm{EW}}$ & $\beta_{S4,\mathrm{EW}}$ & $\beta_{S3,\mathrm{EW}}$ \\
\hline
$B \to \pi \pi$ &
$( -17.6^{+1.8}_{-2.3}) + ( -12.4^{+1.3}_{-1.6})\,i$ &
$( 10.8^{+1.6}_{-1.3}) + ( 8.8^{+1.2}_{-1.0})\,i$ &  &  \\

$B \to K K$ &
$( -12.5^{+0.4}_{-0.4}) + ( -8.8^{+0.3}_{-0.4})\,i$ &
$( 10.8^{+0.6}_{-0.6}) + ( 8.8^{+0.3}_{-0.3})\,i$ &  &  \\

$B \to \pi K$ &
$( -17.6^{+1.8}_{-2.3}) + ( -12.4^{+1.3}_{-1.6})\,i$ &  &  &  \\

$B_s \to K K$ &
 & $( 13.0^{+0.8}_{-0.8}) + ( 10.6^{+0.4}_{-0.4})\,i$ &  &  \\

$B_s \to \pi \pi$ &
 & $( 9.1^{+0.6}_{-0.6}) + ( 7.4^{+0.3}_{-0.3})\,i$ &  &  \\

\hline
$B \to \pi \eta_q$ &
$( -17.6^{+1.8}_{-2.3}) + ( -12.4^{+1.3}_{-1.6})\,i$ &
$( 10.8^{+1.6}_{-1.3}) + ( 8.8^{+1.2}_{-1.0})\,i$ &
$( -7.7^{+8.7}_{-8.9}) + ( -12.8^{+7.7}_{-8.1})\,i$ &
$( 24.2^{+3.5}_{-2.8}) + ( 16.9^{+2.3}_{-1.8})\,i$ \\

$B \to \pi \eta_q^\prime$ &
$( -17.7^{+1.8}_{-2.3}) + ( -12.4^{+1.3}_{-1.6})\,i$ &
$( 10.8^{+1.6}_{-1.3}) + ( 8.8^{+1.2}_{-1.0})\,i$ &
$( -7.7^{+8.7}_{-8.9}) + ( -12.8^{+7.7}_{-8.1})\,i$ &
$( 24.2^{+3.5}_{-2.8}) + ( 16.9^{+2.3}_{-1.8})\,i$ \\

$B \to \pi \eta_s$ &
 &  &
$( -7.7^{+8.7}_{-8.9}) + ( -12.8^{+7.7}_{-8.1})\,i$ &
$( 24.2^{+3.5}_{-2.8}) + ( 16.9^{+2.3}_{-1.8})\,i$ \\

$B \to \pi \eta_s^\prime$ &
 &  &
$( -7.7^{+8.7}_{-8.9}) + ( -12.8^{+7.7}_{-8.1})\,i$ &
$( 24.2^{+3.5}_{-2.8}) + ( 16.9^{+2.3}_{-1.8})\,i$ \\

$B \to \eta_q \pi$ &
$( -17.8^{+1.9}_{-2.3}) + ( -12.5^{+1.3}_{-1.7})\,i$ &
$( 10.9^{+1.6}_{-1.3}) + ( 8.9^{+1.2}_{-1.0})\,i$ &  &  \\

$B \to \eta_q^\prime \pi$ &
$( -18.2^{+1.9}_{-2.4}) + ( -12.8^{+1.4}_{-1.7})\,i$ &
$( 11.2^{+1.6}_{-1.3}) + ( 9.1^{+1.2}_{-1.0})\,i$ &  &  \\

$B \to K \eta_q$ &
 &  &  &
$( 17.0^{+1.1}_{-1.1}) + ( 11.9^{+0.5}_{-0.5})\,i$ \\

$B \to K \eta_q^\prime$ &
 &  &  &
$( 17.0^{+1.1}_{-1.1}) + ( 11.9^{+0.5}_{-0.5})\,i$ \\

$B \to \eta_q K$ &
$( -17.8^{+1.9}_{-2.3}) + ( -12.5^{+1.3}_{-1.7})\,i$ &
 &  &  \\

$B \to \eta_q^\prime K$ &
$( -18.2^{+1.9}_{-2.4}) + ( -12.8^{+1.4}_{-1.7})\,i$ &
 &  &  \\

$B \to K \eta_s$ &
$( -12.5^{+0.4}_{-0.4}) + ( -8.8^{+0.3}_{-0.4})\,i$ &
 &  &
$( 17.0^{+1.1}_{-1.1}) + ( 11.9^{+0.5}_{-0.5})\,i$ \\

$B \to K \eta_s^\prime$ &
$( -12.5^{+0.4}_{-0.4}) + ( -8.8^{+0.3}_{-0.4})\,i$ &
 &  &
$( 17.0^{+1.1}_{-1.1}) + ( 11.9^{+0.5}_{-0.5})\,i$ \\

$B \to \pi \eta_s$ &
 &  &
$( -7.7^{+8.7}_{-8.9}) + ( -12.8^{+7.7}_{-8.1})\,i$ &  \\

$B \to \pi \eta_s^\prime$ &
 &  &
$( -7.7^{+8.7}_{-8.9}) + ( -12.8^{+7.7}_{-8.1})\,i$ &  \\

$B_s \to \pi \eta_q$ &
 & $( 7.6^{+0.5 }_{-0.5}) + (6.2^{+0.3}_{-0.3})\,i$ &
$( -5.3^{+6.0}_{-6.0}) + ( -8.8^{+5.3}_{-5.3})\,i$ &  \\

$B_s \to \pi \eta_q^\prime$ &
 & $(6.2^{+0.4}_{-0.4}) + ( 5.0^{+0.2}_{-0.2})\,i$ &
$( -4.3^{+4.9}_{-4.9}) + ( -7.2^{+4.4}_{-4.4})\,i$ &  \\

$B_s \to \pi \eta_s$ &
 &  &
$( 5.4^{+6.2}_{-6.2}) + ( 9.0^{+5.5}_{-5.4})\,i$ &  \\

$B_s \to \pi \eta_s^\prime$ &
 &  &
$( -6.6^{+7.6}_{-7.6}) + ( -11.1^{+6.6}_{-6.7})\,i$ &  \\

\hline

$B \to \eta_q \eta_q^\prime$ &
 & $( 10.9^{+1.6}_{-1.3}) + ( 8.9^{+1.2}_{-1.0})\,i$ &
$( -7.56^{+8.66}_{-8.94}) + ( -12.76^{+7.72}_{-8.08})\,i$ & \\

$B \to \eta_q^\prime \eta_q$ &
 & $( 11.2^{+1.6}_{-1.3}) + ( 9.1^{+1.2}_{-1.0})\,i$ &
$( -7.85^{+8.94}_{-9.12}) + ( -13.10^{+7.93}_{-8.12})\,i$ & \\

$B \to \eta_q \eta_s^\prime$ &
 &  &
$( -7.56^{+8.66}_{-8.94}) + ( -12.76^{+7.72}_{-8.08})\,i$ & \\






\hline

$B_s \to \eta_q \eta_q^\prime$ &
 & $( 5.10^{+0.38}_{-0.37}) + ( 4.15^{+0.24}_{-0.23})\,i$ &
$( -3.6^{+4.1}_{-4.1}) + ( -6.0^{+3.6}_{-3.6})\,i$ & \\

$B_s \to \eta_q^\prime \eta_q$ &
 & $( 5.10^{+0.38}_{-0.37}) + ( 4.15^{+0.24}_{-0.23})\,i$ &
$( -3.6^{+4.1}_{-4.1}) + ( -6.0^{+3.6}_{-3.6})\,i$ & \\

$B_s \to \eta_q \eta_s^\prime$ &
 &  &
$( -5.4^{+6.2}_{-6.3}) + ( -9.1^{+5.5}_{-5.6})\,i$ & \\



$B_s \to \eta_q^\prime \eta_s$ &
 &  &
$( 3.6^{+4.2}_{-4.2}) + ( 6.1^{+3.7}_{-3.7})\,i$ & \\



\hline

$B \to \eta_q \eta_q$ &
 & $( 10.9^{+1.6}_{-1.3}) + ( 8.9^{+1.2}_{-1.0})\,i$ &
$( -7.56^{+8.66}_{-8.94}) + ( -12.76^{+7.72}_{-8.08})\,i$ & \\

$B \to \eta_q \eta_s$ &
 &  &
$( -7.56^{+8.66}_{-8.94}) + ( -12.76^{+7.72}_{-8.08})\,i$ & \\



\hline

$B_s \to \eta_q \eta_q$ &
 & $( 6.2^{+0.5}_{-0.5}) + ( 5.1^{+0.3}_{-0.3})\,i$ &
$( -4.36^{+4.97}_{-5.01}) + ( -7.30^{+4.38}_{-4.45})\,i$ & \\

$B_s \to \eta_q \eta_s$ &
 &  &
$( 4.46^{+5.13}_{-5.09}) + ( 7.47^{+4.52}_{-4.51})\,i$ & \\



\hline

$B \to \eta_q^\prime \eta_q^\prime$ &
 & $( 11.2^{+1.6}_{-1.3}) + ( 9.1^{+1.2}_{-1.0})\,i$ &
$( -7.85^{+8.94}_{-9.12}) + ( -13.10^{+7.93}_{-8.12})\,i$ & \\

$B \to \eta_q^\prime \eta_s^\prime$ &
 &  &
$( -7.85^{+8.94}_{-9.12}) + ( -13.10^{+7.93}_{-8.12})\,i$ & \\



\hline

$B_s \to \eta_q^\prime \eta_q^\prime$ &
 & $( 4.2^{+0.4}_{-0.3}) + ( 3.4^{+0.2}_{-0.2})\,i$ &
$( -7.57^{+8.65}_{-8.83}) + ( -12.67^{+7.67}_{-8.01})\,i$ & \\

$B_s \to \eta_q^\prime \eta_s^\prime$ &
 &  &
$( -4.46^{+5.07}_{-5.13}) + ( -7.47^{+4.49}_{-4.55})\,i$ & \\



\hline
\end{tabular}
\end{adjustbox}

\caption{Fitted values of the QCDF sub-amplitudes $\beta_{3,\mathrm{EW}}$, $\beta_{4,\mathrm{EW}}$, $\beta_{S4,\mathrm{EW}}$ and $\beta_{S3,\mathrm{EW}}$. All values are given in units of $10^{-2}$.
\label{tab:beta_ew_channels}}
\end{table}

\subsection{Comparison to other studies}

Let us now compare our results against other studies in the literature. In particular, flavour-$SU(3)$ breaking effects in non-leptonic $B$ decays into two light pseudoscalars have been incorporated previously in ref.~\cite{BurgosMarcos:2025xja}. While both groups implement $SU(3)$-breaking in a very similar manner and find a very good quality of fit in the $SU(3)$-broken case, there are also some differences between the two setups which we analyze in the following. 

The first difference is that the analysis in ref.~\cite{BurgosMarcos:2025xja} excludes all channels involving final states with $\eta$ and/or $\eta'$ mesons. Secondly, it should be noticed that this previous analysis performs the comparison to data in the $\lambda_u-\lambda_c$ convention for the CKM matrix elements. The full transformation dictionary between the QCDF parameterization as well as the $\lambda_u-\lambda_c$ and the $\lambda_u-\lambda_t$ basis was also introduced in detail in ref.~\cite{Huber:2021cgk}. We find that the former basis is less practical for disentangling the individual sub-amplitudes when making use of the (N)NLO results calculated within the QCDF framework. Another difference is in the experimental results used during the $\chi^2$ fit; in particular, this study makes use of the ratios shown in table~\ref{tab:corr_ratios}. Notice that the results reported there are single experimental measurements and not the world averages of the results from different experiments. The expected advantage is that these ratios account for the correlations between numerator and denominator. To avoid biases, we decided to perform the fit to the latest world averages of the relevant experimental measurements, although we acknowledge that the different averaging groups provide correlations between the various reported observables. It is interesting to see from table~\ref{tab:corr_ratios} that the measured correlated ratios of branching fractions are well consistent with the corresponding theoretical estimations based on our fit. This also supports our treatment of the experimental data during the $\chi^2$ fit.

\begin{table}[t]
\renewcommand{\arraystretch}{1.6} 
\setlength{\tabcolsep}{13.5pt} 
\centering
\begin{tabular}{|c|c|c|c|}
\hline
Channel & Experimental data & Fit result & Pull \\
\hline

$\frac{f_s}{f_d}\frac{\mathcal{B}(B_s^0 \to \pi^+\pi^-)}
{\mathcal{B}(B^0 \to K^+\pi^-)}$ 
& $(9.15 \pm 0.71 \pm 0.83)\times10^{-3}$ 
& $(9.52^{+5.76}_{-4.62})\times10^{-3}$ 
& $\phantom{+}0.06$ \\

$\frac{f_s}{f_d}\frac{\mathcal{B}(B_s^0 \to K^-\pi^+)}
{\mathcal{B}(B^0 \to K^+\pi^-)}$ 
& $0.074 \pm 0.006 \pm 0.006$ 
& $0.068^{+0.025}_{-0.017}$ 
& $-0.22$ \\

$\frac{f_s}{f_d}\frac{\mathcal{B}(B_s^0 \to K^+K^-)}
{\mathcal{B}(B^0 \to K^+\pi^-)}$ 
& $0.316 \pm 0.009 \pm 0.019$ 
& $0.330^{+0.132}_{-0.092}$ 
& $\phantom{+}0.10$ \\

$\frac{\mathcal{B}(B^+ \to \bar{K}^0 K^+)}
{\mathcal{B}(B^+ \to K^0 \pi^+)}$ 
& $0.064 \pm 0.009 \pm 0.004$ 
& $0.056^{+0.009}_{-0.007}$ 
& $-0.63$ \\

$\frac{\mathcal{B}(B^0 \to \pi^+\pi^-)}
{\mathcal{B}(B^0 \to K^+\pi^-)}$ 
& $0.262 \pm 0.009 \pm 0.017$ 
& $0.268^{+0.113}_{-0.082}$ 
& $\phantom{+}0.05$ \\

$\frac{\mathcal{B}(B^0 \to K^+K^-)}
{\mathcal{B}(B^0 \to K^+\pi^-)}$ 
& $(3.98 \pm 0.65 \pm 0.42)\times10^{-3}$ 
& $(4.88^{+3.87}_{-3.28})\times10^{-3}$ 
& $\phantom{+}0.23$ \\

\hline
\end{tabular}
\caption{Measured ratios of branching fractions and the corresponding theoretical estimates based on our fit, together with the resulting pulls.
\label{tab:corr_ratios}}
\end{table}

Concerning the individual TDA sub-amplitudes, the authors of ref.~\cite{BurgosMarcos:2025xja} conclude a large disparity between $\tilde{\alpha}^u_4$ and $\tilde{\alpha}^c_4$, \footnote{These parameters were introduced in ref.~\cite{Huber:2021cgk} as $c_{3}^{\mathrm{QCDF},u}/A_{M_1 M_2}$ and $c_{3}^{\mathrm{QCDF},c}/A_{M_1 M_2}$.} where $\tilde{\alpha}^p_4= \alpha_4^{p}+ \beta_3^{p} -\frac{1}{2}( \alpha_{4,\mathrm{EW}}^{p}  + \beta_{3,\mathrm{EW}}^{p})$, with $p=u,c$. In particular, they quote that $\left|\tilde{\alpha}_4^c/\tilde{\alpha}_4^u\right|
\in [10^{-4},10^{-2}]$, which indicates that the $u$-penguin coefficient $\tilde{\alpha}_4^u$ is much larger than its $c$-penguin counterpart  $\tilde{\alpha}_4^c$. In contrast, based on our best-fit point we find that the current data is compatible with the constraint, $|\tilde{T}_P|<0.02$, as given in eq.~\eqref{eq:tilde_conditions}. With eq.~\eqref{eq:Tree_dict}, this in turn implies that
\begin{eqnarray}
|\tilde{T}_P|=|\tilde{\alpha}_4^u-\tilde{\alpha}_4^c|<0.02,    
\end{eqnarray}
as obtained at the NLO in QCDF~\cite{Beneke:1999br,Beneke:2000ry,Beneke:2001ev,Beneke:2003zv}. Thus, in our case there is no large disparity between $\tilde{\alpha}^u_4$ and $\tilde{\alpha}^c_4$ as indicated by the fitted values of $\tilde{T}_P$ and $\tilde{P}$ given in table~\ref{tab:TDA_bestfitpoint}.

Based on the information provided in ref.~\cite{BurgosMarcos:2025xja}, we can infer that $\tilde{\alpha}_1$ is of $\mathcal{O}(1)$, which, together with the result $\left|\alpha_{4,\mathrm{EW}}^c/\tilde{\alpha}_1\right|
\in [10^{-4},10^{-2}]$, suggests a smaller value for $|\alpha_{4,\mathrm{EW}}^c|$ compared to $|\alpha_{4,\mathrm{EW}}^c|=0.186\pm0.002$ we get in the present study. Another difference is in their result $|\tilde{b}^c_4/\tilde{b}^u_4|\in [10^{-4},10^{-2}]$, which indicates that $\tilde{b}^c_4$ is much smaller than $\tilde{b}^u_4$. However, in our case these amplitudes are contained in $T_{PA}$ inside eq.~\eqref{eq:Tree_dict}, where once more the QCDF-based assumption of $\beta_4^u -\frac{1}{2}\beta_{4,\mathrm{EW}}^u=\beta_4^c -\frac{1}{2}\beta_{4,\mathrm{EW}}^c$ is supported by our good fit to the experimental data. 

Finally, it should be noted that while our previous analysis~\cite{Huber:2021cgk}  in the exact $SU(3)$ limit already showed reasonable agreement with experiment, the inclusion of moderate $SU(3)$-breaking effects in the present work leads to an improved description of the data and a more reliable determination of the topological amplitudes. This stands in contrast to other recent analyses~\cite{Berthiaume:2023kmp,Bhattacharya:2025wcq} that, by employing a reduced set of amplitudes, obtain the need for abnormally large (numerically $1000\%$) flavour-$SU(3)$ breaking effects to describe the $\Delta S=0$ and $\Delta S=1$ channels simultaneously. Our results demonstrate that once the full twenty amplitude basis and the flavour-$SU(3)$ breaking effects from transition form factors, decay constants and phase space factors are taken into account, all the data can be described without inconsistency.

\section{Phenomenological implications}
\label{sec:pheno}

In this section, we address the phenomenological implications of the results from our fit as shown in tables~\ref{tab:resultsBRI}~--~\ref{tab:resultsMixCPIII}, focusing on several key observables that lie at the core of two-body charmless $B\to PP$ decays.

\begin{table}[t] 
\renewcommand{\arraystretch}{1.3} 
\setlength{\tabcolsep}{20pt} 
\centering 
\begin{tabular}{|l|c|c|c|} 
\hline 
\multicolumn{4}{|c|}{Branching Fractions I} \\ 
\hline 
Channel & Fit result [$10^{-6}$] & Experiment [$10^{-6}$] & Pull [$\sigma$] \\ 
\hline 
$B^- \to \pi^0 \pi^-$ & $5.3738^{+1.3656}_{-1.1793}$ & $5.31 \pm 0.26$ & 0.046 \\ $B^- \to K^0 K^-$ & $1.3335^{+0.0996}_{-0.0921}$ & $1.32 \pm 0.17$ & 0.069 \\ $\bar{B}^0 \to \pi^+ \pi^-$ & $5.3533^{+1.3631}_{-1.1854}$ & $5.37 \pm 0.20$ & 0.012 \\ 
$\bar{B}^0 \to \pi^0 \pi^0$ & $1.5027^{+0.4190}_{-0.3509}$ & $1.55 \pm 0.17$ & 0.105 \\ 
$\bar{B}^0 \to K^+ K^-$ & $0.0977^{+0.0649}_{-0.0394}$ & $0.078 \pm 0.015$ & 0.296 \\ $\bar{B}^0 \to K^0 \bar K^0$ & $1.2245^{+0.1394}_{-0.1289}$ & $1.21 \pm 0.16$ & 0.068 \\ 
$\bar{B}_s \to \pi^0 K^0$ & $1.3048^{+0.2707}_{-0.2336}$ & -- & -- \\ 
$\bar{B}_s \to \pi^- K^+$ & $5.7076^{+0.9206}_{-0.8239}$ & $5.90 \pm 0.70$ & 0.166 \\ $B^- \to \pi^0 K^-$ & $13.2864^{+3.0063}_{-2.6580}$ & $13.20 \pm 0.40$ & 0.028 \\ $B^- \to \pi^- \bar K^0$ & $24.0024^{+2.9280}_{-2.3080}$ & $23.90 \pm 0.60$ & 0.034 \\ 
$\bar{B}^0 \to \pi^+ K^-$ & $19.9871^{+4.7900}_{-4.2332}$ & $20.00 \pm 0.40$ & 0.003 \\ $\bar{B}^0 \to \pi^0 \bar K^0$ & $10.2119^{+1.9219}_{-1.5219}$ & $10.10 \pm 0.40$ & 0.057 
\\ $\bar{B}_s \to \pi^+ \pi^-$ & $0.7952^{+0.3738}_{-0.2890}$ & $0.72 \pm 0.10$ & 0.194 
\\ $\bar{B}_s \to \pi^0 \pi^0$ & $0.3976^{+0.1869}_{-0.1445}$ & $< 7.70$ & -- \\ $\bar{B}_s \to K^+ K^-$ & $27.6593^{+5.9902}_{-5.3600}$ & $27.20 \pm 2.30$ & 0.072 \\ $\bar{B}_s \to K^0 \bar K^0$ & $18.0978^{+2.8146}_{-2.5834}$ & $17.60 \pm 3.10$ & 0.119 \\ 
\hline 
\end{tabular} 
\caption{Our fitting results and the experimental data of branching fractions for decays with no $\eta$ or $\eta'$ mesons in the final state, together with the resulting pulls. Upper limits correspond to $90\%$ C.L.
\label{tab:resultsBRI}}
\end{table}

\begin{table}[htbp] 
\renewcommand{\arraystretch}{1.3} 
\setlength{\tabcolsep}{20.5pt} 
\centering \begin{tabular}{|l|c|c|c|} 
\hline 
\multicolumn{4}{|c|}{Branching Fractions II} \\ \hline Channel & Fit result [$10^{-6}$] & Experiment [$10^{-6}$] & Pull [$\sigma$] \\ 
\hline 
$B^- \to \eta \pi^-$ & $3.9738^{+1.0222}_{-0.8559}$ & $4.02 \pm 0.27$ & 0.044 \\ $\bar{B}^0 \to \eta \pi^0$ & $0.3760^{+0.2314}_{-0.1529}$ & $0.41 \pm 0.17$ & 0.118 \\ 
$\bar{B}_s \to \eta K^0$ & $0.8322^{+0.1816}_{-0.1553}$ & -- & -- \\ 
$B^- \to \eta K^-$ & $2.5890^{+1.1184}_{-0.8719}$ & $2.40 \pm 0.40$ & 0.159 \\ $\bar{B}^0 \to \eta \bar K^0$ & $1.3418^{+0.9157}_{-0.6291}$ & $1.23^{+0.27}_{-0.24}$ & 0.117 \\ 
$\bar{B}_s \to \eta \pi^0$ & $0.9244^{+1.5723}_{-0.6541}$ & $<1000$ & -- \\ 
\hline 
$B^- \to \eta' \pi^-$ & $2.5436^{+0.8560}_{-0.5528}$ & $2.70 \pm 0.90$ & 0.126 \\ $\bar{B}^0 \to \eta' \pi^0$ & $4.8107^{+4.5518}_{-2.7193}$ & $1.20 \pm 0.60$ & 0.786 \\ 
$\bar{B}_s \to \eta' K^0$ & $2.0434^{+0.3788}_{-0.3307}$ & $<8.16$~\cite{Belle:2022zay} & -- \\ 
$B^- \to \eta' K^-$ & $71.5236^{+13.1335}_{-11.3445}$ & $70.40 \pm 2.50$ & 0.084 \\ $\bar{B}^0 \to \eta' \bar K^0$ & $65.8116^{+9.7872}_{-8.6365}$ & $66.00 \pm 4.00$ & 0.018 \\ 
$\bar{B}_s \to \eta' \pi^0$ & $33.4106^{+54.7752}_{-24.9864}$ & -- & -- \\ 
\hline 
\end{tabular} 
\caption{Our fitting results and the experimental data of branching fractions for decays with one $\eta$ or $\eta'$ meson in the final state, together with the resulting pulls. The other captions are the same as in table~\ref{tab:resultsBRI}. \label{tab:resultsBRII}}
\end{table}

\begin{table}[htbp]
\renewcommand{\arraystretch}{1.3}
\setlength{\tabcolsep}{22pt}
\centering
\begin{tabular}{|l|c|c|c|}
\hline
\multicolumn{4}{|c|}{Branching Fractions III} \\
\hline
Channel & Fit result [$10^{-6}$] & Experiment [$10^{-6}$] & Pull [$\sigma$] \\
\hline
$\bar{B}^0 \to \eta \eta$
& $0.6103^{+0.4590}_{-0.3000}$
& $<1.0$
& -- \\

$\bar{B}_s \to \eta \eta$
& $6.2049^{+5.2315}_{-3.7727}$
& $<143$~\cite{Belle:2021tsq}
& -- \\
\hline

$\bar{B}^0 \to \eta' \eta'$
& $4.5537^{+4.2729}_{-2.6016}$
& $<1.70$
& -- \\

$\bar{B}_s \to \eta' \eta'$
& $53.1927^{+74.8192}_{-38.6425}$
& $33.0 \pm 7.0$
& -- \\
\hline

$\bar{B}^0 \to \eta' \eta$
& $3.3237^{+3.6718}_{-1.9904}$
& $<1.2$
& -- \\

$\bar{B}_s \to \eta' \eta$
& $97.0778^{+101.0321}_{-\phantom{1}62.2639}$
& $<65$~\cite{Belle:2021bxn}
& -- \\
\hline
\end{tabular}
\caption{Our fitting results and the experimental data of branching fractions for decays with two $\eta^{(\prime)}$ mesons in the final state, together with the resulting pulls. The other captions are the same as in table~\ref{tab:resultsBRI}.
\label{tab:resultsBRIII}}
\end{table}

\begin{table}[t] 
\renewcommand{\arraystretch}{1.3} 
\setlength{\tabcolsep}{20pt} 
\centering 
\begin{tabular}{|l|c|c|c|} 
\hline 
\multicolumn{4}{|c|}{Direct CP Asymmetries I} \\ \hline Channel & Fit result [$10^{-2}$] & Experiment [$10^{-2}$] & Pull [$\sigma$] \\ 
\hline 
$B^- \to \pi^0 \pi^-$ & $-2.4414^{+1.5688}_{-1.5706}$ & $-1.20 \pm 3.10$ & 0.357 \\ $B^- \to K^0 K^-$ & $9.9042^{+20.4979}_{-21.1721}$ & $4.00 \pm 14.0$ & 0.233 \\ $\bar{B}^0 \to \pi^+ \pi^-$ & $35.9107^{+4.6619}_{-4.3762}$ & $31.4 \pm 3.0$ & 0.814 \\ 
$\bar{B}^0 \to \pi^0 \pi^0$ & $33.4359^{+7.1074}_{-7.2281}$ & $23.0 \pm 18.0$ & 0.538 \\ 
$\bar{B}^0 \to K^+ K^-$ & $-30.9353^{+55.4875}_{-46.6351}$ & -- & -- \\ $\bar{B}^0 \to K^0 \bar K^0$ & $-4.3341^{+3.7669}_{-3.9982}$ & $-60.0 \pm 70.0$ & 0.794 \\ $\bar{B}_s \to \pi^0 K^0$ & $42.0670^{+4.0089}_{-3.8294}$ & -- & -- \\ 
$\bar{B}_s \to \pi^- K^+$ & $21.5213^{+2.0595}_{-2.0289}$ & $22.4 \pm 1.2$ & 0.369 \\ $B^- \to \pi^0 K^-$ & $2.7391^{+2.8502}_{-2.7146}$ & $2.70 \pm 1.20$ & 0.013 \\ $B^- \to \pi^- \bar K^0$ & $-0.3975^{+0.8215}_{-0.8240}$ & $-2.30 \pm 1.40$ & 1.171 \\ $\bar{B}^0 \to \pi^+ K^-$ & $-8.3535^{+0.7598}_{-0.7682}$ & $-8.31 \pm 0.31$ & 0.053 \\ 
$\bar{B}^0 \to \pi^0 \bar K^0$ & $-20.4839^{+1.6261}_{-1.6508}$ & $-1.00 \pm 13.0$ & 1.487 \\ 
$\bar{B}_s \to \pi^+ \pi^-$ & $2.5832^{+4.8704}_{-4.6516}$ & -- & -- \\ $\bar{B}_s \to \pi^0 \pi^0$ & $2.5832^{+4.8704}_{-4.6516}$ & -- & -- \\ 
$\bar{B}_s \to K^+ K^-$ & $-11.7899^{+2.0334}_{-2.3355}$ & $-16.2 \pm 3.5$ & 1.048 \\ $\bar{B}_s \to K^0 \bar K^0$ & $0.4511^{+0.4045}_{-0.3964}$ & -- & -- \\ 
\hline 
\end{tabular} 
\caption{Our fitting results and the experimental data of direct CP asymmetries for decays without $\eta^{(\prime)}$ mesons in the final state, together with the resulting pulls. The other captions are the same as in table~\ref{tab:resultsBRI}.
\label{tab:resultsACPI}}
\end{table}

\begin{table}[htbp] 
\renewcommand{\arraystretch}{1.3} 
\setlength{\tabcolsep}{20.5pt} 
\centering 
\begin{tabular}{|l|c|c|c|} 
\hline 
\multicolumn{4}{|c|}{Direct CP Asymmetries II} \\ 
\hline Channel & Fit result [$10^{-2}$] & Experiment [$10^{-2}$] & Pull [$\sigma$] \\ \hline 
$B^- \to \eta \pi^-$ & $-11.1594^{+11.0944}_{-12.5383}$ & $-14.0 \pm 7.0$ & 0.198 \\ $\bar{B}^0 \to \eta \pi^0$ & $-23.9428^{+64.0877}_{-48.9835}$ & -- & -- \\ 
$\bar{B}_s \to \eta K^0$ & $-5.7331^{+4.3017}_{-4.2344}$ & $<0.10$ & -- \\ 
$B^- \to \eta K^-$ & $-38.0846^{+11.7891}_{-16.0498}$ & $-37.0 \pm 8.0$ & 0.060 \\ $\bar{B}^0 \to \eta \bar K^0$ & $6.0055^{+12.3185}_{-16.9289}$ & -- & -- \\ $\bar{B}_s \to \eta \pi^0$ & $5.7625^{+36.9675}_{-46.0481}$ & -- & -- \\ 
\hline 
$B^- \to \eta' \pi^-$ & $7.9717^{+28.2470}_{-35.1341}$ & $6.0 \pm 16.0$ & 0.051 \\ 
$\bar{B}^0 \to \eta' \pi^0$ & $-7.2495^{+72.2262}_{-64.7704}$ & -- & -- \\ 
$\bar{B}_s \to \eta' K^0$ & $-35.6340^{+7.2975}_{-6.7990}$ & -- & -- \\ 
$B^- \to \eta' K^-$ & $0.6529^{+1.7892}_{-1.7327}$ & $0.40 \pm 1.10$ & 0.120 \\ $\bar{B}^0 \to \eta' \bar K^0$ & $3.9775^{+0.6452}_{-0.6199}$ & $8.0 \pm 4.0$ & 0.993 \\ 
$\bar{B}_s \to \eta' \pi^0$ & $-0.3001^{+12.4792}_{-12.3931}$ & -- & -- \\ 
\hline 
\end{tabular} 
\caption{Our fitting results and the experimental data of direct CP asymmetries for decays with one $\eta$ or $\eta'$ meson in the final state, together with the resulting pulls. The other captions are the same as in table~\ref{tab:resultsBRI}.
\label{tab:resultsACPII}}
\end{table}

\begin{table}[htbp]
\renewcommand{\arraystretch}{1.3} 
\setlength{\tabcolsep}{22pt} 
\centering
\begin{tabular}{|l|c|c|c|}
\hline
 \multicolumn{4}{|c|}{Direct CP Asymmetries III} \\
\hline
Channel & Fit result [$10^{-2}$] & Experiment [$10^{-2}$] & Pull [$\sigma$]  \\
\hline
$\bar{B}^0 \to \eta \eta$              & $0.4^{+43.3}_{-44.5}$  & -- & -- \\
$\bar{B}_s \to \eta \eta$              & $-10.9^{+\phantom{1}7.6}_{-10.1}$ & -- & -- \\
\hline
$\bar{B}^0 \to \eta' \eta'$            & $1.5^{+68.0}_{-69.6}$ & -- & -- \\
$\bar{B}_s \to \eta' \eta'$            & $0.04^{+9.57}_{-9.48}$ & -- & -- \\
\hline
$\bar{B}^0 \to \eta' \eta$             & $-1.8^{+64.6}_{-63.8}$ & -- & -- \\
$\bar{B}_s \to \eta' \eta$             & $0.38^{+6.57}_{-6.40}$  & -- & -- \\
\hline
\end{tabular}
\caption{Our fitting results and the experimental data of direct CP asymmetries for decays with two $\eta^{(\prime)}$ mesons in the final state, together with the resulting pulls. The other captions are the same as in table~\ref{tab:resultsBRI}.
\label{tab:resultsACPIII}}
\end{table}

\begin{table}[htbp] 
\renewcommand{\arraystretch}{1.3} 
\setlength{\tabcolsep}{20pt} 
\centering 
\begin{tabular}{|l|c|c|c|} 
\hline 
\multicolumn{4}{|c|}{Mixing-Induced CP Asymmetries I} \\ 
\hline 
Channel & Fit result [$10^{-2}$] & Experiment [$10^{-2}$] & Pull [$\sigma$] \\ 
\hline 
$\bar{B}^0 \to \pi^+ \pi^-$ & $-65.4123^{+5.2767}_{-4.9069}$ & $-66.6 \pm 2.9$ & 0.363 \\ 
$\bar{B}^0 \to \pi^0 \pi^0$ & $48.3618^{+7.6002}_{-7.4294}$ & -- & -- \\ 
$\bar{B}^0 \to K^+ K^-$ & $70.7086^{+21.4976}_{-53.3108}$ & -- & -- \\ 
$\bar{B}^0 \to K^0 \bar K^0$ & $-13.1626^{+11.5952}_{-11.2942}$ & $-80.0 \pm 50.0$ & 1.336 \\ 
\hline 
$\bar{B}_s \to \pi^0 K^0$ & $36.0626^{+6.3772}_{-6.7687}$ & -- & -- \\ 
$\bar{B}_s \to \pi^- K^+$ & $-97.5845^{+0.4910}_{-0.4304}$ & -- & -- \\ 
$\bar{B}^0 \to \pi^+ K^-$ & $-56.4665^{+1.1926}_{-1.1926}$ & -- & -- \\ 
$\bar{B}^0 \to \pi^0 \bar K^0$ & $77.5335^{+2.4528}_{-2.2164}$ & $64.0 \pm 13.0$ & 1.039 \\ 
\hline 
$\bar{B}_s \to \pi^+ \pi^-$ & $-6.5202^{+5.1912}_{-5.5795}$ & -- & -- \\ 

$\bar{B}_s \to \pi^0 \pi^0$ & $-6.5202^{+5.1912}_{-5.5795}$ & -- & -- \\ 
$\bar{B}_s \to K^+ K^-$ & $16.0802^{+1.7755}_{-1.6590}$ & $14.0 \pm 3.0$ & 0.729 \\ $\bar{B}_s \to K^0 \bar K^0$ & $0.1429^{+0.1732}_{-0.1284}$ & -- & -- \\ 
\hline 
\end{tabular} 
\caption{Our fitting results and the experimental data of mixing-induced CP asymmetries for decays without $\eta^{(\prime)}$ mesons in the final state, together with the resulting pulls. The other captions are the same as in table~\ref{tab:resultsBRI}.
\label{tab:resultsMixCPI}}
\end{table}

\begin{table}[htbp] 
\renewcommand{\arraystretch}{1.3} 
\setlength{\tabcolsep}{21pt} 
\centering 
\begin{tabular}{|l|c|c|c|} 
\hline 
\multicolumn{4}{|c|}{Mixing-Induced CP Asymmetries II} \\ 
\hline 
Channel & Fit result [$10^{-2}$] & Experiment [$10^{-2}$] & Pull [$\sigma$] \\ 
\hline 
$\bar{B}^0 \to \eta \pi^0$ & $-47.7436^{+61.6033}_{-37.2269}$ & -- & -- \\ 
$\bar{B}_s \to \eta K^0$ & $56.1392^{+7.8549}_{-8.8944}$ & $57.0 \pm 10.0$ & 0.085 \\ $\bar{B}^0 \to \eta \bar K^0$ & $95.3159^{+3.0308}_{-3.9466}$ & -- & -- \\ 
$\bar{B}_s \to \eta \pi^0$ & $19.0501^{+30.4357}_{-43.2701}$ & -- & -- \\ 
\hline 
$\bar{B}^0 \to \eta' \pi^0$ & $-18.5105^{+75.7292}_{-57.7093}$ & -- & -- \\ $\bar{B}_s \to \eta' K^0$ & $31.9321^{+8.5757}_{-8.9752}$ & -- & -- \\ 
$\bar{B}^0 \to \eta' \bar K^0$ & $70.1850^{+1.1782}_{-1.1808}$ & $64.0 \pm 5.0$ & 1.239 \\ 
$\bar{B}_s \to \eta' \pi^0$ & $-2.0753^{+12.0910}_{-12.5058}$ & -- & -- \\ \hline \end{tabular} 
\caption{Our fitting results and the experimental data of mixing-induced CP asymmetries for decays with one $\eta$ or $\eta'$ meson in the final state, together with the resulting pulls. The other captions are the same as in table~\ref{tab:resultsBRI}.
\label{tab:resultsMixCPII}}
\end{table}

\begin{table}[htbp]
\renewcommand{\arraystretch}{1.3} 
\setlength{\tabcolsep}{22pt} 
\centering
\begin{tabular}{|l|c|c|c|}
\hline
 \multicolumn{4}{|c|}{Mixing-Induced CP Asymmetries III} \\
\hline
Channel & Fit result [$10^{-2}$] & Experiment [$10^{-2}$] & Pull [$\sigma$]  \\
\hline
$\bar{B}^0 \to \eta \eta$              & $-67.4530^{+34.0045}_{-21.4451}$ & -- & -- \\
$\bar{B}_s \to \eta \eta$              & $-14.6707^{+\phantom{0}7.3128}_{-10.6542}$ & -- & -- \\
\hline
$\bar{B}^0 \to \eta' \eta'$            & $1.0663^{+68.4755}_{-68.9586}$ & -- & -- \\
$\bar{B}_s \to \eta' \eta'$            & $-0.1777^{+9.5365}_{-9.4603}$   & -- & -- \\
\hline
$\bar{B}^0 \to \eta' \eta$             & $0.4957^{+63.2652}_{-64.0992}$ & -- & -- \\
$\bar{B}_s \to \eta' \eta$             & $0.4061^{+6.4154}_{-6.2142}$   & -- & -- \\
\hline
\end{tabular}
\caption{Our fitting results and the experimental data of mixing-induced CP asymmetries for decays with two $\eta^{(\prime)}$ mesons in the final state, together with the resulting pulls. The other captions are the same as in table~\ref{tab:resultsBRI}.
\label{tab:resultsMixCPIII}}
\end{table}

\subsection{\texorpdfstring{The $B\to K\pi$, $\pi\pi$ and $KK$ puzzles}{The B to Kpi, pipi, KK puzzles}}

Using the best‑fit point obtained in section~\ref{sec:ampresults}, we compute the following quantities related to $B\to K\pi$, $\pi\pi$ and $KK$ decays~\cite{Gronau:2005kz,Buras:2003yc,Mishima:2004um,Fleischer:2007mq,Bell:2015koa,Fleischer:2018bld,Grossman:2024amc,Amhis:2022hpm}:
\begin{align}
\Delta A_{CP} &= A_{CP}(B^- \to \pi^0 K^-) - A_{CP}(\bar{B}^0 \to \pi^+ K^-) = (11.1^{+2.9}_{-2.9})\%,\\[4pt]
R &= \frac{\mathrm{Br}(\bar{B}^0 \to \pi^+ K^-)}{\mathrm{Br}(B^- \to \pi^- \bar{K}^0)}\cdot\frac{\tau_{B^-}}{\tau_{B^0}} = 0.90^{+0.24}_{-0.22},\\[4pt]
R_c &= \frac{2\,\mathrm{Br}(B^- \to \pi^0 K^-)}{\mathrm{Br}(B^- \to \pi^- \bar{K}^0)} = 1.11^{+0.28}_{-0.25},\\[4pt]
R_n &= \frac{1}{2}\frac{\mathrm{Br}(\bar{B}^0 \to \pi^+ K^-)}{\mathrm{Br}(\bar{B}^0 \to \pi^0 \bar{K}^0)} = 0.98^{+0.31}_{-0.25},\\[4pt]
R_{+-}^{\pi\pi} &= \frac{2\,\mathrm{Br}(B^- \to \pi^- \pi^0)}{\mathrm{Br}(\bar{B}^0 \to \pi^+ \pi^-)}\cdot\frac{\tau_{B^0}}{\tau_{B^-}} = 1.86^{+0.76}_{-0.54},\\[4pt]
R_{00}^{\pi\pi} &= \frac{2\,\mathrm{Br}(\bar{B}^0 \to \pi^0 \pi^0)}{\mathrm{Br}(\bar{B}^0 \to \pi^+ \pi^-)} = 0.56^{+0.24}_{-0.17},\\[4pt]
R_{KK}^{ss}  
&= \frac{\mathrm{Br}(\bar{B}_s\to K^0\bar{K}^0)}{\mathrm{Br}(\bar{B}_s \to K^+K^-)} 
= 0.65^{+0.20}_{-0.14},\\[4pt]
R_{KK}^{sd} 
&= \left|\frac{V_{td}}{V_{ts}}\right|^2\cdot 
\frac{\mathrm{Br}(\bar{B}_s \to K^0\bar{K}^0)}{\mathrm{Br}(\bar{B}^0 \to K^0\bar{K}^0)} \cdot 
\frac{\tau_{B^0}}{\tau_{B_s}} 
= 0.62^{+0.12}_{-0.11}.
\end{align}
Specifically, the ratios $R$, $R_c$ and $R_n$ are derived from the branching fractions of $B\to K\pi$ decays, while $R_{+-}^{\pi\pi}$ and $R_{00}^{\pi\pi}$ are derived from $B\to\pi\pi$ decays. Using the world averages provided by PDG~\cite{ParticleDataGroup:2024cfk} and HFLAV~\cite{HeavyFlavorAveragingGroupHFLAV:2024ctg}, the experimental values of these ratios are found to be in good agreement with our results. In particular, the CP difference $\Delta A_{CP}$, which has been at the heart of the so-called ``$K\pi$ puzzle'' for more than a decade, is reproduced within $1\sigma$ of the experimental value, $\Delta A_{\text{CP}}^{\text{exp}} = (11.0 \pm 1.2)\%$. Our results of the ratios $R$, $R_c$ and $R_n$ also match the data perfectly within uncertainties, with $R^{\text{exp}} = 0.89 \pm 0.03$, $R_c^{\text{exp}} = 1.10 \pm 0.04$, and $R_n^{\text{exp}} = 0.97 \pm 0.04$. For the $B\to\pi\pi$ system, the colour-allowed ratio $R_{+-}^{\pi\pi}$ is sensitive to the relative strength of the $\Delta I=3/2$ tree amplitude, while the colour-suppressed ratio $R_{00}^{\pi\pi}$ is particularly sensitive to the interference between tree and penguin amplitudes. Our theoretical predictions for these ratios are in good agreement with the experimental averages, $R_{+-}^{\pi\pi}=1.83\pm0.11$ and $R_{00}^{\pi\pi}=0.58\pm0.07$, indicating that the relative phases and magnitudes of the fitted topological amplitudes are realistic. Furthermore, we extend our analysis to the $B\to KK$ system characterized by the last two ratios, which provide complementary tests of isospin and $U$-spin symmetries, respectively. The fact that our fitting results are consistent with the current experimental data, $R_{KK}^{ss}=0.65\pm0.13$ and $R_{KK}^{sd}=0.61\pm0.13$, supports the validity of these subgroups of $SU(3)$ in these decays.

\subsection{\texorpdfstring{The $B\to K\pi$ sum rule}{The B to Kpi sum rule}}

A more refined test of the $B\to K\pi$ system is provided by the sum rule~\cite{Gronau:1998ep,Gronau:2005kz,Bell:2015koa}
\begin{align}
\Delta_{\text{SR}} &= A_{CP}(B^-\to\pi^0 K^-) \frac{2\mathrm{Br}(B^-\to\pi^0 K^-)}{\mathrm{Br}(\bar{B}^0\to\pi^+ K^-)} \frac{\tau_{B^0}}{\tau_{B^-}} + A_{CP}(\bar{B}^0\to\pi^0 \bar{K}^0) \frac{2\mathrm{Br}(\bar{B}^0\to\pi^0 \bar{K}^0)}{\mathrm{Br}(\bar{B}^0\to\pi^+ K^-)} \nonumber \\
& - A_{CP}(\bar{B}^0\to\pi^+ K^-) - A_{CP}(B^-\to\pi^- \bar{K}^0) \frac{\mathrm{Br}(B^-\to\pi^- \bar{K}^0)}{\mathrm{Br}(\bar{B}^0\to\pi^+ K^-)} \frac{\tau_{B^0}}{\tau_{B^-}}.
\end{align}
Our fit yields $\Delta_{\text{SR}} = -0.09^{+0.03}_{-0.03}$. Using the PDG world averages, the experimental result reads $\Delta_{\text{SR}}^{\text{exp}} = 0.13 \pm 0.14$. Recently, the Belle II collaboration also reported that  $\Delta_{\text{SR}}^{\text{Belle II}} = 0.03 \pm 0.13(\text{stat.}) \pm 0.04(\text{syst.})$~\cite{Belle-II:2023ksq}, with the experimental correlations taken into account properly. It can be seen that our result agrees well with both measurements within uncertainties and is compatible with zero, as expected within the SM~\cite{Gronau:1998ep,Gronau:2005kz,Bell:2015koa}. 

\subsection{Role of mixing‑induced CP asymmetries}

The mixing‑induced CP asymmetries $S_f$ can provide additional powerful probes of the interference between decay amplitudes with different weak phases. In our fit, we have included the available data on $S_f$ for several channels.\footnote{A first measurement of the mixing-induced CP asymmetry in the $B^0 \to \pi^0 \pi^0$ channel has recently been reported by the Belle II collaboration at Rencontres de Moriond 2026, with $S_{\pi^0\pi^0}^\mathrm{Belle~II}=0.61^{+0.75}_{-0.79}(\mathrm{stat.})\pm0.11(\mathrm{syst.})$~\cite{DeLaMotte:Moriond2026}. This result is not included in our analysis but is found to be consistent with our fitted result, $S_{\pi^0\pi^0}=0.484^{+0.076}_{-0.074}$, within uncertainties. }. As can be seen from tables~\ref{tab:resultsMixCPI}~--~\ref{tab:resultsMixCPIII}, the theoretical values obtained from our best‑fit point are in excellent agreement with the experimental measurements. For instance, we obtain $S_{\pi^+\pi^-} = -0.654^{+0.053}_{-0.050}$, $S_{K^+K^-} = 0.161^{+0.018}_{-0.017}$, $S_{\eta K^0} = 0.561^{+0.079}_{-0.089}$, all being consistent with the corresponding experimental data, $S_{\pi^+\pi^-}^{\text{exp}} = -0.666 \pm 0.029$, $S_{K^+K^-}^{\text{exp}} = 0.140 \pm 0.030$, $S_{\eta K^0}^{\text{exp}} = 0.57 \pm 0.10$, for the $\bar{B}^0\to \pi^+\pi^-$, $\bar{B}_s\to K^+K^-$, and $\bar{B}_s\to \eta K^0$ decays. This level of agreement demonstrates that our extracted amplitudes capture correctly the interplay of tree and penguin contributions, and that the mixing‑induced asymmetries do not exhibit any anomalous behaviours. 

\subsection{Scrutinizing the EWP--tree relations}

An important aspect of our analysis is the treatment of independent topological amplitudes. In contrast to some recent studies~\cite{Berthiaume:2023kmp,Bhattacharya:2025wcq,Bhattacharya:2025rrv} that reduce the number of independent amplitudes by imposing the EWP--tree relations~\cite{Gronau:1998fn,Neubert:1998pt,Neubert:1998jq}, we retain all the twenty independent topological amplitudes in our fit. This allows us to perform a model-independent test of the validity of the isospin-based EWP--tree relations revisited recently in ref.~\cite{Bhattacharya:2025rrv}.

In fact, the EWP--tree relations can be derived under the strict leading-order (LO) QCDF limit, with two non-trivial implicit assumptions: (i) all non-factorizable QCD corrections, including vertex, penguin, hard-spectator scattering, and weak annihilation contributions, are neglected; (ii) the strong-interaction dynamics of EWP amplitudes are identical to those of tree amplitudes, such that their hadronic matrix elements share a common factorizable structure. Under these assumptions, the EWP amplitudes are strictly proportional to the tree amplitudes, with the proportionality constant determined solely by the Wilson coefficients of EWP and tree operators. Explicitly, using the fitted values of the tree amplitudes $T$ and $C$, together with the NLO Wilson coefficients given at the scale $\mu=m_{b}$, $C_{1}=-0.190$, $C_{2}=1.081$, $C_{9}=-1.254 \cdot \alpha$, $C_{10}=0.223 \cdot \alpha$, with $\alpha=1/129$~\cite{Beneke:2001ev}, we can obtain the EWP amplitudes through the EWP--tree relations~\cite{Bhattacharya:2025rrv}
\begin{equation} 
P_{EW} = P_{T} = -\frac{3}{2} \frac{C_{9}+C_{10}}{C_{1}+C_{2}} C, \qquad 
P_{EW}^{C} = P_{C} = -\frac{3}{2} \frac{C_{9}+C_{10}}{C_{1}+C_{2}} T . 
\end{equation}
Here $P_{EW}$ and $P_{EW}^{C}$ denote the colour-allowed and colour-suppressed EWP amplitudes respectively, which correspond to the parameters $P_{T}$ and $P_{C}$ in our TDA parametrization. The common proportionality factor reads
\begin{equation}
r \equiv -\frac{3}{2} \frac{C_{9}+C_{10}}{C_{1}+C_{2}}=0.0135 .
\end{equation}

If the EWP--tree relations held exactly as derived in the factorized limit, we would expect $|P_{T}| \approx r |C|$ and $|P_{C}| \approx r |T|$, with both amplitudes carrying the same strong phases consistent with the real proportionality constant $r$. However, our fitted values $\tilde{P}_{T}=(-0.245^{+0.002}_{-0.002})+(-0.133^{+0.003}_{-0.003})i$ and $\tilde{P}_{C}=(0.179^{+0.002}_{-0.002})+(0.154^{+0.004}_{-0.004})i$ are orders of magnitude larger than the factorized expectations of $r \tilde{C}$ and $r \tilde{T}$. Numerically, we find that $|P_{T}/C| \approx 0.24$ and $|P_{C}/T| \approx 0.45$, exceeding the expected factorized ratio $r \approx 0.0135$ by factors of $20~-~30$. Such a feature was also found in ref.~\cite{Bhattacharya:2025wcq} once the EWP--tree relations are relaxed. Moreover, the fitted EWP amplitudes carry large non-zero strong phases, which are entirely absent in the naive factorized approximation.

This severe breakdown of the EWP--tree relations can be quantitatively understood from the scaling of non-factorizable corrections in the QCDF framework~\cite{Beneke:1999br,Beneke:2000ry,Beneke:2001ev,Beneke:2003zv}. In QCDF power counting, the LO factorizable tree amplitudes $T$ and $C$ are of $\mathcal{O}(1)$ (normalized to the colour-allowed tree amplitude), while the LO factorizable EWP amplitudes are naturally suppressed to $\mathcal{O}(r) \approx 10^{-2}$ by the ratio of EWP-to-tree Wilson coefficients. In contrast, the non-factorizable corrections from vertex, penguin, hard-spectator scattering, and weak annihilation contributions are generally of $\mathcal{O}(\alpha_s/\pi) \approx 10^{-1}$ for both tree and EWP amplitudes, where $\alpha_s$ is the strong coupling constant at the $b$-quark scale. For the tree amplitudes, this corresponds to a relative correction of only $\sim 10~-~30\%$, which is a sub-leading perturbative correction that does not alter the leading factorizable behavior. For the EWP amplitudes, however, the $\mathcal{O}(10^{-1})$ non-factorizable corrections are more than an order of magnitude larger than the $\mathcal{O}(10^{-2})$ LO factorizable contribution, and thus become the dominant component of the full EWP amplitude. In addition, the non-factorizable hard-scattering amplitudes contain non-trivial phases from the imaginary parts of the loop diagrams, which generate the large strong phases observed in our fitted $P_T$ and $P_C$ parameters. Still, a more refined analysis of the dynamical mechanism behind the enhancement of the EWP amplitudes has to be carried out in the future.

Our findings clearly demonstrate that the naive EWP--tree relations are badly broken, rendering the simple proportionality to tree amplitudes invalid, and reinforce the necessity of treating the EWP amplitudes as fully independent parameters in the TDA analyses~\cite{He:2018php,beneketalk,Shi:2025eyp}. Consequently, any analysis that imposes the EWP--tree relations a priori runs the risk of omitting the dominant dynamical component of these amplitudes.

\subsection{Pure annihilation decay modes}

Our fit also provides interesting information on pure annihilation decay modes, such as $\bar{B}_s\to\pi^+\pi^-$ and $\bar{B}^0\to K^+K^-$. Although these channels have not yet been measured with high precision, our prediction for their branching fractions can be already compared with the current data. The fitted bare annihilation parameters, such as $\tilde{E}$, $\tilde{A}$, are found to be of $\mathcal{O}(10)$, which can be translated into the physical amplitudes with expected sizes after being multiplied by the channel‑dependent factor $B_{M_1M_2} \approx 4.4 \times 10^{-3} A_{M_1M_2}$. As can be seen from table~\ref{tab:resultsBRI}, the resulting branching fractions for these pure annihilation decay modes are all compatible with the current experimental measurements. This is in line with the earlier QCDF‑based analyses~\cite{Wang:2013fya,Chang:2014yma,Bobeth:2014rra}, which found that the annihilation parameters need to be sizable and non-universal to accommodate the data.

\subsection{Other decay modes and observables}

By checking the resulting pulls for the observables listed in tables~\ref{tab:resultsBRI}~--~\ref{tab:resultsMixCPIII}, one can see that there exist some small deviations between our fitted results and the experimental data for the direct CP asymmetries $A_{CP}(B^- \to \pi^-\bar{K}^0)$, $A_{CP}(\bar{B}^0 \to \pi^0\bar{K}^0)$, and $A_{CP}(\bar{B}_s \to K^+ K^-)$, as well as for the mixing-induced CP asymmetries $S_{\pi^0 \bar{K}^0}$ and $S_{\eta^\prime \bar{K}^0}$, which are, however, all below $1.5\sigma$.

Finally, with our best-fit point at hand, we also present in tables~\ref{tab:resultsBRI}~--~\ref{tab:resultsMixCPIII} our predictions for the observables of some decay modes, for which there are no measurements yet. We therefore expect that the future data from LHCb and Belle II experiments will be helpful in further refining our analysis.

\section{Conclusion}
\label{sec:conclusion}

The phenomenology of non-leptonic $B$ decays will remain in the focus of contemporary and future flavour physics programs, with the goal of further improving our qualitative and quantitative understanding of quark flavour mixing and CP violation. Given the complicated structure of the dynamics inside and between the purely hadronic initial and final states, it has remained notoriously difficult to describe the entire set of experimental data from first field-theoretic principles in a convincing manner. This circumstance motivated performing global fits to data in various parametrizations and under different assumptions.

In the present work, we have performed a global analysis of two-body charmless non-leptonic $B \to PP$ decays. In our parametrization, we have implemented flavour-$SU(3)$ breaking at the level of transition form factors, decay constants, and phase-space factors. During the fitting procedure, we find that the minimum of the $\chi^2$-function turns out to be largely degenerate. Yet, we find a point in parameter space that describes the data in a very satisfactory manner. When translated to the QCD-factorization amplitudes, the central values resemble many features of the dynamical predictions obtained within the QCDF framework at NNLO, for instance the colour-allowed tree amplitude and the real part of the colour-suppressed tree amplitude. Also the small differences between up- and charm-initiated amplitudes that~--~based on the QCDF experience~--~are used as an assumption, get confirmed by the good quality of the fit. The EWP amplitudes come out quite sizable at first glance. However, we give a power-counting-based argument for why their magnitude can be roughly one order larger compared to the LO expectation. Still, it will be interesting to carry out a dedicated study in the future that reveals the dynamical mechanism behind this enhancement. On the other hand, for the annihilation amplitudes, which are analyzed in great detail, we do not find any strong indications that they are numerically enhanced beyond the naive $\Lambda_{\textrm{QCD}}/m_b$ scaling.

Finally, we have addressed a number of phenomenological applications, among which are two variants of the $K\pi$ puzzle, the role of mixing-induced CP asymmetries, the EWP--tree relations, and the pure annihilation modes. The good quality of our fit also implies that the robust relations for charmless non-leptonic $B \to PP$ decays are very well described by our obtained amplitudes. Last but not least, our results are used to predict yet unmeasured observables in certain channels, and it will be exciting to compare them with future data.

\acknowledgments

We would like to thank Anshika Bansal, Aritra Biswas, Thorsten Feldmann, Thomas Mannel, Di Wang, Wei Wang, and Fu-Sheng Yu for useful discussions. This research was supported by the Deutsche Forschungsgemeinschaft (DFG, German Research Foundation) under grant 396021762 --- TRR 257 ``Particle Physics Phenomenology after the Higgs Discovery''. We also acknowledge support from the Deutsche Forschungsgemeinschaft (DFG, German Research Foundation) under Germany's Excellence Strategy~--~Cluster of Excellence ``Color meets Flavor'', EXC 3107~--~Project-ID 533766364. G.~T.-X.\ received support from the European Union's Horizon 2020 research and innovation programme under the Marie Sklodowska-Curie grant agreement No 945422. This work is also supported by the National Natural Science Foundation of China under Grant Nos.~12475094 and 12135006. It has also has been partially supported by STFC consolidated grant ST/X000664/1. The calculations of this work were partially performed on the computer cluster at KIT in Karlsruhe and the HPC cluster {\texttt{OMNI}} at Siegen University.

\begin{appendix}

\section{Amplitude coefficients}
\label{app:ampcoeffs}

Here we present the amplitude coefficients that we relegated from section~\ref{sec:physamps}.

\begin{table}[htbp]
\centering
\setlength{\tabcolsep}{0.015\textwidth} 
\renewcommand{\arraystretch}{1.22}
\begin{tabular}{|c c | c c c c c c c c c c|}
\hline
Channel & $M_1 M_2$
& $\tilde T$ & $\tilde C$ & $\tilde A$ & $\tilde E$
& $\tilde T_{PA}$ & $\tilde T_{ES}$ & $\tilde T_{AS}$ & $\tilde T_{SS}$
& $\tilde T_P$ & $\tilde T_S$ \\
\hline

\multirow{2}{*}{$B^- \to \pi^- \pi^0$}
& $\pi^- \pi^0$
& 0 & $\frac{1}{\sqrt{2}}$ & $-\frac{1}{\sqrt{2}}$ & 0
& 0 & 0 & 0 & 0 & $-\frac{1}{\sqrt{2}}$ & 0 \\
& $\pi^0 \pi^-$
& $\frac{1}{\sqrt{2}}$ & 0 & $\frac{1}{\sqrt{2}}$ & 0
& 0 & 0 & 0 & 0 & $\frac{1}{\sqrt{2}}$ & 0 \\
\hline

\multirow{2}{*}{$\bar{B}^0 \to \pi^+ \pi^-$}
& $\pi^+ \pi^-$
& 1 & 0 & 0 & 0
& 1 & 0 & 0 & 0 & 1 & 0 \\
& $\pi^- \pi^+$
& 0 & 0 & 0 & 1
& 1 & 0 & 0 & 0 & 0 & 0 \\
\hline

\multirow{2}{*}{$\bar{B}^0 \to \pi^0 \pi^0$}
& $\pi^0 \pi^0$
& 0 & $-\frac12$ & 0 & $\frac12$
& 1 & 0 & 0 & 0 & $\frac12$ & 0 \\
& $\pi^0 \pi^0$
& 0 & $-\frac12$ & 0 & $\frac12$
& 1 & 0 & 0 & 0 & $\frac12$ & 0 \\
\hline

\multirow{2}{*}{$B^- \to K^0 K^-$}
& $K^0 K^-$
& 0 & 0 & 0 & 0
& 0 & 0 & 0 & 0 & 0 & 0 \\
& $K^- K^0$
& 0 & 0 & 1 & 0
& 0 & 0 & 0 & 0 & 1 & 0 \\
\hline

\multirow{2}{*}{$\bar{B}^0 \to K^+ K^-$}
& $K^+ K^-$
& 0 & 0 & 0 & 0
& 1 & 0 & 0 & 0 & 0 & 0 \\
& $K^- K^+$
& 0 & 0 & 0 & 1
& 1 & 0 & 0 & 0 & 0 & 0 \\
\hline

\multirow{2}{*}{$\bar{B}^0 \to \bar K^0 K^0$}
& $\bar K^0 K^0$
& 0 & 0 & 0 & 0
& 1 & 0 & 0 & 0 & 1 & 0 \\
& $K^0 \bar K^0$
& 0 & 0 & 0 & 0
& 1 & 0 & 0 & 0 & 0 & 0 \\
\hline

\multirow{2}{*}{$\bar{B}_s \to \pi^- K^+$}
& $\pi^- K^+$
& 0 & 0 & 0 & 0
& 0 & 0 & 0 & 0 & 0 & 0 \\
& $K^+ \pi^-$
& 1 & 0 & 0 & 0
& 0 & 0 & 0 & 0 & 1 & 0 \\
\hline

\multirow{2}{*}{$\bar{B}_s \to \pi^0 K^0$}
& $\pi^0 K^0$
& 0 & 0 & 0 & 0
& 0 & 0 & 0 & 0 & 0 & 0 \\
& $K^0 \pi^0$
& 0 & $\frac{1}{\sqrt{2}}$ & 0 & 0
& 1 & 0 & 0 & 0 & $-\frac{1}{\sqrt{2}}$ & 0 \\
\hline

\multirow{2}{*}{$B^- \to \pi^- \bar{K}^0$}
& $\pi^- \bar{K}^0$
& 0 & 0 & 1 & 0
& 0 & 0 & 0 & 0 & 1 & 0 \\
& $\bar{K}^0 \pi^-$
& 0 & 0 & 0 & 0
& 0 & 0 & 0 & 0 & 0 & 0 \\
\hline

\multirow{2}{*}{$B^- \to \pi^0 K^-$}
& $\pi^0 K^-$
& $\frac{1}{\sqrt{2}}$ & 0 & $\frac{1}{\sqrt{2}}$ & 0
& 0 & 0 & 0 & 0 & $\frac{1}{\sqrt{2}}$ & 0 \\
& $K^- \pi^0$
& 0 & $\frac{1}{\sqrt{2}}$ & 0 & 0
& 0 & 0 & 0 & 0 & 0 & 0 \\
\hline

\multirow{2}{*}{$\bar{B}^0 \to \pi^+ K^-$}
& $\pi^+ K^-$
& 1 & 0 & 0 & 0
& 0 & 0 & 0 & 0 & 1 & 0 \\
& $K^- \pi^+$
& 0 & 0 & 0 & 0
& 0 & 0 & 0 & 0 & 0 & 0 \\
\hline

\multirow{2}{*}{$\bar{B}^0 \to \pi^0 \bar K^0$}
& $\pi^0 \bar K^0$
& 0 & 0 & 0 & 0
& 0 & 0 & 0 & 0 & $-\frac{1}{\sqrt{2}}$ & 0 \\
& $\bar K^0 \pi^0$
& 0 & $\frac{1}{\sqrt{2}}$ & 0 & 0
& 0 & 0 & 0 & 0 & 0 & 0 \\
\hline

\multirow{2}{*}{$\bar{B}_s \to K^+ K^-$}
& $K^+ K^-$
& 1 & 0 & 0 & 0
& 1 & 0 & 0 & 0 & 1 & 0 \\
& $K^- K^+$
& 0 & 0 & 0 & 1
& 1 & 0 & 0 & 0 & 0 & 0 \\
\hline

\multirow{2}{*}{$\bar{B}_s \to \bar K^0 K^0$}
& $\bar K^0 K^0$
& $0$ & $0$ & $0$ & $0$
& $1$ & $0$ & $0$ & $0$ & $0$ & $0$ \\
& $K^0 \bar K^0$
& $0$ & $0$ & $0$ & $0$
& $1$ & $0$ & $0$ & $0$ & $1$ & $0$ \\
\hline

\multirow{2}{*}{$\bar{B}_s \to \pi^+ \pi^-$}
& $\pi^+ \pi^-$
& 0 & 0 & 0 & 0
& 1 & 0 & 0 & 0 & 0 & 0 \\
& $\pi^- \pi^+$
& 0 & 0 & 0 & 1
& 1 & 0 & 0 & 0 & 0 & 0 \\
\hline

\multirow{2}{*}{$\bar{B}_s \to \pi^0 \pi^0$}
& $\pi^0 \pi^0$
& $0$ & $0$ & $0$ & $0$
& $\frac12$ & 0 & 0 & 0 & 0 & 0 \\
& $\pi^0 \pi^0$
& 0 & 0 & 0 & $\frac12$
& 1 & 0 & 0 & 0 & 0 & 0 \\
\hline
\end{tabular}
\caption{Amplitudes for the tree sector with no $\eta^{(\prime)}$ mesons in the final state. \label{tab:tree-without eta}}
\end{table}

\begin{table}[htbp]
\centering
\setlength{\tabcolsep}{0.016\textwidth} 
\renewcommand{\arraystretch}{1.3}
\begin{tabular}{|c c | c c c c c c c c c c|}
\hline
Channel & $M_1 M_2$
& $\tilde T$ & $\tilde C$ & $\tilde A$ & $\tilde E$
& $\tilde T_{PA}$ & $\tilde T_{ES}$ & $\tilde T_{AS}$ & $\tilde T_{SS}$ & $\tilde T_P$ & $\tilde T_S$ \\
\hline

\multirow{2}{*}{$B^- \to \pi^- \eta_q$}
& $\pi^- \eta_q$
& 0 & $\tfrac{1}{\sqrt{2}}$ & $\tfrac{1}{\sqrt{2}}$ & 0
& 0 & $\sqrt{2}$ & 0 & 0 & $\tfrac{1}{\sqrt{2}}$ & $\sqrt{2}$ \\
& $\eta_q \pi^-$
& $\tfrac{1}{\sqrt{2}}$ & 0 & $\tfrac{1}{\sqrt{2}}$ & 0
& 0 & 0 & 0 & 0 & $\tfrac{1}{\sqrt{2}}$ & 0 \\
\hline

\multirow{2}{*}{$B^- \to \pi^- \eta_s$}
& $\pi^- \eta_s$
& 0 & 0 & 0 & 0
& 0 & 1 & 0 & 0 & 0 & 1 \\
& $\eta_s \pi^-$
& 0 & 0 & 0 & 0
& 0 & 0 & 0 & 0 & 0 & 0 \\
\hline

\multirow{2}{*}{$\bar{B}^0 \to \pi^0 \eta_q$}
& $\pi^0 \eta_q$
& 0 & $-\tfrac{1}{2}$ & 0 & $\tfrac{1}{2}$
& 0 & 0 & 1 & 0 & $-\tfrac{1}{2}$ & $-1$ \\
& $\eta_q \pi^0$
& 0 & $\tfrac{1}{2}$ & 0 & $\tfrac{1}{2}$
& 0 & 0 & 0 & 0 & $-\tfrac{1}{2}$ & 0 \\
\hline

\multirow{2}{*}{$\bar{B}^0 \to \pi^0 \eta_s$}
& $\pi^0 \eta_s$
& 0 & 0 & 0 & 0
& 0 & 0 & $\tfrac{1}{\sqrt{2}}$ & 0 & 0 & $-\tfrac{1}{\sqrt{2}}$ \\
& $\eta_s \pi^0$
& 0 & 0 & 0 & 0
& 0 & 0 & 0 & 0 & 0 & 0 \\
\hline

\multirow{2}{*}{$\bar{B}_s \to K^0 \eta_q$}
& $K^0 \eta_q$
& 0 & $\tfrac{1}{\sqrt{2}}$ & 0 & 0
& 0 & 0 & 0 & 0 & $\tfrac{1}{\sqrt{2}}$ & $\sqrt{2}$ \\
& $\eta_q K^0$
& 0 & 0 & 0 & 0
& 0 & 0 & 0 & 0 & 0 & 0 \\
\hline

\multirow{2}{*}{$\bar{B}_s \to K^0 \eta_s$}
& $K^0 \eta_s$
& 0 & 0 & 0 & 0
& 0 & 0 & 0 & 0 & 0 & 1 \\
& $\eta_s K^0$
& 0 & 0 & 0 & 0
& 0 & 0 & 0 & 0 & 1 & 0 \\
\hline


\multirow{2}{*}{$B^- \to K^- \eta_q$}
& $K^- \eta_q$
& 0 & $\tfrac{1}{\sqrt{2}}$ & 0 & 0
& 0 & $\sqrt{2}$ & 0 & 0 & 0 & $\sqrt{2}$ \\
& $\eta_q K^-$
& $\tfrac{1}{\sqrt{2}}$ & 0 & $\tfrac{1}{\sqrt{2}}$ & 0
& 0 & 0 & 0 & 0 & $\tfrac{1}{\sqrt{2}}$ & 0 \\
\hline

\multirow{2}{*}{$B^- \to K^- \eta_s$}
& $K^- \eta_s$
& 0 & 0 & 1 & 0
& 0 & 1 & 0 & 0 & 1 & 1 \\
& $\eta_s K^-$
& 0 & 0 & 0 & 0
& 0 & 0 & 0 & 0 & 0 & 0 \\
\hline

\multirow{2}{*}{$\bar{B}^0 \to \bar{K}^0 \eta_q$}
& $\bar{K}^0 \eta_q$
& 0 & $\tfrac{1}{\sqrt{2}}$ & 0 & 0
& 0 & 0 & 0 & 0 & 0 & $\sqrt{2}$ \\
& $\eta_q \bar{K}^0$
& 0 & 0 & 0 & 0 
& 0 & 0 & 0 & 0 & $\tfrac{1}{\sqrt{2}}$ & 0 \\
\hline

\multirow{2}{*}{$\bar{B}^0 \to \bar{K}^0 \eta_s$}
& $\bar{K}^0 \eta_s$
& 0 & 0 & 0 & 0
& 0 & 0 & 0 & 0 & 1 & 1 \\
& $\eta_s \bar{K}^0$
& 0 & 0 & 0 & 0
& 0 & 0 & 0 & 0 & 0 & 0 \\
\hline

\multirow{2}{*}{$\bar{B}_s \to \pi^0 \eta_q$}
& $\pi^0 \eta_q$
& 0 & 0 & 0 & $\tfrac{1}{2}$
& 0 & 0 & 1 & 0 & 0 & 0 \\
& $\eta_q \pi^0$
& 0 & 0 & 0 & $\tfrac{1}{2}$
& 0 & 0 & 0 & 0 & 0 & 0 \\
\hline

\multirow{2}{*}{$\bar{B}_s \to \pi^0 \eta_s$}
& $\pi^0 \eta_s$
& 0 & 0 & 0 & 0
& 0 & 0 & $\tfrac{1}{\sqrt{2}}$ & 0 & 0 & 0 \\
& $\eta_s \pi^0$
& 0 & $\tfrac{1}{\sqrt{2}}$ & 0 & 0
& 0 & 0 & 0 & 0 & 0 & 0 \\
\hline
\end{tabular}
\caption{Amplitudes for the tree sector with one $\eta$ meson in the final state, in the FKS scheme. The same expressions hold for the cases with one $\eta^{\prime}$ meson. \label{tab:tree-with one eta}}
\end{table}

\begin{table}[htbp]
\centering
\setlength{\tabcolsep}{0.019\textwidth} 
\renewcommand{\arraystretch}{1.3}
\begin{tabular}{|c c|cccccccccc|}
\hline
Channel & $M_1 M_2$
& $\tilde T$ & $\tilde C$ & $\tilde A$ & $\tilde E$
& $\tilde T_{PA}$ & $\tilde T_{ES}$ & $\tilde T_{AS}$ & $\tilde T_{SS}$
& $\tilde T_P$ & $\tilde T_S$ \\
\hline

$B^0 \to \eta_q \eta_q$
& $\eta_q\eta_q$
& 0 & 1 & 0 & 1
& 2 & 0 & 2 & 4
& 1 & 2 \\
\hline

\multirow{2}{*}{$B^0 \to \eta_q \eta_s$}
& $\eta_q\eta_s$
& 0 & 0 & 0 & 0
& 0 & 0 & $\tfrac{1}{\sqrt2}$ & $\sqrt2$
& 0 & $\tfrac{1}{\sqrt2}$ \\
& $\eta_s\eta_q$
& 0 & 0 & 0 & 0
& 0 & 0 & 0 & $\sqrt{2}$ & 0 & 0 \\
\hline

$B^0 \to \eta_s \eta_s$
& $\eta_s\eta_s$
& 0 & 0 & 0 & 0
& 2 & 0 & 0 & 2
& 0 & 0 \\
\hline

$\bar{B}_s \to \eta_q \eta_q$
& $\eta_q\eta_q$
& 0 & 0 & 0 & 1
& 2 & 0 & 2 & 4
& 0 & 0 \\
\hline

\multirow{2}{*}{$\bar{B}_s \to \eta_q \eta_s$}
& $\eta_q \eta_{s}$
& 0 & 0 & 0 & 0
& 0 & 0 & $\tfrac{1}{\sqrt{2}}$ & $\sqrt{2}$ & 0 & 0 \\
& $\eta_{s} \eta_q$
& 0 & $\tfrac{1}{\sqrt{2}}$ & 0 & 0
& 0 & 0 & 0 & $\sqrt{2}$ & 0 & $\sqrt{2}$ \\
\hline

$\bar{B}_s \to \eta_s \eta_s$
& $\eta_s\eta_s$
& 0 & 0 & 0 & 0
& 2 & 0 & 0 & 2
& 2 & 2 \\
\hline
\end{tabular}
\caption{Amplitudes for the tree sector with two $\eta$ mesons in the final state, in the FKS scheme. The same expressions hold for the cases with two $\eta^{\prime}$ mesons in the final state.}
\end{table}

\begin{table}[htbp]
\centering
\setlength{\tabcolsep}{0.019\textwidth} 
\renewcommand{\arraystretch}{1.3}
\begin{tabular}{|c c | c c c c c c c c c c|}
\hline
Channel & $M_1 M_2$
& $\tilde T$ & $\tilde C$ & $\tilde A$ & $\tilde E$
& $\tilde T_{PA}$ & $\tilde T_{ES}$ & $\tilde T_{AS}$ & $\tilde T_{SS}$ & $\tilde T_P$ & $\tilde T_S$ \\
\hline

\multirow{2}{*}{$\bar{B}^0 \to \eta_q \eta_{q'}$}
& $\eta_q \eta_{q'}$
& 0 & $\tfrac{1}{2}$ & 0 & $\tfrac{1}{2}$
& 1 & 0 & 1 & 2 & $\tfrac{1}{2}$ & 1 \\
& $\eta_{q'} \eta_q$
& 0 & $\tfrac{1}{2}$ & 0 & $\tfrac{1}{2}$
& 1 & 0 & 1 & 2 & $\tfrac{1}{2}$ & 1 \\
\hline

\multirow{2}{*}{$\bar{B}^0 \to \eta_q \eta_{s'}$}
& $\eta_q \eta_{s'}$
& 0 & 0 & 0 & 0
& 0 & 0 & $\tfrac{1}{\sqrt{2}}$ & $\sqrt{2}$ & 0 & $\tfrac{1}{\sqrt{2}}$ \\
& $\eta_{s'} \eta_q$
& 0 & 0 & 0 & 0
& 0 & 0 & 0 & $\sqrt{2}$ & 0 & 0 \\
\hline

\multirow{2}{*}{$\bar{B}^0 \to \eta_s \eta_{q'}$}
& $\eta_s \eta_{q'}$
& 0 & 0 & 0 & 0
& 0 & 0 & 0 & $\sqrt{2}$ & 0 & 0 \\
& $\eta_{q'} \eta_s$
& 0 & 0 & 0 & 0
& 0 & 0 & $\tfrac{1}{\sqrt{2}}$ & $\sqrt{2}$ & 0 & $\tfrac{1}{\sqrt{2}}$ \\
\hline

\multirow{2}{*}{$\bar{B}^0 \to \eta_s \eta_{s'}$}
& $\eta_s \eta_{s'}$
& 0 & 0 & 0 & 0
& 1 & 0 & 0 & 1 & 0 & 0 \\
& $\eta_{s'} \eta_s$
& 0 & 0 & 0 & 0
& 1 & 0 & 0 & 1 & 0 & 0 \\
\hline

\multirow{2}{*}{$\bar{B}_s \to \eta_q \eta_{q'}$}
& $\eta_q \eta_{q'}$
& 0 & 0 & 0 & $\tfrac{1}{2}$
& 1 & 0 & 1 & 2 & 0 & 0 \\
& $\eta_{q'} \eta_q$
& 0 & 0 & 0 & $\tfrac{1}{2}$
& 1 & 0 & 1 & 2 & 0 & 0 \\
\hline

\multirow{2}{*}{$\bar{B}_s \to \eta_q \eta_{s'}$}
& $\eta_q \eta_{s'}$
& 0 & 0 & 0 & 0
& 0 & 0 & $\tfrac{1}{\sqrt{2}}$ & $\sqrt{2}$ & 0 & 0 \\
& $\eta_{s'} \eta_q$
& 0 & $\tfrac{1}{\sqrt{2}}$ & 0 & 0
& 0 & 0 & 0 & $\sqrt{2}$ & 0 & $\sqrt{2}$ \\
\hline

\multirow{2}{*}{$\bar{B}_s \to \eta_s \eta_{q'}$}
& $\eta_s \eta_{q'}$
& 0 & $\tfrac{1}{\sqrt{2}}$ & 0 & 0
& 0 & 0 & 0 & $\sqrt{2}$ & 0 & $\sqrt{2}$ \\
& $\eta_{q'} \eta_s$
& 0 & 0 & 0 & 0
& 0 & 0 & $\tfrac{1}{\sqrt{2}}$ & $\sqrt{2}$ & 0 & 0 \\
\hline

\multirow{2}{*}{$\bar{B}_s \to \eta_s \eta_{s'}$}
& $\eta_s \eta_{s'}$
& 0 & 0 & 0 & 0
& 1 & 0 & 0 & 1 & 1 & 1 \\
& $\eta_{s'} \eta_s$
& 0 & 0 & 0 & 0
& 1 & 0 & 0 & 1 & 1 & 1 \\
\hline
\end{tabular}
\caption{Amplitudes for the tree sector with one $\eta$ and one $\eta^{\prime}$ meson in the final state, in the FKS scheme. \label{tab:tree-with eta and eta'}}
\end{table}

\begin{table}[htbp]
\centering
\setlength{\tabcolsep}{0.013\textwidth} 
\renewcommand{\arraystretch}{1.22}
\begin{tabular}{|c c | c c c c c c c c c c|}
\hline
Channel & $M_1 M_2$
& $\tilde{P}_T$ & $\tilde{P}_C$ & $\tilde{P}_{TA}$ & $\tilde{P}$
& $\tilde{P}_{TE}$ & $\tilde{P}_{A}$ & $\tilde{P}_{AS}$
& $\tilde{P}_{ES}$ & $\tilde{P}_{SS}$ & $\tilde{S}$ \\
\hline

\multirow{2}{*}{$B^- \to \pi^- \pi^0$}
& $\pi^- \pi^0$
& $0$ & $-\tfrac{1}{\sqrt2}$ & $\tfrac{1}{\sqrt2}$ & $\tfrac{1}{\sqrt2}$
& $0$ & $0$ & $0$ & $0$ & $0$ & $0$ \\
& $\pi^0 \pi^-$
& $-\tfrac{1}{\sqrt2}$ & $0$ & $-\tfrac{1}{\sqrt2}$ & $-\tfrac{1}{\sqrt2}$
& $0$ & $0$ & $0$ & $0$ & $0$ & $0$ \\
\hline

\multirow{2}{*}{$\bar{B}^0 \to \pi^+ \pi^-$}
& $\pi^+ \pi^-$
& $-1$ & $0$ & $0$ & $-1$ 
& $0$ & $-1$ & $0$ & $0$ & $0$ & $0$ \\
& $\pi^- \pi^+$
& $0$ & $0$ & $0$ & $0$ 
& $-1$ & $-1$ & $0$ & $0$ & $0$ & $0$ \\
\hline

\multirow{2}{*}{$\bar{B}^0 \to \pi^0 \pi^0$}
& $\pi^0 \pi^0$
& $0$ & $\tfrac{1}{2}$ & $0$ & $-\tfrac{1}{2}$ 
& $-\tfrac{1}{2}$ & $-1$ & $0$ & $0$ & $0$ & $0$ \\
& $\pi^0 \pi^0$
& $0$ & $\tfrac{1}{2}$ & $0$ & $-\tfrac{1}{2}$ 
& $-\tfrac{1}{2}$ & $-1$ & $0$ & $0$ & $0$ & $0$ \\
\hline

\multirow{2}{*}{$B^- \to K^0 K^-$}
& $K^0 K^-$
& $0$ & $0$ & $0$ & $0$
& $0$ & $0$ & $0$ & $0$ & $0$ & $0$ \\
& $K^- K^0$
& $0$ & $0$ & $-1$ & $-1$
& $0$ & $0$ & $0$ & $0$ & $0$ & $0$ \\
\hline

\multirow{2}{*}{$\bar{B}^0 \to K^+ K^-$}
& $K^+ K^-$
& $0$ & $0$ & $0$ & $0$
& $0$ & $-1$ & $0$ & $0$ & $0$ & $0$ \\
& $K^- K^+$
& $0$ & $0$ & $0$ & $0$
& $-1$ & $-1$ & $0$ & $0$ & $0$ & $0$ \\
\hline

\multirow{2}{*}{$\bar{B}^0 \to \bar{K}^0 K^0$}
& $\bar{K}^0 K^0$
& $0$ & $0$ & $0$ & $-1$
& $0$ & $-1$ & $0$ & $0$ & $0$ & $0$ \\
& $K^0 \bar{K}^0$
& $0$ & $0$ & $0$ & $0$
& $0$ & $-1$ & $0$ & $0$ & $0$ & $0$ \\
\hline

\multirow{2}{*}{$\bar{B}_s \to \pi^- K^+$}
& $\pi^- K^+$
& $0$ & $0$ & $0$ & $0$
& $0$ & $0$ & $0$ & $0$ & $0$ & $0$ \\
& $K^+ \pi^-$
& $-1$ & $0$ & $0$ & $-1$
& $0$ & $0$ & $0$ & $0$ & $0$ & $0$ \\
\hline

\multirow{2}{*}{$\bar{B}_s \to \pi^0 K^0$}
& $\pi^0 K^0$
& $0$ & $0$ & $0$ & $0$
& $0$ & $0$ & $0$ & $0$ & $0$ & $0$ \\
& $K^0 \pi^0$
& $0$ & $-\tfrac{1}{\sqrt2}$ & $0$ & $\tfrac{1}{\sqrt2}$
& $0$ & $0$ & $0$ & $0$ & $0$ & $0$ \\
\hline


\multirow{2}{*}{$B^- \to \pi^- \bar{K}^0$}
& $\pi^- \bar{K}^0$
& $0$ & $0$ & $-1$ & $-1$
& $0$ & $0$ & $0$ & $0$ & $0$ & $0$ \\
& $\bar{K}^0 \pi^-$
& $0$ & $0$ & $0$ & $0$
& $0$ & $0$ & $0$ & $0$ & $0$ & $0$ \\
\hline

\multirow{2}{*}{$B^- \to \pi^0 K^-$}
& $\pi^0 K^-$
& $-\frac{1}{\sqrt{2}}$ & $0$ & $-\frac{1}{\sqrt{2}}$ & $-\frac{1}{\sqrt{2}}$
& $0$ & $0$ & $0$ & $0$ & $0$ & $0$ \\
& $K^- \pi^0$
& $0$ & $-\frac{1}{\sqrt{2}}$ & $0$ & $0$
& $0$ & $0$ & $0$ & $0$ & $0$ & $0$ \\
\hline

\multirow{2}{*}{$\bar{B}^0 \to \pi^+ K^-$}
& $\pi^+ K^-$
& $-1$ & $0$ & $0$ & $-1$
& $0$ & $0$ & $0$ & $0$ & $0$ & $0$ \\
& $K^- \pi^+$
& $0$ & $0$ & $0$ & $0$
& $0$ & $0$ & $0$ & $0$ & $0$ & $0$ \\
\hline

\multirow{2}{*}{$\bar{B}^0 \to \pi^0 \bar K^0$}
& $\pi^0 \bar K^0$
& $0$ & $0$ & $0$ & $\frac{1}{\sqrt{2}}$
& $0$ & $0$ & $0$ & $0$ & $0$ & $0$ \\
& $\bar K^0 \pi^0$
& $0$ & $-\frac{1}{\sqrt{2}}$ & $0$ & $0$
& $0$ & $0$ & $0$ & $0$ & $0$ & $0$ \\
\hline

\multirow{2}{*}{$\bar{B}_s \to K^+ K^-$}
& $K^+ K^-$
& $-1$ & $0$ & $0$ & $-1$
& $0$ & $-1$ & $0$ & $0$ & $0$ & $0$ \\
& $K^- K^+$
& $0$ & $0$ & $0$ & $0$
& $-1$ & $-1$ & $0$ & $0$ & $0$ & $0$ \\
\hline

\multirow{2}{*}{$\bar{B}_s \to \bar K^0 K^0$}
& $\bar K^0 K^0$
& $0$ & $0$ & $0$ & $0$
& $0$ & $-1$ & $0$ & $0$ & $0$ & $0$ \\
& $K^0 \bar K^0$
& $0$ & $0$ & $0$ & $-1$
& $0$ & $-1$ & $0$ & $0$ & $0$ & $0$ \\
\hline

\multirow{2}{*}{$\bar{B}_s \to \pi^+ \pi^-$}
& $\pi^+ \pi^-$
& $0$ & $0$ & $0$ & $0$
& $0$ & $-1$ & $0$ & $0$ & $0$ & $0$ \\
& $\pi^- \pi^+$
& $0$ & $0$ & $0$ & $0$
& $-1$ & $-1$ & $0$ & $0$ & $0$ & $0$ \\
\hline

\multirow{2}{*}{$\bar{B}_s \to \pi^0 \pi^0$}
& $\pi^0 \pi^0$
& $0$ & $0$ & $0$ & $0$
& $-\tfrac{1}{2}$ & $-1$ & $0$ & $0$ & $0$ & $0$ \\
& $\pi^0 \pi^0$
& $0$ & $0$ & $0$ & $0$
& $-\tfrac{1}{2}$ & $-1$ & $0$ & $0$ & $0$ & $0$ \\
\hline
\end{tabular}
\caption{Amplitudes for the penguin sector with no $\eta^{(\prime)}$ mesons in the final state. \label{tab:penguin-without eta}}
\end{table}

\begin{table}[htbp]
\centering
\setlength{\tabcolsep}{0.0115\textwidth} 
\renewcommand{\arraystretch}{1.3}
\begin{tabular}{|c c | c c c c c c c c c c|}
\hline
Channel & $M_1 M_2$
& $\tilde{P}_T$ & $\tilde{P}_C$ & $\tilde{P}_{TA}$ & $\tilde{P}$
& $\tilde{P}_{TE}$ & $\tilde{P}_{A}$ & $\tilde{P}_{AS}$
& $\tilde{P}_{ES}$ & $\tilde{P}_{SS}$ & $\tilde{S}$ \\
\hline

\multirow{2}{*}{$B^- \to \pi^- \eta_q$}
& $\pi^- \eta_q$
& $0$ & $-\tfrac{1}{\sqrt2}$ & $-\tfrac{1}{\sqrt2}$ & $-\tfrac{1}{\sqrt2}$
& $0$ & $0$ & $0$ & $-\sqrt2$ & $0$ & $-\sqrt2$ \\
& $\eta_q \pi^-$
& $-\tfrac{1}{\sqrt2}$ & $0$ & $-\tfrac{1}{\sqrt2}$ & $-\tfrac{1}{\sqrt2}$
& $0$ & $0$ & $0$ & $0$ & $0$ & $0$ \\
\hline

\multirow{2}{*}{$B^- \to \pi^- \eta_s$}
& $\pi^- \eta_s$
& $0$ & $0$ & $0$ & $0$
& $0$ & $0$ & $0$ & $-1$ & $0$ & $-1$ \\
& $\eta_s \pi^-$
& $0$ & $0$ & $0$ & $0$
& $0$ & $0$ & $0$ & $0$ & $0$ & $0$ \\
\hline

\multirow{2}{*}{$\bar{B}^0 \to \pi^0 \eta_q$}
& $\pi^0 \eta_q$
& $0$ & $\tfrac{1}{2}$ & $0$ & $\tfrac{1}{2}$
& $-\tfrac{1}{2}$ & $0$ & $-1$ & $0$ & $0$ & $1$ \\
& $\eta_q \pi^0$
& $0$ & $-\tfrac{1}{2}$ & $0$ & $\tfrac{1}{2}$
& $-\tfrac{1}{2}$ & $0$ & $0$ & $0$ & $0$ & $0$ \\
\hline

\multirow{2}{*}{$\bar{B}^0 \to \pi^0 \eta_s$}
& $\pi^0 \eta_s$
& $0$ & $0$ & $0$ & $0$
& $0$ & $0$ & $-\tfrac{1}{\sqrt{2}}$ & $0$ & $0$ & $\tfrac{1}{\sqrt{2}}$ \\
& $\eta_s \pi^0$
& $0$ & $0$ & $0$ & $0$
& $0$ & $0$ & $0$ & $0$ & $0$ & $0$ \\
\hline

\multirow{2}{*}{$\bar{B}_s \to K^0 \eta_q$}
& $K^0 \eta_q$
& $0$ & $-\tfrac{1}{\sqrt2}$ & $0$ & $-\tfrac{1}{\sqrt2}$
& $0$ & $0$ & $0$ & $0$ & $0$ & $-\sqrt2$ \\
& $\eta_q K^0$
& $0$ & $0$ & $0$ & $0$
& $0$ & $0$ & $0$ & $0$ & $0$ & $0$ \\
\hline

\multirow{2}{*}{$\bar{B}_s \to K^0 \eta_s$}
& $K^0 \eta_s$
& $0$ & $0$ & $0$ & $0$
& $0$ & $0$ & $0$ & $0$ & $0$ & $-1$ \\
& $\eta_s K^0$
& $0$ & $0$ & $0$ & $-1$
& $0$ & $0$ & $0$ & $0$ & $0$ & $0$ \\
\hline

\multirow{2}{*}{$B^- \to K^- \eta_q$}
& $K^- \eta_q$
& $0$ & $-\tfrac{1}{\sqrt2}$ & $0$ & $0$
& $0$ & $0$ & $0$ & $-\sqrt2$ & $0$ & $-\sqrt2$ \\
& $\eta_q K^-$
& $-\tfrac{1}{\sqrt2}$ & $0$ & $-\tfrac{1}{\sqrt2}$ & $-\tfrac{1}{\sqrt2}$
& $0$ & $0$ & $0$ & $0$ & $0$ & $0$ \\
\hline

\multirow{2}{*}{$B^- \to K^- \eta_s$}
& $K^- \eta_s$
& $0$ & $0$ & $-1$ & $-1$
& $0$ & $0$ & $0$ & $-1$ & $0$ & $-1$ \\
& $\eta_s K^-$
& $0$ & $0$ & $0$ & $0$
& $0$ & $0$ & $0$ & $0$ & $0$ & $0$ \\
\hline

\multirow{2}{*}{$\bar{B}^0 \to \bar K^0 \eta_q$}
& $\bar K^0 \eta_q$
& $0$ & $-\tfrac{1}{\sqrt2}$ & $0$ & $0$
& $0$ & $0$ & $0$ & $0$ & $0$ & $-\sqrt2$ \\
& $\eta_q \bar K^0$
& $0$ & $0$ & $0$ & $-\tfrac{1}{\sqrt2}$
& $0$ & $0$ & $0$ & $0$ & $0$ & $0$ \\
\hline

\multirow{2}{*}{$\bar{B}^0 \to \bar K^0 \eta_s$}
& $\bar K^0 \eta_s$
& $0$ & $0$ & $0$ & $-1$
& $0$ & $0$ & $0$ & $0$ & $0$ & $-1$ \\
& $\eta_s \bar K^0$
& $0$ & $0$ & $0$ & $0$
& $0$ & $0$ & $0$ & $0$ & $0$ & $0$ \\
\hline

\multirow{2}{*}{$\bar{B}_s \to \pi^0 \eta_q$}
& $\pi^0 \eta_q$
& $0$ & $0$ & $0$ & $0$
& $-\tfrac{1}{2}$ & $0$ & $-1$ & $0$ & $0$ & $0$ \\
& $\eta_q \pi^0$
& $0$ & $0$ & $0$ & $0$
& $-\tfrac{1}{2}$ & $0$ & $0$ & $0$ & $0$ & $0$ \\
\hline

\multirow{2}{*}{$\bar{B}_s \to \pi^0 \eta_s$}
& $\pi^0 \eta_s$
& $0$ & $0$ & $0$ & $0$
& $0$ & $0$ & $-\tfrac{1}{\sqrt{2}}$ & $0$ & $0$ & $0$ \\
& $\eta_s \pi^0$
& $0$ & $-\tfrac{1}{\sqrt{2}}$ & $0$ & $0$
& $0$ & $0$ & $0$ & $0$ & $0$ & $0$ \\
\hline

\end{tabular}
\caption{Amplitudes for the penguin sector with one $\eta$ meson in the final state. The same expressions hold for the cases with one $\eta^{\prime}$ meson. \label{tab:penguin-with one eta}}
\end{table}

\begin{table}[htbp]
\centering
\setlength{\tabcolsep}{0.014\textwidth} 
\renewcommand{\arraystretch}{1.3}
\begin{tabular}{|c c|cccccccccc|}
\hline
Channel & $M_1 M_2$
& $\tilde{P}_T$ & $\tilde{P}_C$ & $\tilde{P}_{TA}$ & $\tilde{P}$
& $\tilde{P}_{TE}$ & $\tilde{P}_{A}$ & $\tilde{P}_{AS}$
& $\tilde{P}_{ES}$ & $\tilde{P}_{SS}$ & $\tilde{S}$ \\
\hline

$\bar{B}^0 \to \eta_q \eta_q$
& $\eta_q\eta_q$
& $0$ & $-1$ & $0$ & $-1$
& $-1$ & $-2$ & $-2$ & $0$ & $-4$ & $-2$ \\
\hline

\multirow{2}{*}{$\bar{B}^0 \to \eta_q \eta_s$}
& $\eta_q \eta_s$
& $0$ & $0$ & $0$ & $0$ 
& $0$ & $0$ & $-\tfrac{1}{\sqrt2}$ & $0$ & $-\sqrt2$ & $-\tfrac{1}{\sqrt2}$ \\
& $\eta_s \eta_q$
& $0$ & $0$ & $0$ & $0$ 
& $0$ & $0$ & $0$ & $0$ & $-\sqrt2$ & $0$ \\
\hline

$\bar{B}^0 \to \eta_s \eta_s$
& $\eta_s\eta_s$
& $0$ & $0$ & $0$ & $0$
& $0$ & $-2$ & $0$ & $0$ & $-2$ & $0$ \\
\hline

$\bar{B}_s \to \eta_q \eta_q$
& $\eta_q\eta_q$
& $0$ & $0$ & $0$ & $0$
& $-1$ & $-2$ & $-2$ & $0$ & $-4$ & $0$ \\
\hline

\multirow{2}{*}{$\bar{B}_s \to \eta_q \eta_s$}
& $\eta_q \eta_s$
& $0$ & $0$ & $0$ & $0$ 
& $0$ & $0$ & $-\tfrac{1}{\sqrt2}$ & $0$ & $-\sqrt2$ & $0$ \\
& $\eta_s \eta_q$
& $0$ & $-\tfrac{1}{\sqrt2}$ & $0$ & $0$ 
& $0$ & $0$ & $0$ & $0$ & $-\sqrt2$ & $-\sqrt2$ \\
\hline

$\bar{B}_s \to \eta_s \eta_s$
& $\eta_s\eta_s$
& $0$ & $0$ & $0$ & $-2$
& $0$ & $-2$ & $0$ & $0$ & $-2$ & $-2$ \\
\hline
\end{tabular}
\caption{Amplitudes for the penguin sector with two $\eta$ mesons in the final state. The same expressions hold for the cases with two $\eta^{\prime}$ mesons in the final state.}
\end{table}

\begin{table}[htbp]
\centering
\setlength{\tabcolsep}{0.014\textwidth} 
\renewcommand{\arraystretch}{1.3}
\begin{tabular}{|c c|cccccccccc|}
\hline
Channel & $M_1M_2$
& $\tilde{P}_T$ & $\tilde{P}_C$ & $\tilde{P}_{TA}$ & $\tilde{P}$
& $\tilde{P}_{TE}$ & $\tilde{P}_{A}$ & $\tilde{P}_{AS}$
& $\tilde{P}_{ES}$ & $\tilde{P}_{SS}$ & $\tilde{S}$ \\
\hline


\multirow{2}{*}{$\bar{B}^0\to \eta_q \eta_q'$}
& $\eta_q \eta_q'$ 
& $0$ & $-\tfrac12$ & $0$ & $-\tfrac12$ 
& $-\tfrac12$ & $-1$ & $-1$ & $0$ & $-2$ & $-1$ \\
& $\eta_q' \eta_q$
& $0$ & $-\tfrac12$ & $0$ & $-\tfrac12$ 
& $-\tfrac12$ & $-1$ & $-1$ & $0$ & $-2$ & $-1$\\
\hline

\multirow{2}{*}{$\bar{B}^0\to \eta_q \eta_s'$}
& $\eta_q \eta_s'$ 
& $0$ & $0$ & $0$ & $0$ 
& $0$ & $0$ & $-\tfrac{1}{\sqrt2}$ & $0$ & $-\sqrt2$ & $-\tfrac{1}{\sqrt2}$ \\
& $\eta_s' \eta_q$
& $0$ & $0$ & $0$ & $0$ 
& $0$ & $0$ & $0$ & $0$ & $-\sqrt2$ & $0$ \\
\hline

\multirow{2}{*}{$\bar{B}^0\to \eta_s \eta_q'$}
& $\eta_s \eta_q'$ 
& $0$ & $0$ & $0$ & $0$ 
& $0$ & $0$ & $0$ & $0$ & $-\sqrt2$ & $0$ \\
& $\eta_q'\eta_s$
& $0$ & $0$ & $0$ & $0$ 
& $0$ & $0$ & $-\tfrac{1}{\sqrt2}$ & $0$ & $-\sqrt2$ & $-\tfrac{1}{\sqrt2}$ \\
\hline

\multirow{2}{*}{$\bar{B}^0\to \eta_s \eta_s'$}
& $\eta_s \eta_s'$ 
& $0$ & $0$ & $0$ & $0$ 
& $0$ & $-1$ & $0$ & $0$ & $-1$ & $0$ \\
& $\eta_s' \eta_s$
& $0$ & $0$ & $0$ & $0$ 
& $0$ & $-1$ & $0$ & $0$ & $-1$ & $0$ \\
\hline


\multirow{2}{*}{$\bar{B}_s\to \eta_q \eta_q'$}
& $\eta_q \eta_q'$ 
& $0$ & $0$ & $0$ & $0$ 
& $-\tfrac12$ & $-1$ & $-1$ & $0$ & $-2$ & $0$ \\
& $\eta_q' \eta_q$
& $0$ & $0$ & $0$ & $0$ 
& $-\tfrac12$ & $-1$ & $-1$ & $0$ & $-2$ & $0$ \\
\hline

\multirow{2}{*}{$\bar{B}_s\to \eta_q \eta_s'$}
& $\eta_q \eta_s'$ 
& $0$ & $0$ & $0$ & $0$ 
& $0$ & $0$ & $-\tfrac{1}{\sqrt2}$ & $0$ & $-\sqrt2$ & $0$ \\
& $\eta_s' \eta_q$
& $0$ & $-\tfrac{1}{\sqrt2}$ & $0$ & $0$ 
& $0$ & $0$ & $0$ & $0$ & $-\sqrt2$ & $-\sqrt2$ \\
\hline

\multirow{2}{*}{$\bar{B}_s\to \eta_s \eta_q'$}
& $\eta_s \eta_q'$ 
& $0$ & $-\tfrac{1}{\sqrt2}$ & $0$ & $0$ 
& $0$ & $0$ & $0$ & $0$ & $-\sqrt2$ & $-\sqrt2$ \\
& $\eta_q' \eta_s$
& $0$ & $0$ & $0$ & $0$ 
& $0$ & $0$ & $-\tfrac{1}{\sqrt2}$ & $0$ & $-\sqrt2$ & $0$ \\
\hline

\multirow{2}{*}{$\bar{B}_s\to \eta_s \eta_s'$}
& $\eta_s \eta_s'$ 
& $0$ & $0$ & $0$ & $-1$ 
& $0$ & $-1$ & $0$ & $0$ & $-1$ & $-1$ \\
& $\eta_s' \eta_s$
& $0$ & $0$ & $0$ & $-1$ 
& $0$ & $-1$ & $0$ & $0$ & $-1$ & $-1$ \\
\hline
\end{tabular}
\caption{Amplitudes for the penguin sector with one $\eta$ and one $\eta^{\prime}$ meson in the final state, in the FKS scheme. \label{tab:penguin-with eta and eta'}}
\end{table}

\end{appendix}

\newpage

\bibliographystyle{JHEP}
\bibliography{paper}

\end{document}